\documentclass[amsmath,amssymb,reprint,floatfix]{revtex4-1}

\usepackage{amsmath}
\usepackage{amssymb}
\usepackage{graphicx}
\usepackage{dcolumn}
\usepackage{bm}

\usepackage[utf8]{inputenc}
\usepackage[T1]{fontenc}
\usepackage{mathptmx}
\usepackage{etoolbox}

\usepackage{subcaption}
\usepackage{xcolor}
\usepackage{tikz}
\usetikzlibrary{cd}
\usepackage{pgfplots}
\pgfplotsset{compat=1.18}
\usepackage{mathtools}
\usepackage{amsthm}
\usepackage{float}

\def\OToNa{{a}}
\def\OToNb{{b}}
\def\OToNc{{{c}}}
\def\OToNd{{{d}}}
\def\OToNe{{{e}}}
\def\OToNf{{{f}}}
\def\OToNg{{{g}}}
\def\OToNh{{{h}}}

\def\OneToNa{{{\underline{a}}}}
\def\OneToNb{{{\underline{b}}}}
\def\OneToNc{{{\underline{c}}}}
\def\OneToNd{{{\underline{d}}}}
\def\OneToNe{{{\underline{e}}}}
\def\OneToNf{{{\underline{f}}}}

\def\OToThreea{{{\mu}}}
\def\OToThreeb{{{\nu}}}
\def\OToThreec{{{\rho}}}

\def\OneToThreea{{{\underline{\mu}}}}

\def\Map{{\eta}}

\def\FrameName{{coordinate time }}

\definecolor{MacroparticleLine}{HTML}{BB4430}
\definecolor{FoliationOne}{HTML}{6592A4}
\definecolor{FoliationTwo}{HTML}{985D80}

\makeatletter
\def\@email#1#2{%
 \endgroup
 \patchcmd{\titleblock@produce}
 {\frontmatter@RRAPformat}
 {\frontmatter@RRAPformat{\produce@RRAP{*#1\href{mailto:#2}{#2}}}\frontmatter@RRAPformat}
 {}{}
}%
\makeatother
\begin{document}

\title[Moment tracking and their coordinate transformations for macroparticles]{Moment tracking and their coordinate transformations for macroparticles with an application to plasmas around black holes}

\author{Alexander Warwick}
 \email{a.warwick@lancaster.ac.uk, orcid:0009-0007-7300-9646}
\affiliation{ 
Physics Department, Lancaster University, Lancaster, LA1 4YB, UK.}
\affiliation{
The Cockcroft Institute, Daresbury Laboratory, Daresbury, WA4 4AD, UK.
}

\author{Jonathan Gratus}%
 \email{j.gratus@lancaster.ac.uk, orcid:0000-0003-1597-6084}
\affiliation{ 
Physics Department, Lancaster University, Lancaster, LA1 4YB, UK.}
\affiliation{
The Cockcroft Institute, Daresbury Laboratory, Daresbury, WA4 4AD, UK.
}

\date{\today}

\begin{abstract}
  Particle-in-cell codes usually represent large groups of particles as a single macroparticle. These codes are computationally efficient but lose information about the internal structure of the macroparticle. To improve the accuracy of these codes, this work presents a method in which, as well as tracking the macroparticle, the moments of the macroparticle are also tracked. Although the equations needed to track these moments are known, the coordinate transformations for moments where the space and time coordinates are mixed cannot be calculated using the standard method for representing moments. These coordinate transformations are important in astrophysical plasma, where there is no preferred coordinate system. This work uses the language of Schwartz distributions to calculate the coordinate transformations of moments. Both the moment tracking and coordinate transformation equations are tested by modelling the motion of uncharged particles in a circular orbit around a black hole in both Schwarzschild and Kruskal-Szekeres coordinates. Numerical testing shows that the error in tracking moments is small, and scales quadratically. This error can be improved by including higher order moments. By choosing an appropriate method for using these moments to deposit the charge back onto the grid, a full particle-in-cell code can be developed.
\end{abstract}

\maketitle

\section{Introduction}

\begin{figure}
	\centering
	\includegraphics[width=0.8\linewidth]{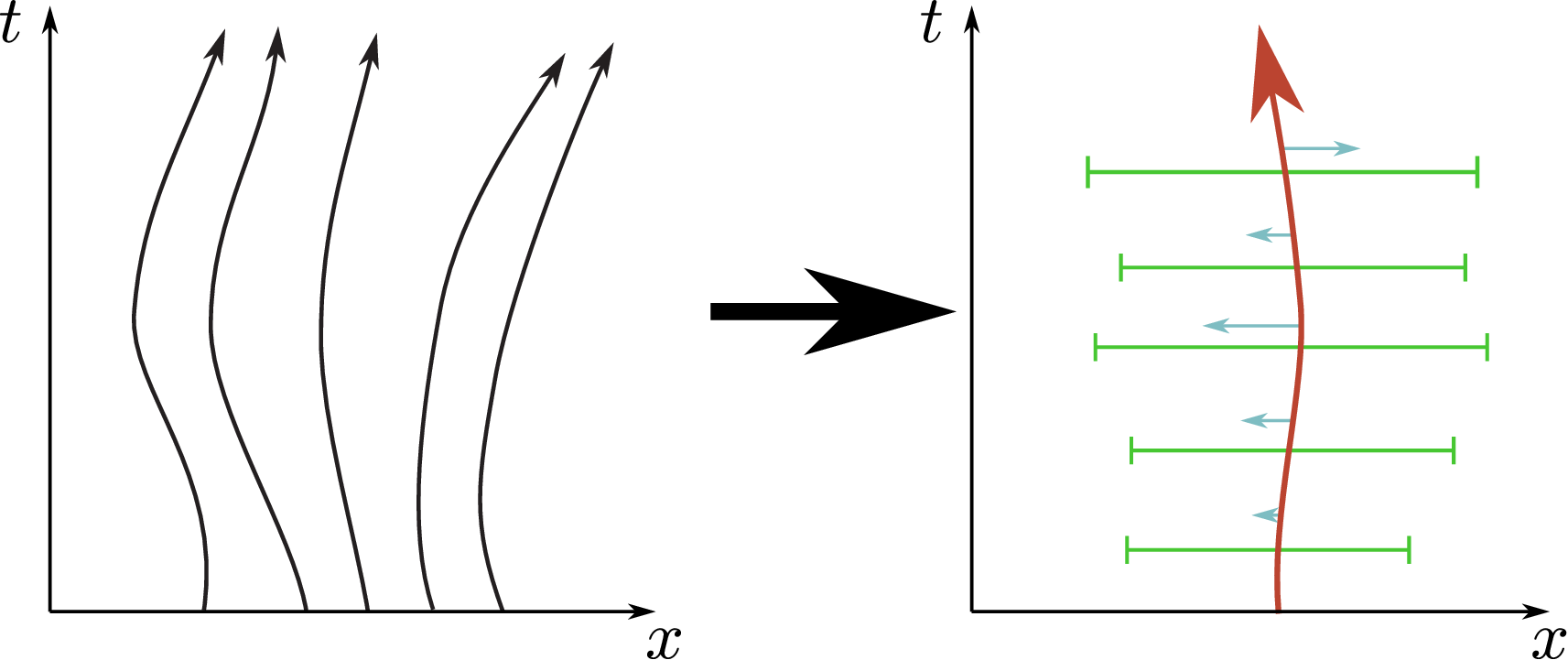}
	\caption{Tracking several individual particles compared to a macroparticle with moments. The 5 particles (the black lines) in the left diagram are replaced by a single macroparticle (the orange line) in the right diagram. By tracking the moments, quantities such as the difference between the centre of charge of the 5 particles and the position of the macroparticle (the first order moment, represented by the horizontal blue arrows), and the variance in position of the particles (the second order moment, represented by the green error bars) can be tracked.}
	\label{fig:MacroVsSuperMacro}
\end{figure}

In numerical simulations involving the dynamics of a large number of particles, for example, in a plasma, it is impossible to track the dynamics of every ion and electron. In particle-in-cell (PIC) codes, the plasma is modelled using macroparticles, where each simulated macroparticle represents a large number of actual particles. To improve the accuracy of a PIC code, there are two options: use more macroparticles, or give each macroparticle more information. A macroparticle typically has only a position and a velocity. This paper presents a different method, in which a macroparticle represents the moments of a group of particles \cite{dymnikovMomentMethodDynamics1978, channellMomentApproachCharged1983, channellMomentCodeBEDLAM1985, applebyMerlinFlexibleFeaturerich2022, shadwickGeneralMomentModel1999} (figure \ref{fig:MacroVsSuperMacro}). Such a model may be more efficient in cases where a large number of particles can be accurately modelled by only a small number of macroparticles and their moments, and the electromagnetic field does not vary much across the extent of the macroparticle. Additionally, in cases where the dominating interactive force can be calculated from the Li\'{e}nard-Wiechert potential, the electromagnetic fields can be calculated directly from the moments of the macroparticle \cite{ellisElectromagneticFieldsMoving1966}. \\

\begin{table*}
	\resizebox{\textwidth}{!}{%
		\begin{tabular}{|c|c|c|c|}
			\hline
			\textbf{\begin{tabular}[c]{@{}c@{}}Order of \\ moments tracked\end{tabular}} & \textbf{Information known}                     & \textbf{\begin{tabular}[c]{@{}c@{}}Number of\\ differential equations\\ to solve\end{tabular}} & \textbf{\begin{tabular}[c]{@{}c@{}} Number of \\ differential equations \\ to solve relative to \\ a standard macro-particle \end{tabular}} \\ \hline
				Monopole (zeroth order) & $\mathbf{X}^\mu$,$\mathbf{U}^\mu$ & 6 & 1
			\\ \hline
			Dipole (first order) & $\mathbf{X}^\mu$,$\mathbf{U}^\mu$, $V^a$ & 12 & 2
			\\ \hline
			Quadrupole (second order) & $\mathbf{X}^\mu$,$\mathbf{U}^\mu$, $V^a$, $V^{ab}$ & 33 & 5.5
			\\ \hline
			Octopole (third order) & $\mathbf{X}^\mu$,$\mathbf{U}^\mu$, $V^a$, $V^{ab}$, $V^{abc}$ & 89 &14.83
			\\ \hline
			Hexadecapole (fourth order) & $\mathbf{X}^\mu$,$\mathbf{U}^\mu$, $V^a$, $V^{ab}$, $V^{abc}$, $V^{abcd}$ & 215 &35.83
			\\ \hline
			
		\end{tabular}%
	}
	\caption{The amount of information in a macroparticle that also tracks moments compared to a standard macroparticle. This increased information results in more memory usage and a larger number of first order differential equations to solve. Since there are 6 differential equations per time step for a standard macroparticle, the fourth column is the third column divided by six.}
	\label{table:NumberOfMoments}
\end{table*}

There are several existing methods for tracking moments: the code MERLIN implements a transition matrix approach for particle accelerators \cite{applebyMerlinFlexibleFeaturerich2022} and a continuous model has been developed through a Hamiltonian approach \cite{shadwickGeneralMomentModel1999}. Moment tracking can also be done continuously by differentiating the definition of a moment and using the Vlasov equation \cite{channellMomentCodeBEDLAM1985, ackermannEfficientTimeIntegration2006}. In plasma physics, the concept of moment tracking is often used to describe hybrid-Vlasov approaches, where individual species of the plasma may be modelled through their bulk properties \cite{palmrothVlasovMethodsSpace2018, masonImplicitMomentParticle1981, bellFastelectronTransportHighintensity1997}. Moments in the context of hybrid-Vlasov approaches are constructed by integrating over velocity space only and can be interpreted as physical quantities such as temperature and pressure. In this work, moments shall be constructed by integrating over both velocity and position space, giving related, but different quantities. The transport of moments is also used in wider fields where the Liouville equation holds, such as particle nucleation \cite{hulburtProblemsParticleTechnology1964}, crystal growth \cite{grubinTransportLiouvilleEquation1993}, and nuclear collisions \cite{stroscioMomentequationRepresentationDissipative1986}. \\

To perfectly track the moments, an infinite set of moments is required. This is because, as is shown in this paper, higher order moments generate lower order moments. Tracking more moments is more computationally expensive, as these moments come with additional computational work (both more differential equations to solve each time step and more memory usage per macroparticle). This increased computational work is balanced by reducing the total number of macroparticles in the simulation. At the quadrupole (second) order there are 33 equations to solve each time step (21 quadrupole moments, 6 dipole (first order) moments, 3 components of velocity and 3 components of position). This expands to 215 equations if the expansion is carried out to the hexadecapole level (fourth order) (table \ref{table:NumberOfMoments}). To make moment tracking computationally feasible, a \emph{truncation} is required. This truncation is the highest order of moments considered, above which the contribution from higher order moments are neglected. Care must be taken to ensure the truncation is of as low an order as possible to minimise computational load, whilst also ensuring that neglecting the higher order moments does not significantly impact accuracy. This article calculates all quantities to quadrupole order, although all results presented can be generalised to arbitrary order. \\

\begin{figure}[tb]
	\begin{center}
    \resizebox{\linewidth}{!}{
		\begin{tikzpicture}[auto, transform shape]			
				\begin{axis}[
					no marks,
					xmin = 0,
					xmax = 1,
					ymin=0,
					ymax=1,
					axis x line = middle,
					axis y line = middle,
          ticks = none,
					xlabel = {$R$},
					ylabel = {$T$},
          every axis plot/.append style={ultra thick},
          clip mode = individual
					]
					\begin{scope}[samples=350]
						
						\pgfplotsset{cycle list shift=-1}
            \addplot+[dashed, color=black] {x};
						\addplot+[color=FoliationOne] {0.4*x};
						\addplot+[color=FoliationOne] {0.8*x} node [below, pos=0.7, rotate = 90, color=black]{$t = \text{constant}$};
						\addplot[domain=0:3,smooth,variable=\t, color = FoliationTwo] plot ({\t}, {0.3});
            \addplot[domain=0:3,smooth,variable=\t, color = FoliationTwo] plot ({\t}, {0.6});

            \addplot[domain=0:3,smooth,variable=\t, color=MacroparticleLine] plot ({0.4*sinh(\t)},{0.4*cosh(\t)});
            \addplot[domain=0:3,smooth,variable=\t, color=MacroparticleLine] plot ({0.4*cosh(\t)},{0.4*sinh(\t)});
            
					\end{scope}

        \draw (1, 0.3) node[right, color=black] {$T = \text{constant}$};
        \draw (1, 0.915) node[right, color=black] {$r = $ constant} ;
        \draw (1, 0.4) node[right, color=black] {$t =\text{constant}$}; 
        \draw (1, 1) node[right, color=black] {Event horizon}; 
   
				\end{axis}

		\end{tikzpicture}
 }
		\caption{A spacetime diagram in Kruskal-Szekeres coordinates of a particle travelling at a constant $r$ in Schwarzschild coordinates (the orange hyperbola). The diagonal blue lines are time slicings in Schwarzschild coordinates (constant $t$), and the horizontal purple lines are time slicings in Kruskal-Szekeres coordinates (constant $T$). The dashed black line is the event horizon. The moments in a given coordinate system have no components in the direction of the time slicing in that coordinate system.}
		\label{fig:DifferentFoliations}
	\end{center}
\end{figure}
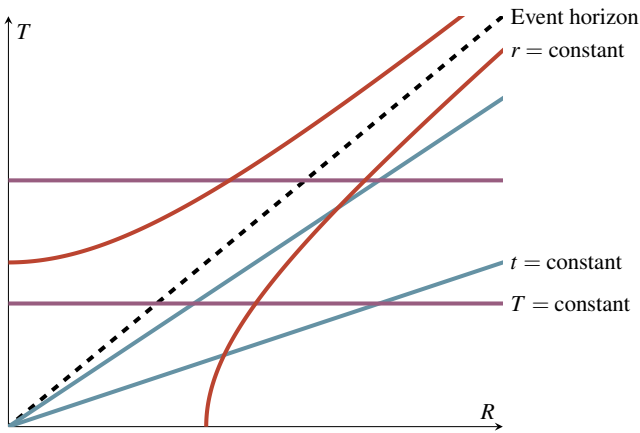

Recently there has been a focus on using PIC codes to model the dynamics of plasma around black holes \cite{ceruttiSIMULATIONSPARTICLEACCELERATION2013, crinquandParticleAccelerationKerr2021, crinquandMultidimensionalSimulationsErgospheric2020, parfreyFirstPrinciplesPlasmaSimulations2019}. Such plasmas may be important in active galactic nuclei, pulsars and gamma-ray bursts \cite{parfreyFirstPrinciplesPlasmaSimulations2019, rosenbergOnsetPlasmoidReconnection2021, philippovAbinitioPulsarMagnetosphere2018}. Ref. \cite{nishikawaPICMethodsAstrophysics2021} contains a full review of these studies. In such systems, there is not a preferred choice of coordinate system i.e. when modelling a static uncharged black hole, there is a choice to work in Schwarzschild coordinates, or Kruskal-Szekeres coordinates, amongst others. Because there is a choice in coordinate systems, it is useful to be able to transform between different coordinate systems, especially where the time and space coordinates are mixed together (such as the coordinate transform between Kruskal-Szekeres and Schwarzschild coordinates, shown in figure \ref{fig:DifferentFoliations}). Coordinate transformations that mix space and time coordinates also appear in particle accelerators. When simulating the motion of a linearly accelerating bunch in a particle accelerator, the transformation into the instantaneous rest frame of an accelerating bunch mixes space and time coordinates (this transformation is similar to the one shown in figure \ref{fig:DifferentFoliations}), so the spacetime coordinate transformations presented in this work are necessary. \\

\begin{figure}[tb]
	\centering
	\includegraphics[width=0.5\linewidth]{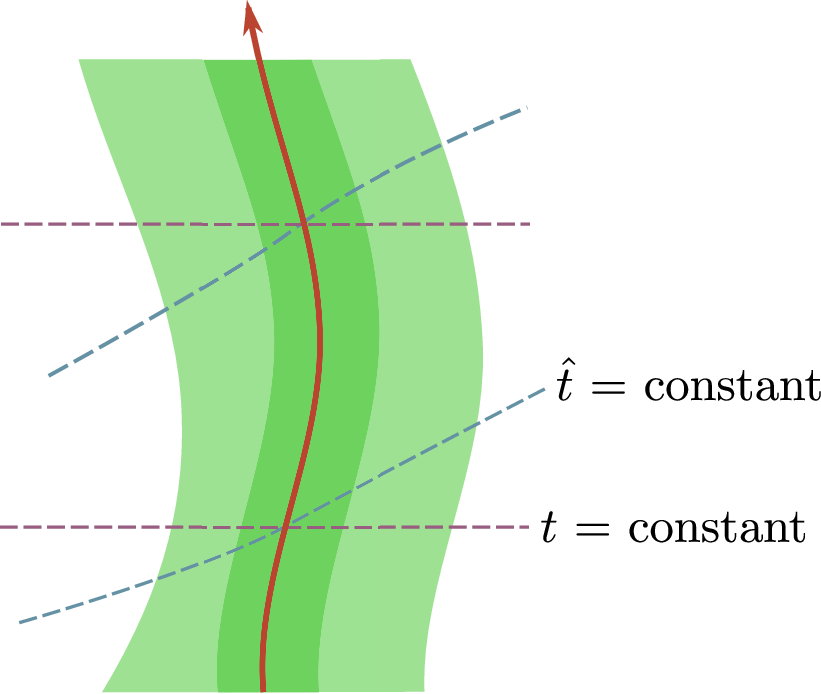}
	\caption{An example of two different time slicings. When taking moments, the time slicing is the hypersurface the moments are integrated over. By using a different time slicing (e.g. either the horizontal dashed purple lines $t$ or the curved dashed blue lines $\hat{t}$), the distance between the centre of the macroparticle (the orange line) and the nearby particles the macroparticle is representing (the green contours) is different, so different moments will be found.}
	\label{fig:differentfoliations2}
\end{figure}

 The moments of a macroparticle depend on the choice of time slicing (figure \ref{fig:differentfoliations2}). In all coordinate systems considered in this article, the global \emph{coordinate time} will be used as the time slicing. This time slicing is a foliation given by spatial hypersurfaces of constant coordinate time. In general, different coordinate systems will give different time slicings. This means when transforming between coordinate systems that mix temporal and spatial coordinates, for example, transforming between Schwarzschild and Kruskal-Szekeres coordinates, the time slicing will change (figure \ref{fig:DifferentFoliations}). This time slicing may be the global Killing timelike vector in relativistic scenarios, or the lab time in particle accelerators. Another possible time slicing is the backward light cone of an observer, which is the frame used when making astrophysical observations. A choice of time slicing commonly used in modelling plasma around black holes is found in the \emph{fiducial observer} (FIDO) scheme, where the time slicing is given relative to local moving observers \cite{thorneElectrodynamicsCurvedSpacetime1982,dodinVlasovEquationCollisionless2010}. A fourth possible time slicing is to take all the vectors orthogonal to the velocity of the world line. By using geodesics to propagate these to the world line, these can be used for the Dixon representation of a multipole \cite{gratusTensorialRepresentationDistributional2023}. \\

One of the key results of this article is the formula for the coordinate transformation of moments between coordinate systems that mix space and time coordinates. The standard integral representation of moments cannot transform moments between coordinate systems where the time slicing changes. Simply transforming the moments into the new coordinate system does not take into account the change in time slicing. If one were to use the standard representation of moments, an `effective' coordinate transformation can be performed. This is the process of reconstructing the distribution function from the moments, transforming the distribution function into the new coordinate system and retaking the moments in the new coordinate system. In order to find the coordinate transformations that change the time slicing, a different representation of moments must be used, in terms of Schwartz distributions \cite{gratusCorrectUnusualCoordinate2018, gratusDistributionalStressEnergy2020}. In this representation of moments, the moments are transformed into the new coordinate system, then \emph{projected} onto the new time slicing to find the full coordinate transformation. \\

In a standard particle-in-cell code there are three stages: the particle pusher, charge and current deposition, and the field updater. This work will focus only on the particle pusher stage of the PIC code, showing how tracking the moments of the group of particles represented by the macroparticle gives extra structure to the information within the macroparticle. There will be no consideration of particle-particle interactions. Particles will only be affected by external fields. This approach means it is not meaningful to talk about the cells at this stage. Future work will consider how to deposit the charge and current associated with the extra structure of the moments onto the grid. This future work will examine the appropriate cell size needed for this deposition. \\

This article calculates the moment tracking equations and coordinate transformations for arbitrary coordinate systems. The structure of this article is as follows:
Section \ref{sec:Vlasov} introduces the Vlasov equation, which describes the dynamics of collisionless charged particles. Section \ref{sec:Moments} introduces moments, and the existing methods for modelling them. Section \ref{sec:EllisMultipoles} introduces a distributional representation of moments in terms of derivatives of Dirac delta functions and uses this representation to show the time evolution of moments. Section \ref{sec:CoordinateTransforms} finds the coordinate transformations for the moments, which is the main result of this work. Section \ref{sec:ComputationalValidation} shows a computational model of uncharged particles circling a black hole to validate the theory at a proof-of-concept level. Section \ref{sec:CoordinateFree} uses the language of differential geometry and de Rham currents to present a geometric interpretation of the moment differential equations and finds the coordinate transformations of the moments through this language. Section \ref{sec:FullPicCode} discusses the considerations that will need to be made in the development of a method that uses the moments to deposit charge and current onto the grid, in order to develop a full PIC code. We conclude in section \ref{sec:Conclusion} by discussing other interesting direction of future investigation. The appendix contains the more technical proofs needed to establish the theory. \\

\subsection{Notation of indices}

To study the dynamics of plasma, calculations are performed in 7 dimensional phase space. Because of this, several different summation conventions for indices are needed: summations over just space, just velocity, space and velocity, time and space but not velocity, and space, time and velocity. In order to account for all these different summations, this article uses the convention that Latin indices $\OToNa, \OToNb, \OToNc$ represent summations over $0, \ldots, 6$; Greek indices $\OToThreea, \OToThreeb, \OToThreec$ represent summations from $0,1,2,3$; and an underlined index means there is no summation over the 0 index, i.e. $\OneToNa = 1, \ldots, 6$ and $\OneToThreea = 1,2,3$. The coordinates used are $(t, x^\OneToThreea, u^\OneToThreea)$, where $t$ is the global time, $x^\OneToThreea$ is a space coordinate, and $u^\OneToThreea$ is a spatial component of the 4-velocity. The notation $(\xi^\OToNa)$ will be used to represent a general coordinate, such that
\begin{equation}
	\xi^0 = t, \quad \xi^\OneToThreea = x^\OneToThreea, \quad \xi^{\OneToThreea + 3} = u^\OneToThreea
	\label{eq:CoordinateSystem}
\end{equation}
i.e. $\xi^{4} = u^1$. Dimensions are used such that $c=G = 1$. The notation $f|_p$ will be used to represent the evaluation of a function at a point, i.e. $f|_p$ is $f$ evaluated at $p$. After their introduction, the arguments of functions will not be written, unless needed for emphasis.

\section{The Vlasov equation}
\label{sec:Vlasov}

\subsection{7 Dimensional phase space}

To calculated the complex dynamics of plasmas, calculations are performed in 7 dimensional phase space. These dimensions are time, 3 spatial dimensions and 3 proper velocity (4-velocity) dimensions, where $t$ is the global \emph{coordinate time}. As previously stated, these will be represented in coordinates as $(t, x^\OneToThreea, u^\OneToThreea)$. The time component of 4-velocity, $u^0(t,x^\OneToThreea,u^\OneToThreea)$, is a function in phase space, defined as the positive solution to the quadratic equation 
\begin{equation}
g_{\OToThreea \OToThreeb} u^\OToThreea u^\OToThreeb = -1
\label{eq:u0equation}
\end{equation} 
where $g(t, x^\OneToThreea)$ is the metric with signature $(-,+,+,+)$. In the specific example of Minkowski spacetime with Cartesian coordinates, $u^0$ is the Lorentz factor in special relativity, given by 
\begin{equation}
u^0 = \Big(1 + \sum_{\OneToThreea = 1}^3 u^\OneToThreea u^\OneToThreea\Big)^{\frac{1}{2}}.
\end{equation}

\subsection{The Vlasov equation}

Consider the dynamics of a group of charged particles, each with charge $q$ and mass $m$ with distribution function $f(t, x^\OneToThreea, u^\OneToThreea)$. By assuming these particles are collisionless, they can be modelled by the Vlasov equation. To find the Vlasov equation, consider an arbitrary spacetime with metric $g_{\OToThreea \OToThreeb}(t, x^\OneToThreea)$, Christoffel symbols $\Gamma^{\OToThreea}_{\OToThreeb \OToThreec}(t, x^\OneToThreea)$ and electromagnetic 2-form $F_{\OToThreea \OToThreeb}(t, x^\OneToThreea)$. Let $C(t)$ be a world line parameterised by $t$, the same parameter as the global time $\xi^0$. Let $\Map(t)$ be the prolongation of $C(t)$, that is, a curve such that
\begin{equation}
    	\Map^0(t) = C^0(t) = t, \quad \Map^\OneToThreea(t) = C^\OneToThreea(t), \quad \Map^{\OneToThreea + 3}(t) = \frac{dC^\OneToThreea(t)}{dt}.
\end{equation} 
For a given value of $t$, $\Map(t)$ is a point in 7-dimensional phase space. In terms of coordinate functions $(t, x^\OneToThreea, u^\OneToThreea)$, this can be represented as  
\begin{equation}
  \begin{gathered}
    	\Map^0(t) = C^0(t) = t, \quad \Map^\OneToThreea(t) = C^\OneToThreea(t) = x^\OneToThreea|_{\Map(t)}, \\ \Map^{\OneToThreea + 3} = \frac{dC^\OneToThreea(t)}{dt} = \left. \frac{u^\OneToThreea}{u^0} \right|_{\Map(t)}.
     \label{eq:EtaDefinition}
  \end{gathered}
\end{equation} 

In general, $\Map(t)$ is a curve in position-velocity phase space. In this work $\Map(t)$ will be the curve moments are taken around. In particle accelerators, the ideal orbit is a natural choice for $\Map(t)$. This choice does not exist in plasmas; in these cases choices for $\Map(t)$ could be the position of the macroparticle, or a trajectory based on the initial centre of charge of the macroparticle. \\

The Vlasov vector field, $W$, is the integral curves of $\Map$, defined as 
\begin{equation}
	W^\OToNa|_{\Map(t)} = \frac{d}{dt} \Map^\OToNa(t).
\end{equation}	
Finding these derivatives gives 7 ODEs,
\begin{equation}
  \begin{gathered}
  W^0|_{\Map} = \left. \frac{dC^0(t)}{dt} \right|_{\Map} = 1, \quad 
	W^{\OneToThreea}|_{\Map} = \frac{dC^\OneToThreea(t)}{dt} = \left.\frac{u^\OneToThreea}{u^0} \right|_{\Map}, \\
	W^{\OneToThreea + 3}|_{\Map} = \frac{d^2 C^\OneToThreea(t)}{dt^2}. \label{eq:GeneralW} 
  \end{gathered}
\end{equation}
The acceleration can be found using the pregeodesic equation combined with the (pre-)Lorentz force,
\begin{multline}
  	\left. \frac{d^2 C^\OToThreea(t)}{dt^2} \right|_{\Map} + \left. \left( \Gamma^\OToThreea_{\OToThreeb \OToThreec} \frac{dC^\OToThreeb (t)}{dt} \frac{dC^\OToThreec (t)}{dt} \right) \right|_{\Map} \\ = \left. \left( \kappa \frac{dC^\OToThreea(t)}{dt} \right) \right|_{\Map} + \left. \left( \frac{q}{m} \frac{1}{u^0} F_{\OToThreeb \OToThreec} g^{\OneToThreea \OToThreeb} \frac{d C^{\OToThreec}(t)}{dt} \right) \right|_{\Map} 
\end{multline}
where the $1/u^0$ term in the Lorentz force arises when \FrameName time is the parameterisation, rather than proper time, and $\kappa(t)$ arises from the parameterisation. $\kappa(t)$ can be found by noting $C^0 = t$, hence $d^2 C^0/dt^2 = 0$, solving this gives
\begin{equation}
  \kappa = \Gamma^0_{\OToThreeb \OToThreec} \frac{dC^\OToThreeb (t)}{dt} \frac{dC^\OToThreec (t)}{dt} - \frac{q}{m} \frac{1}{u^0} F_{\OToThreeb \OToThreec} g^{0 \OToThreeb} \frac{d C^{\OToThreec}(t)}{dt}.
\end{equation}
From this, solving equation \eqref{eq:GeneralW} for all $\Map$ gives
\begin{equation}
\begin{gathered}
  W^0 = 1, W^{\OneToThreea} = \frac{u^\OneToThreea}{u^0}, \\ W^{\OneToThreea + 3} = \begin{multlined}[t] - \Gamma^\OneToThreea_{\OToThreeb \OToThreec} \frac{u^\OToThreeb}{u^0} \frac{u^\OToThreec}{u^0} + \frac{q}{m} \frac{1}{u^0} F_{\OToThreeb \OToThreec} g^{\OneToThreea \OToThreeb} \frac{u^\OToThreec}{u^0} \\ + \Gamma^0_{\OToThreeb \OToThreec} \frac{u^\OToThreeb}{u^0} \frac{u^\OToThreec}{u^0} \frac{u^\OneToThreea}{u^0} - \frac{q}{m} \frac{1}{u^0} F_{\OToThreeb \OToThreec} g^{0 \OToThreeb} \frac{u^\OToThreec}{u^0} \frac{u^\OneToThreea}{u^0}.
  \end{multlined}
  \label{eq:VlasovVectorField}
\end{gathered}
\end{equation}
The Vlasov equation describes the motion of the whole particle distribution,
\begin{equation}
	W^\OToNa \partial_{\OToNa} f = 0.
\end{equation}		
where $\partial_\OToNa = \partial/\partial \xi^\OToNa$. \\

This article will work exclusively in \emph{coordinate time frames}: these are frames where the parameterisation of $\Map$ is the same as the time coordinate. In any coordinate time frame, the Vlasov vector field has the form of equation \eqref{eq:VlasovVectorField}. This formulation of the Vlasov vector field is distinct from the $3+1$ formalism used in other approaches for simulating plasma in general relativity \cite{thorneElectrodynamicsCurvedSpacetime1982,dodinVlasovEquationCollisionless2010}, although the moment tracking and coordinate transformations presented in this work can be used in both formalisms.
	
\section{Moments}
\label{sec:Moments}

The collective properties of a group of charged particles with distribution function $f$ can be modelled by their moments. This article makes the choice that $f$ is a scalar density of weight 1. This choice means that the moments are affected by the measure from a curved spacetime through a change in $f$. The zeroth order moment, known as the monopole, is defined as
\begin{equation}
	q = \int f(t,x,u) \, d^3 x \; d^3 u
\end{equation}
and corresponds to the total charge of the macroparticle. The first order moment, known as the dipole moment, is given by
\begin{equation}
  \begin{gathered}
      V^{\OneToThreea}(t) = \int \Big(x^\OneToThreea - \Map^\OneToThreea (t) \Big) f(t,x,u) \, d^3 x \; d^3 u, \\ \quad V^{\OneToThreea + 3}(t) = \int \Big(u^\OneToThreea - \frac{dC^{\OneToThreea}}{dt} (t) \Big) f(t,x,u) \, d^3 x \; d^3 u.
  \end{gathered}
\end{equation}
These can be combined into a single equation,
\begin{equation}
	V^{\OneToNa} (t) = \int \Big(\xi^\OneToNa - \Map^\OneToNa (t) \Big) f(t,x,u) \, d^3 x \; d^3 u
\end{equation}
where $\xi^\OneToNa$ and $\Map^\OneToNa$ are defined by \eqref{eq:CoordinateSystem} and \eqref{eq:EtaDefinition} respectively. The dipole moment corresponds to the deviation of the centre of charge from the centre of the macroparticle. If the initial centre of the macroparticle is chosen such that the dipole moments are initially zero, then whilst the centre of the macroparticle will obey the Lorentz force equation, the centre of charge will not \cite{ginzburgSimulationCollisionlessUltrarelativistic2016}. Hence, the two paths will diverge. The dipole moment represents the difference between these.\\

This formalism can be generalised to the $n$th order moment,
\begin{multline}
  	V^{\OneToNa_{1}, \ldots \OneToNa_{n}} (t) = \frac{1}{n!} \int \bigg( \Big(\xi^{\OneToNa_{1}} - \Map^{\OneToNa_{1}}(t)\Big) \ldots \Big(\xi^{\OneToNa_{n}} - \Map^{\OneToNa_{n}}(t) \Big) \bigg) \\ \times f (t,x,u) \, d^3 x \; d^3 u
	\label{eq:GeneralMoment}
\end{multline}
where $V$ is totally symmetric and $n!$ is required for counting due to the symmetry. Note the sums are only over space and velocity coordinates, and there are no corresponding `time' moments. The moments in this work are integrated over both position and momentum space, in contrast to the conventional moment equations for plasmas, which are just integrated over velocity space \cite{palmrothVlasovMethodsSpace2018}. The naming convention for multipoles scales as $2^k$, so the zeroth order moment is the monopole, then the dipole, quadrupole, octopole etc. (table \ref{table:NumberOfMoments}). The quadrupole moment corresponds to the variance of the macroparticle; the octopole moment corresponds to the skew of the macroparticle, and the hexadecapole moment corresponds to the kurtosis of the macroparticle.

\subsection{Dynamics of the moments}
\label{subsec:DymnikovPerelshtein}

When parameterising with a global time, the dynamics of the moments can be calculated by differentiating \eqref{eq:GeneralMoment} with respect to time \cite{dymnikovMomentMethodDynamics1978, channellMomentApproachCharged1983},
\begin{equation}
	\frac{dV^{\OneToNa_{1}, \ldots \OneToNa_{n}}}{dt} =\frac{1}{n!} \frac{d}{dt} \int (\xi^{\OneToNa_{1}} - \Map^{\OneToNa_{1}}) \ldots (\xi^{\OneToNa_{n}} - \Map^{\OneToNa_{n}}) f d^3x d^3u.
	\label{eq:StandardMoments}
\end{equation}
By using the Vlasov equation and Taylor expanding around $\Map$, the dynamics can be found. As an example, the differential equation for the dipole is
\begin{equation}
	\frac{dV^\OneToNa}{dt} = \sum_{m=1}^\infty \frac{1}{m!} V^{\OneToNb_1 \ldots \OneToNb_m} \partial_{\OneToNb_1} \ldots \partial_{\OneToNb_m} W^\OneToNa |_{\Map}.
\end{equation} 
This differential equation is an infinite series of higher order moments, and a choice must be made to truncate this series at some point. Truncating at quadrupole (second) order, the differential equations for the dipole and quadrupole are given by
\begin{equation}
	\begin{gathered}
		\frac{d V^{\OneToNa}}{dt} = V^{\OneToNb} \partial_\OneToNb W^\OneToNa|_{\Map} + \frac{1}{2} V^{\OneToNb \OneToNc} \partial_\OneToNb \partial_\OneToNc W^\OneToNa|_{\Map}, \\
		\frac{d V^{\OneToNa \OneToNb}}{dt} = V^{\OneToNc \OneToNb} \partial_\OneToNc W^\OneToNa|_{\Map} + V^{\OneToNa \OneToNc} \partial_\OneToNc W^\OneToNb|_{\Map}.
	\end{gathered}
\label{eq:MomentTransportEquations}
\end{equation}
The dipole generated from the quadrupole can clearly be seen. All moment tracking methods suffer from this error - higher order moments generate lower order moments. The importance of the choice of truncation on the accuracy of the model is discussed in section \ref{sec:ComputationalValidation}.

\subsection{Coordinate transformations of moments where the time slicing is preserved}
\label{subsec:ExistingCoordTransforms}

As moments are defined through integrals, they are highly coordinate dependent objects. For coordinate transformations where the time coordinate is unchanged, this definition of the moments can be used to find the coordinate transformation. Consider a new coordinate system denoted by hatted coordinates $(\hat{t},\hat{\xi}^{\hat{\OneToNa}})$ where $\hat{t}= t$ (since the time coordinate stays the same) and use hatted indices to represent a summation in the new coordinate system. The distribution function $f$ transforms as a scalar density \footnote{An integral over the unit hyperboloid must contain the measure $\text{det(g)}/u^0$. There is a choice of making either $f$ a density of weight 1, or to put the measure in the integrands. This article uses the convention that $f$ is a scalar density of weight 1} of weight $1$, i.e. 
\begin{equation}
	\hat{f} = f \frac{d^7 \xi}{d^7 \hat{\xi}}
\end{equation}
where a hatted variable represents that variable in the new coordinate system, and $d^7 \xi/d^7 \hat{\xi}$ is the Jacobian. By expanding $\xi$ in terms of $\hat{\xi}$ around $\hat{\Map}$, 
\begin{equation}
	{\xi}^{\OneToNa} = \sum_{n=0}^{\infty} \frac{1}{n!} \left(\hat{\xi}^{\OneToNb_{1}} - \hat{\Map}^{\OneToNb_{1}} \right) \ldots \left(\hat{\xi}^{\OneToNb_{n}} - \hat{\Map}^{\OneToNb_{n}} \right) \frac{\partial}{\partial \hat{\xi}^{\OneToNb_{1}}} \ldots \frac{\partial}{\partial \hat{\xi}^{\OneToNb_{n}}} {{\xi}}^\OneToNa|_{{\Map}}, 
\end{equation}
and noting $\hat{\Map}(\hat{t}) = \Map(t)$, the coordinate transformations can be found. These are infinite sums in terms of all higher order moments. Truncating the expansion at quadrupole order, the coordinate transformations up to the quadrupole order for the dipole and quadrupole are given by
\begin{equation}
	\begin{gathered}
		\hat{V}^\OneToNa = V^{\OneToNb} \frac{\partial \hat{\xi}^{\OneToNa}}{\partial \xi^\OneToNb}+ \frac{1}{2} V^{\OneToNb \OneToNc} \frac{\partial^2 \hat{\xi}^\OneToNa}{\partial \xi^\OneToNb \partial \xi^\OneToNc} \\
		\hat{V}^{\OneToNa \OneToNb} = V^{\OneToNc \OneToNd} \frac{\partial \hat{\xi}^{\OneToNa}}{\partial \xi^\OneToNc} \frac{\partial \hat{\xi}^{\OneToNb}}{\partial \xi^\OneToNd}.
 \label{eq:MomentCoordTransforms}
	\end{gathered}
\end{equation}

For coordinate transformations that mix the space and time coordinates, this method for coordinate transformations will not work. The difficulty in transforming coordinates using this method arises from the important dependence on the choice of time. In order to find the moments, the coordinate systems defines a \emph{time slicing}, which is a hypersurface of constant $t$ (figure \ref{fig:differentfoliations2}). The integrals in the definition for the moments in equation \eqref{eq:GeneralMoment} are over these hypersurfaces. When performing a coordinate transformation that mixes the time and space coordinates, then the time slicing will also change. Changing the time slicing will give different moments. The integral representation of moments in equation \eqref{eq:GeneralMoment} cannot be transformed between coordinate systems that mix time and space coordinates. In this article we present a new representation of moments using Schwartz distributions. These distributions are introduced in section \ref{sec:EllisMultipoles}, and the use of this language to find the coordinate transformations for the moments is presented in section \ref{sec:CoordinateTransforms}.

\section{Schwartz Distributional multipoles and their time evolution}
\label{sec:EllisMultipoles}

\subsection{Ellis multipoles}
\label{sec:EllisMultipolesIntro}

Rather than defining moments explicitly through spatial integrals (equation \eqref{eq:GeneralMoment}), an alternative representation of moments is through derivatives of Dirac delta functions. This means that rather than modelling a macroparticle as just a Dirac delta function and using a shape function to deposit charge (the cloud-in-cell approach), the macroparticle is represented as a set of derivatives of Dirac delta functions. By using the language of the Schwartz distributions presented in this section, any coordinate transformation of the moments can be calculated, including those mixing space and time coordinates, which, as previously stated in section \ref{subsec:ExistingCoordTransforms}, cannot be done through equation \eqref{eq:GeneralMoment}. By defining a moment through the method presented below, there is no dependence on the time slicing in the definition of the moment, and as such, the spacetime coordinate transformation can be found. Representing a multipole expansion using derivatives of a Dirac delta function is known as the \emph{Ellis representation} of a multipole \cite{ellisElectromagneticFieldsMoving1966, ellisMotionClassicalParticle1975, gratusDistributionalStressEnergy2020}. It may also be known as the \emph{Schwartz distribution representation} of a multipole. In terms of Dirac delta functions the expansion to the second order (the quadrupole) is
\begin{multline}
	\mathcal{J}^\OToNa = \frac{1}{2} \int_{\mathbb{R}} \dot{\Map}^\OToNa \, V^{\OneToNb \OneToNc} \left( \partial_\OneToNb \partial_\OneToNc \delta^{(6)} (\xi - \Map) \right) dt \\ - \int_{\mathbb{R}} \delta^\OToNa_{\OneToNb} \, X^{\OneToNb \OneToNc} \left( \partial_\OneToNc \delta^{(6)} (\xi-\Map) \right) dt \\
	- \int_{\mathbb{R}} \dot{\Map}^\OToNa \, V^{\OneToNb} \left( \partial_\OneToNb \delta^{(6)}(\xi-\Map) \right) dt + \int_{\mathbb{R}} \delta^\OToNa_{\OneToNb} \, X^{\OneToNb} \, \delta^{(6)}(\xi-\Map) dt \\ + \int_{\mathbb{R}} \dot{\Map}^\OToNa \, q \, \delta^{(6)} (\xi-\Map) dt
	\label{eq:EllisSemiQuadrupole}
\end{multline}
where
\begin{equation}
\delta^{(6)} (\xi-\Map) = \delta(\xi^1 - \Map^1) \ldots \delta(\xi^6 - \Map^6)
\end{equation}
and
\begin{equation}
	\begin{gathered}
		V^{\OneToNb \OneToNc}(t) = \int (\xi^{\OneToNb} - \Map^{\OneToNb}) (\xi^{\OneToNc} - \Map^{\OneToNc}) \, f \, d^6 \xi, \\ \quad V^{\OneToNb}(t) = \int (\xi^{\OneToNb} - \Map^{\OneToNb}) \, f \, d^6 \xi, 
    \quad q = \int f \, d^6 \xi, \\
		X^{\OneToNa \OneToNb}(t) = V^{\OneToNa \OneToNc} \partial_\OneToNc W^\OneToNb, \quad X^{\OneToNa}(t) = V^{\OneToNb} \partial_\OneToNb W^\OneToNa + \frac{1}{2} V^{\OneToNb \OneToNc} \partial_\OneToNb \partial_\OneToNc W^\OneToNa.
	\end{gathered}
\label{eq:VandX}
\end{equation}
where $d^6 \xi = d^3x d^3u$. Note $q$ has no dependence on time due to the conservation of charge. \\

The terms $V^{\OneToNa}, X^{\OneToNa}, V^{\OneToNa \OneToNb}$ and $X^{\OneToNa \OneToNb}$ are called the \emph{components} of a multipole. As this definition involves derivatives of Dirac delta functions, they are defined by their action on test functions $(\phi_0, \ldots, \phi_6)$ which are the components of a covector, and have compact support. Equation \eqref{eq:EllisSemiQuadrupole} acting on $\phi_\OToNa$ gives 
\begin{multline} 
	\int \mathcal{J}^\OToNa \phi_\OToNa d^7 \xi = \frac{1}{2} \int_{\mathbb{R}} \dot{\Map}^\OToNa V^{\OneToNb \OneToNc} \left( \partial_\OneToNc \partial_\OneToNb \phi_\OToNa|_{\Map} \right) dt \\ 
  + \int_{\mathbb{R}} X^{\OneToNa \OneToNb} \left( \partial_\OneToNb \phi_\OneToNa|_{\Map} \right) dt + \int_{\mathbb{R}} \dot{\Map}^\OToNa V^{\OneToNb} \left( \partial_\OneToNb \phi_\OToNa|_{\Map} \right) dt \\ 
  + \int_{\mathbb{R}} X^{\OneToNa} \phi_\OneToNa|_{\Map} dt + \int_{\mathbb{R}} \dot{\Map}^\OToNa \, q \, \phi_\OToNa|_{\Map} dt.
	\label{eq:EllisSemiMultipoleOnTestForm}
\end{multline}
The evaluation of the test form at $\Map$ will not be explicitly written in future representations of $\mathcal{J}^\OToNa$, but is implicitly present. \\

The components in \eqref{eq:VandX} are only over $(1,\ldots,6)$ (there are no terms of the form $V^0$ or $V^{01}$). By writing a multipole in this way the components of a multipole are unique. The components can be extracted by acting $\mathcal{J}^{\OToNa}$ on specific test forms. The details of these test forms and how they isolate the components of $\mathcal{J}^\OToNa$ are outlined in appendix \ref{sec:UniqueComponentsProof}. A discussion on why this means they are unique can be found in ref. \cite{gratusDistributionalStressEnergy2020}. \\

\subsection{Relating the components of Ellis multipoles and moments}

\begin{figure}[htb]
  \centering
  \includegraphics[width=.7\linewidth]{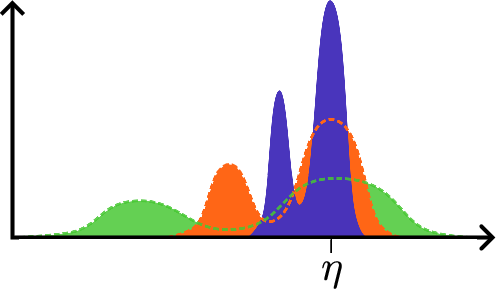}
  \caption{By squeezing the width of a function down whilst increasing its height, such that its area stays the same, a function can be represented by a series of derivatives of Dirac delta functions. The coefficients of this series are the moments of the function.}
  \label{fig:SqueezedForm}
\end{figure}

To show the components of \eqref{eq:EllisSemiMultipoleOnTestForm} are the moments of $f$ (equation \eqref{eq:VandX}), the distribution function $f$ is \emph{squeezed}. This is the process of representing a function using a Dirac delta function and the derivative of Dirac delta functions (figure \ref{fig:SqueezedForm}). The moments of $f$ are the coefficients of this expansion. By doing this, it can be shown that the moments of the distribution function $f$ naturally appear in the components of the Ellis representation of a multipole. \\

Squeezing a distribution can be shown using the language of differential geometry, the process of which can be found in appendix \ref{sec:SqueezedFormProof}.

\subsection{Time evolution of moments}

The evolution of the moments defined by \eqref{eq:EllisSemiQuadrupole} are governed by two conditions: the conservation of charge, described by the equation
\begin{equation}
  \partial_\OToNa \mathcal{J}^\OToNa = 0, \label{eq:ChargeConservation}
\end{equation}
and the Vlasov equation,
\begin{equation}
  \mathcal{J}^\OToNa W^\OToNb - \mathcal{J}^\OToNb W^\OToNa = 0. \label{eq:VlasovEquationCondition}
\end{equation}

The combination of the equations \eqref{eq:ChargeConservation} and \eqref{eq:VlasovEquationCondition} are called the \emph{transport equations}, and are used to find the evolution of the moments in the Ellis representation. The transport equations can also be defined by their actions on test forms $\lambda$ and $\alpha_{\OToNa \OToNb}$,
\begin{equation}
	\int \mathcal{J}^\OToNa \partial_\OToNa \lambda dt = 0, \quad \int \mathcal{J}^\OToNa W^\OToNb \alpha_{\OToNa \OToNb} dt = 0
	\label{eq:TransportEquations}
\end{equation}
where $\alpha_{\OToNa \OToNb}$ is antisymmetric, and both $\lambda$ and $\alpha_{\OToNa \OToNb}$ have compact support. These two conditions give
\begin{equation}
	\begin{gathered}
		\frac{d V^{\OneToNa \OneToNb}}{dt} = X^{\OneToNa \OneToNb} + X^{\OneToNb \OneToNa}, \quad \frac{d V^{\OneToNa}}{dt} = X^{\OneToNa}, \quad \frac{dq}{dt}=0, \\
		X^{\OneToNa \OneToNb} = V^{\OneToNa \OneToNc} \partial_\OneToNc W^\OneToNb, \quad X^{\OneToNa} = V^{\OneToNb} \partial_\OneToNb W^\OneToNa + \frac{1}{2} V^{\OneToNb \OneToNc} \partial_\OneToNb \partial_\OneToNc W^\OneToNa.
	\end{gathered}
\label{eq:EllisTransportSolutions}
\end{equation}
Equation \eqref{eq:MomentTransportEquations} can be found as a trivial corollary of this, so this method can also be used to find the differential equations for the moments. These differential equations are the same as directly differentiating \eqref{eq:GeneralMoment}. 

\begin{proof}
	The proof of this is in appendix \ref{sec:EllisTransportEquationSolution}.
\end{proof}

Whilst both methods achieve the same result, there is a different philosophy to each approach. The method presented in section \ref{subsec:DymnikovPerelshtein} gives an infinite Taylor expansion, and picks the truncation point at the end of the method. Alternatively, the Ellis representation makes the choice of truncation when first performing the multipole expansion to a specific order. This order of expansion is then kept throughout. Whilst this work only carries out this expansion to quadrupole order, it is trivial to extend this result to higher orders. \\

\section{Coordinate transformations of multipoles where the time slicing changes}
\label{sec:CoordinateTransforms}

Having introduced the Schwartz distributional multipole we are now in a position to calculate the coordinate transformations where the time slicing changes. Consider a new coordinate system denoted by hatted coordinates $(\hat{t},\hat{\xi}^{\hat{\OneToNa}})$, where the time coordinate is changed, and use hatted indices to represent a summation in the new coordinate system. To find the transformation rules of a multipole, it is assumed $\mathcal{J}^\OToNa$ transforms as a density, i.e.
\begin{equation}
	\hat{\mathcal{J}}^\OToNb = \mathcal{J}^\OToNa \frac{\partial \hat{\xi}^\OToNb}{\partial \xi^\OToNa} \frac{d^7 \xi}{d^7 \hat{\xi}}
\end{equation}
such that
\begin{equation}
	\int \mathcal{J}^\OToNa \phi_\OToNa d^7 \xi = \int \hat{\mathcal{J}}^\OToNa \hat{\phi}_\OToNa d^7 \hat{\xi}
\end{equation}
is an invariant quantity. \\

Since $\dot{\Map}^\OneToNa$ are functions on a world line, they transform as
\begin{equation}
	\dot{\hat{\Map}}^\OToNa = \frac{\partial \hat{\xi}^\OToNa}{\partial \xi^\OToNb}\frac{d\hat{t}}{dt} \dot{\Map}^\OToNb
\end{equation}
where
\begin{equation}
	\dot{\hat{\Map}}^\OToNa (\hat{t}) = \frac{d \hat{\Map}^\OToNa (\hat{t})}{d \hat{t}}.
\end{equation}

Recall $\phi_\OToNa$ and $\partial_\OToNa$ are tensorial so the transformation rules for these are
\begin{equation}
\hat{\phi}_{\OToNa} = \frac{\partial \xi^\OToNa}{\partial \hat{\xi}^{\OneToNb}} \phi_{\OneToNb}, \quad \frac{\partial}{\partial \hat{\xi}^\OToNa} = \frac{\partial \xi^{\OneToNb}}{\partial \hat{\xi}^{\OToNa}} \frac{\partial}{\partial \xi^{\OneToNb}}. 
\end{equation}
Note that if summations just run over $(1, \ldots 6)$ in the original coordinate system, in general the indices in the new coordinate system will run over $(0, \ldots 6)$. Using this, it can be shown that using these transformation rules gives

\begin{equation}
	\begin{gathered}
		U^{\hat{\OToNb} \hat{\OToNc}} = V^{\OneToNd \OneToNe} A^{\hat{\OToNb}}_{\OneToNd} A^{\hat{\OToNc}}_{\OneToNe}, \\ Y^{\hat{\OToNc} \hat{\OToNd}} = \left(X^{\OneToNa \OneToNb} A^{\hat{\OToNc}}_{\OneToNb} A^{\hat{\OToNd}}_{\OneToNa} + \frac{1}{2} V^{\OneToNe \OneToNf} \dot{{\Map}}^{\OToNa} \partial_\OToNa \left(A^{\hat{\OToNd}}_{\OneToNf} A^{\hat{\OToNc}}_{\OneToNe} \right) \right) \frac{dt}{d \hat{t}}, \\
		U^{\hat{\OToNa}} = V^{\OneToNb} A^{\hat{\OToNa}}_{\OneToNb} + \frac{1}{2} V^{\OneToNb \OneToNc} A^{\hat{\OToNa}}_{\OneToNb \OneToNc}, \\ 
 \begin{multlined}[t]Y^{\hat{\OToNc}} = \left(X^\OneToNd A^{\hat{\OToNc}}_{\OneToNd} + {V}^{\OneToNb} \dot{\Map}^{\OToNa} \partial_\OToNa (A^{\hat{\OToNc}}_{\OneToNb}) \right. \\ \left. + X^{\OneToNa \OneToNb} A^{\hat{\OToNc}}_{\OneToNa \OneToNb} + \frac{1}{2} V^{\OneToNd \OneToNe} \dot{{\Map}}^{\OToNa} \partial_\OToNa \left( A^{\hat{\OToNc}}_{\OneToNd \OneToNe} \right) \right) \frac{d t}{d\hat{t}} \end{multlined}
	\end{gathered}
\label{eq:EllisCoordTransforms}
\end{equation}
where
\begin{equation}
	A^{\hat{\OToNa}}_{\OneToNc} = \frac{\partial \hat{\xi}^\OToNa}{\partial \xi^\OneToNc}, \quad 	A^{\hat{\OToNa}}_{\OneToNb \OneToNc} = \frac{\partial^2 \hat{\xi}^\OToNa}{\partial \xi^\OneToNb \partial \xi^\OneToNc}
\end{equation}
and a hatted index represents a sum over coordinates in the new coordinate system.

\begin{proof}
	The proof of this is in appendix \ref{sec:EllisCoordTransform}.
\end{proof}

Observe that $U^\OToNa, U^{\OToNa \OToNb}, Y^{\OToNa}$ and $Y^{\OToNa \OToNb}$ have indices ranging over $(0,\ldots,6)$, not just $(1, \ldots, 6)$. Multipoles with components that have indices running over $(0 \ldots 6)$ do not have unique components. As shown in section \ref{sec:EllisMultipolesIntro}, for the components of a multipole to be unique, the components must only range over $(1 \ldots 6)$. The reason the multipole contains terms of the form $U^0$ etc. is because it is still adapted to the time slicing $t$ from the original coordinate system. To adapt the components to the new time slicing, and thus find the full coordinate transformation, they are projected onto the new time coordinate $\hat{t}$. Consider differentiating along a world line,
\begin{equation}
  \partial_{\dot{\Map}} = \dot{\Map}^\OToNa \partial_\OToNa = \partial_0 + \dot{\Map}^\OneToNa \partial_\OneToNa.
\end{equation}
By rearranging this for $\partial_0$, these terms are projected onto components along $\partial_\OneToNa$, and components along the world line,
\begin{equation}
	\partial_0 = \partial_{\dot{\Map}} - \dot{\Map}^\OneToNa \partial_\OneToNa.
	\label{eq:EllisProjection}
\end{equation}
Since $U^\OToNa$ are functions along a world line, 
\begin{equation}
  \partial_{\dot{\Map}} U^\OneToNa = \frac{dU^\OneToNa}{dt}
\end{equation}
with similar relations for $U^{\OToNa \OToNb}, Y^\OToNa, Y^{\OToNa \OToNb}$, and $\phi|_{\Map}$. By using this projection, terms like the $U^{00}$ term get projected into terms in $\hat{V}^{\OneToNa \OneToNb}$, $\hat{X}^{\OneToNa \OneToNb}$, $\hat{V}^{\OneToNa}$, and $\hat{X}^{\OneToNa}$. \\

Using this projection allows multipoles to be adapted to the time slicing $\hat{t}$ in the new coordinate system, i.e. adapted such that the components indices only range over $(1 \ldots 6)$, giving the full coordinate transformation. The new moments are

\begin{align}
	\hat{V}^{\OneToNa \OneToNb} &= \begin{multlined}[t] V^{\OneToNc \OneToNd} A^{\hat{\OneToNa}}_{\OneToNc} A^{\hat{\OneToNb}}_{\OneToNd} - \dot{\hat{\Map}}^\OneToNa V^{\OneToNc \OneToNd} A^{\hat{0}}_{\OneToNc} A^{\hat{\OneToNb}}_{\OneToNd} \\ - \dot{\hat{\Map}}^\OneToNb V^{\OneToNc \OneToNd} A^{\hat{\OneToNa}}_{\OneToNc} A^{\hat{0}}_{\OneToNd} + \dot{\hat{\Map}}^\OneToNa \dot{\hat{\Map}}^\OneToNb V^{\OneToNc \OneToNd} A^{\hat{0}}_{\OneToNc} A^{\hat{0}}_{\OneToNd}, \label{eq:QuadrupoleFullTransform} \end{multlined} \\
	\hat{V}^{\OneToNa} &= \begin{multlined}[t]	V^{\OneToNb} A^{\hat{\OneToNa}}_{\OneToNb} - \dot{\hat{\Map}}^\OneToNa V^{\OneToNb} A^{\hat{0}}_{\OneToNb} + \frac{1}{2} V^{\OneToNb \OneToNc} A^{\hat{\OneToNa}}_{\OneToNb \OneToNc} - \frac{1}{2} \dot{\hat{\Map}}^\OneToNa V^{\OneToNb \OneToNc} A^{\hat{0}}_{\OneToNb \OneToNc} 
	\\ - \left(X^{\OneToNa \OneToNb} A^{\hat{\OToNc}}_{\OneToNb} A^{\hat{0}}_{\OneToNa} + \frac{1}{2} V^{\OneToNe \OneToNf} \dot{{\Map}}^{\OToNa} \partial_\OToNa \left(A^{\hat{0}}_{\OneToNf} A^{\hat{\OToNc}}_{\OneToNe} \right) \right) \frac{dt}{d \hat{t}} \\
	+ \frac{1}{2} \ddot{\hat{\Map}}^\OneToNa A^{\hat{0}}_{\OneToNb} A^{\hat{0}}_{\OneToNc} V^{\OneToNb \OneToNc} + \frac{1}{2} \frac{d t}{d\hat{t}} \dot{\hat{\Map}}^\OneToNa \dot{\Map}^{\hat{\OToNd}} \partial_{{\OToNd}} \left(A^{\hat{0}}_{\OneToNb} A^{\hat{0}}_{\OneToNc} \right) V^{\OneToNb \OneToNc} \\ + \frac{1}{2} \frac{d t}{d\hat{t}} \dot{\hat{\Map}}^\OneToNa A^{\hat{0}}_{\OneToNb} A^{\hat{0}}_{\OneToNc} \frac{dV^{\OneToNb \OneToNc}}{dt} \label{eq:DipoleFullTransform}
	\end{multlined}
\end{align}
where
\begin{equation}
	\ddot{\hat{\Map}}^{\OneToNa} = \frac{d^2 \hat{\Map}(\hat{t})}{d\hat{t}^2}.
\end{equation}
Similar terms exist for $X^{\OneToNa \OneToNb}$ and $X^{\OneToNa}$ such that \eqref{eq:EllisTransportSolutions} is still satisfied, and are shown in equation \eqref{eq:FullEllisProjection} in appendix \ref{sec:EllisCoordProjection}. They can also be found using \eqref{eq:VandX} in the new coordinate system.

\begin{proof}
	The proof of this is in appendix \ref{sec:EllisCoordProjection}.
\end{proof}

Equation \eqref{eq:QuadrupoleFullTransform} and equation \eqref{eq:DipoleFullTransform} are a new result, and are numerically tested in the next section to show their validity. In the case there is no change in time coordinate, $A^{\hat{0}}_{\OneToNb} = 0$, this reduces to equation \eqref{eq:MomentCoordTransforms}. This work has transformed between two \FrameName frames, but can be generalised to any frame, not just one where $\dot{\Map}^0 = 1$, with a more general projection
\begin{equation}
	\partial_0 = \frac{1}{\dot{\Map}^0} \partial_{\dot{\Map}} - \frac{\dot{\Map}^\OneToNa}{\dot{\Map}^0} \partial_\OneToNa.
\end{equation}

\section{Computational validation}

\label{sec:ComputationalValidation}

To test the accuracy of the moment tracking and coordinate transformation equations, a computational model was developed. The code tests if the truncation to quadrupole order is acceptable, or if a higher order multipole expansion should be used for practical cases. The results presented here are an example of a coordinate transformation that mixes time and space coordinates, focusing on the particularly challenging case of black holes. This work is not solely limited to plasma around black holes. In particular, the moment tracking can be applied to any plasma. \\

To develop the code, the derivatives of the Vlasov equation were calculated using the symbolic algebra software MAPLE. The simulation itself was written in C++. 
There are 36 first derivatives of the Vlasov equation and 126 second derivatives. Each of these derivatives is very complex, hence the need to calculate them using symbolic algebra software. To give an example of this complexity, in Schwarzschild spacetime, one derivative of the Vlasov field is
\begin{multline}
    \frac{\partial W^{4}}{\partial u_{\theta}} = - \frac{1}{u^0} \left(-\frac{\left(r -r_s \right) r_s u_{\theta}}{r \left(-1+\frac{r_s}{r}\right)}+2 \left(-r +r_s \right) u_{\theta} \right) \\ 
	-\frac{2}{u^0} \frac{\left(\frac{\left(r -r_s \right) r_s u^0 }{2 r^{3}}-\frac{r_s \,u_{r}^{2}}{2 r \left(r -r_s \right)} \right)r^{2} u_{\theta}}{\left(-1-\frac{u_{r}^{2}}{1-\frac{r_s}{r}}-r^{2} u_{\theta}^{2}-r^{2} \left(\sin^{2}\left(\theta \right)\right) u_{\phi}^{2}\right)} \\
    - \frac{2}{u^0} \frac{\left(\left(-r +r_s \right) u_{\theta}^{2}-\left(r -r_s \right) \left(\sin^{2}\left(\theta \right)\right) u_{\phi}^{2}\right) r^{2} u_{\theta}}{\left(-1-\frac{u_{r}^{2}}{1-\frac{r_s}{r}}-r^{2} u_{\theta}^{2}-r^{2} \left(\sin^{2}\left(\theta \right)\right) u_{\phi}^{2}\right)} \\ 
	+ \frac{2}{u^0} \frac{r r_s u_{r}^{2} u_{\theta}}{\left(r -r_s \right) \left(-1-\frac{u_{r}^{2}}{1-\frac{r_s}{r}}-r^{2} u_{\theta}^{2}-r^{2} \left(\sin^{2}\left(\theta \right)\right) u_{\phi}^{2}\right)}. 
\end{multline}

The model uses the forward Euler method to integrate both the particle motion and the moment equations. Particles are updated using the equations
\begin{subequations}
\label{eq:ParticleIterators}
	\begin{align}
		x^\OneToThreea_{\text{new}} & = x^\OneToThreea_{\text{old}} + \frac{u^\OneToThreea_{\text{old}}}{u^0_{\text{old}}} \Delta t, \\
		u^\OneToThreea_{\text{new}} & = u^\OneToThreea_{\text{old}} -\Gamma^\OneToThreea_{\OToThreeb \OToThreec} u^\OToThreeb_{\text{old}} u^\OToThreec_{\text{old}} \, \Delta t,
	\end{align}
\end{subequations} 
where $u^0$ is defined by equation \eqref{eq:u0equation} and $\Delta t$ is the time step size. Moments are updated using the equations
\begin{subequations}
\label{eq:MomentTrackingIterators}
\begin{align}
    V^{\OneToNa}_{\text{new}} & = V^\OneToNb_{\text{old}} \partial_\OneToNb W^{\OneToNa}_{\text{old}} \, \Delta t + \frac{1}{2} V^{\OneToNb \OneToNc}_{\text{old}} \partial_\OneToNb \partial_\OneToNc W^{\OneToNa}_{\text{old}} \, \Delta t, \\
    V^{\OneToNa \OneToNb}_{\text{new}} & = \left( V^{\OneToNa \OneToNc}_{\text{old}} \partial_\OneToNb  W^{\OneToNb}_{\text{old}} + V^{\OneToNc \OneToNb}_{\text{old}} \partial_\OneToNb  W^{\OneToNa}_{\text{old}} \right) \Delta t.
\end{align}
\end{subequations}
Despite the high numerical error associated with the forward Euler method, it is acceptable for use as a test to assess the validity of the equations, as the dominating error is not numerical. \\

In all cases, only uncharged particles will be tracked. The analytical results calculated in the previous sections can be applied to charged particles. To calculate the inter-macroparticle forces needed for a full PIC code, a method to use the moments to deposit the charge from a macroparticle onto the grid must be developed. This deposition process is beyond the scope of this paper. An outline of a method that can be used to deposit the charge and current onto the grid is given in section \ref{sec:FullPicCode}.

\subsection{The spacetimes modelled}

To validate the model, tests were performed in both Schwarzschild and Kruskal-Szekeres coordinates. The coordinates in Schwarzschild coordinates are denoted with lowercase letters $\xi_{(\text{Sw})}^{\OToThreea} = (t,r,\theta,\phi,u_r,u_{\theta}, u_{\phi})$, with metric
\begin{equation}
\begin{gathered}
    g_{00}^{(\text{Sw})} = - \left( 1 - \frac{r_s}{r} \right), \quad  g_{11}^{(\text{Sw})} = \frac{1}{1 - \frac{r_s}{r}} \\ g_{22}^{(\text{Sw})} = r^2, \quad g_{33}^{(\text{Sw})} = r^2 \sin^2(\theta).
\end{gathered}
\end{equation}
where $g_{\OToThreea \OToThreeb}^{(\text{Sw})}$ is the metric in Schwarzschild coordinates.\\

Coordinates in Kruskal-Szekeres will be denoted with a capital letter, such that the coordinates in Kruskal-Szekeres are $\xi^\mu_{(\text{KS})} = (T,R,\Theta,\Phi,U_{R},U_{\Theta},U_{\Phi})$. The transformation from Schwarzschild coordinates to Kruskal-Szekeres coordinates (shown in figure \ref{fig:DifferentFoliations}) is given by
\begin{gather}
  R = \sqrt{\frac{r}{r_s} - 1} \,\exp\! \left( \frac{r}{2 r_s} \right) \,\cosh\! \left( \frac{t}{2 r_s} \right), \\
  T = \sqrt{\frac{r}{r_s} - 1} \,\exp\! \left( \frac{r}{2 r_s} \right) \,\sinh\! \left( \frac{t}{2 r_s} \right), \\
  U^\OneToThreea = u^\OToThreeb \frac{\partial \xi^\OneToThreea_{(\text{KS})}}{\partial \xi_{(\text{Sw})}^\OToThreeb}.
\end{gather} 
where the subscript $(\text{KS})$ indicates the coordinates in Kruskal-Szekeres. The metric in Kruskal-Szekeres coordinates is given by
\begin{equation}
\begin{gathered}
     g_{00}^{(\text{KS})} =  - \frac{4r_s^3}{r} \exp{\left(\frac{-r}{r_s} \right)}, \quad g_{11}^{(\text{KS})} =  \frac{4r_s^3}{r} \exp{\left(\frac{-r}{r_s} \right)} \\ g_{22}^{(\text{KS})} = r^2, \quad g_{33}^{(\text{KS})} = r^2 \sin^2(\theta)
\end{gathered}
\end{equation}
where $r$, the radial coordinate in Schwarzschild coordinates, can be found by the inverse transformation
\begin{equation}
    r= r_s \left(1+ W_0 \left(\frac{R^2 - T^2}{e} \right) \right)
\end{equation}
where $W_0$ is the principal branch of the Lambert $W$ function. \\

The numerical testing performed in this article uses moments over a size that may be considered small on astrophysical scales. This is because for numerical simulations in Kruskal-Szekeres, it is impossible to run the model with large moments, or for large amounts of time. As $R$ is updated, the particle will eventually cross the event horizon (the point $T^2 - R^2 = 1$) due to numerical errors from updating the position of the particle. This will happen with any numerical differential equation solver that overestimates the true value. This is another reason it is important to transform coordinates, so a full PIC simulation can be performed in Schwarzschild coordinates, then transformed to Kruskal-Szekeres at the end, avoiding these numerical issues. Although the variation in the metric over the domain represented by the macroparticle and its moments may be small, this does not mean the derivatives of the metric, which the moments couple to, are small.

\subsection{Computational results}

To test the model, the motion of 200 particles that began normally distributed at $r=30000$ in Schwarzschild coordinates with Schwarzschild radius $r_s = 3000$, and a time step size of $\Delta t = 0.01$ were modelled. These particles were also transformed into Kruskal-Szekeres coordinates. In both spacetimes, the particles were tracked using equation \eqref{eq:ParticleIterators} and the moments taken at $t=10$, and the moments were taken at $t=0$ and the moments tracked using equation \eqref{eq:MomentTrackingIterators}. When taking the moments, recall that $f$ is a density, so the effect of curved spacetime adding a measure to the integrals is included in the definition of $f$. To convert $f$ into a scalar field, $f$ can be divided by $|\text{det}(g)|$, where the lack of square root is because the integral is over six dimensional phase space. \\

By modelling particles at a radius of $10r_s$, the accretion disc of the black hole can be studied. Once the accuracy of the moment tracking model at this distance from the black hole has been established, the accuracy of our model in the extreme environments close to the Schwarzschild radius can be examined. \\ 

\begin{figure*}
	\begin{center}
		\begin{subfigure}[t]{.3\linewidth}
			\centering\includegraphics[width=\linewidth]{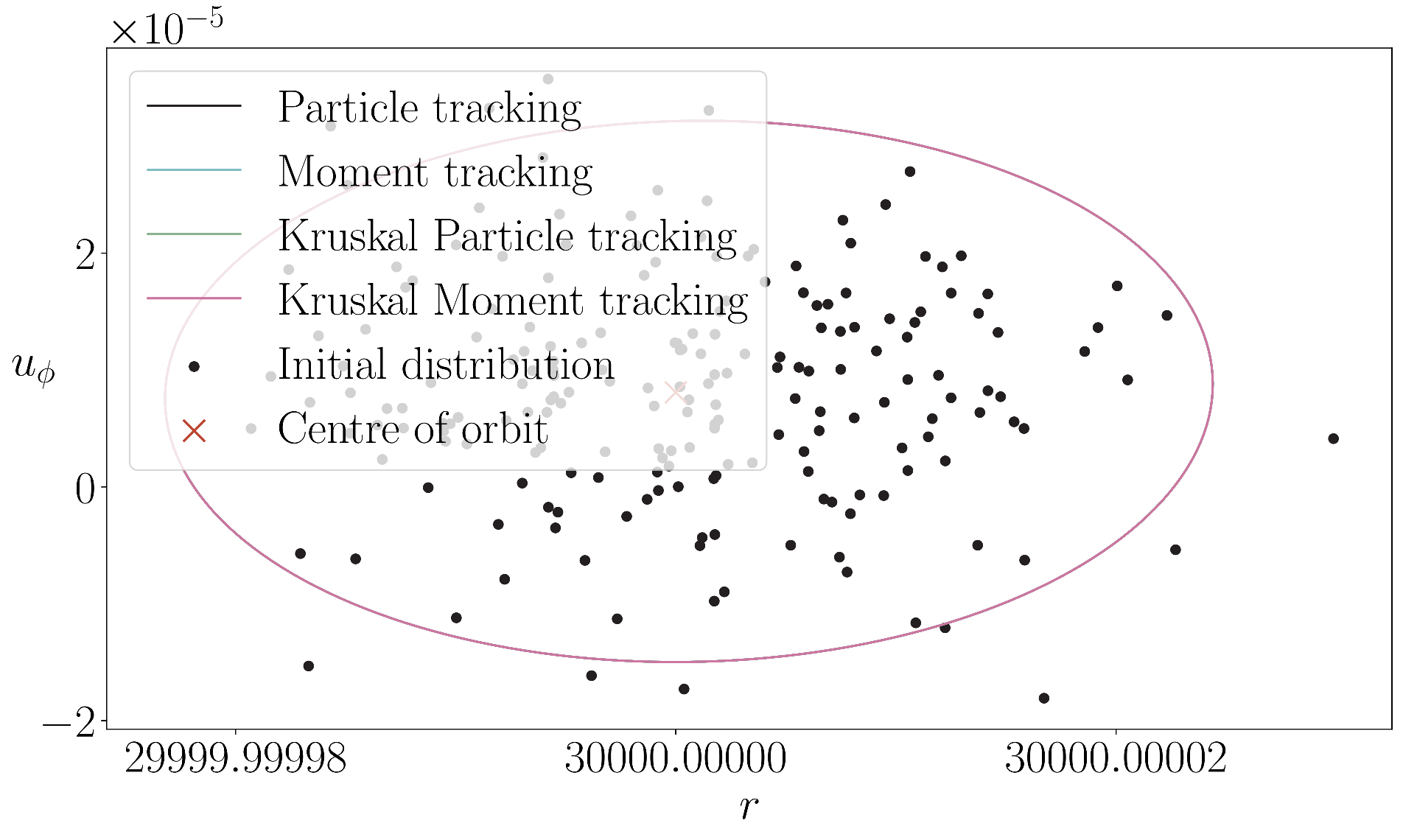}
			\caption{$t=0$, all ellipses overlap.}
			\label{subfig:PhaseSpacet0}
		\end{subfigure}
		\begin{subfigure}[t]{.3\linewidth}
			\centering\includegraphics[width=\linewidth]{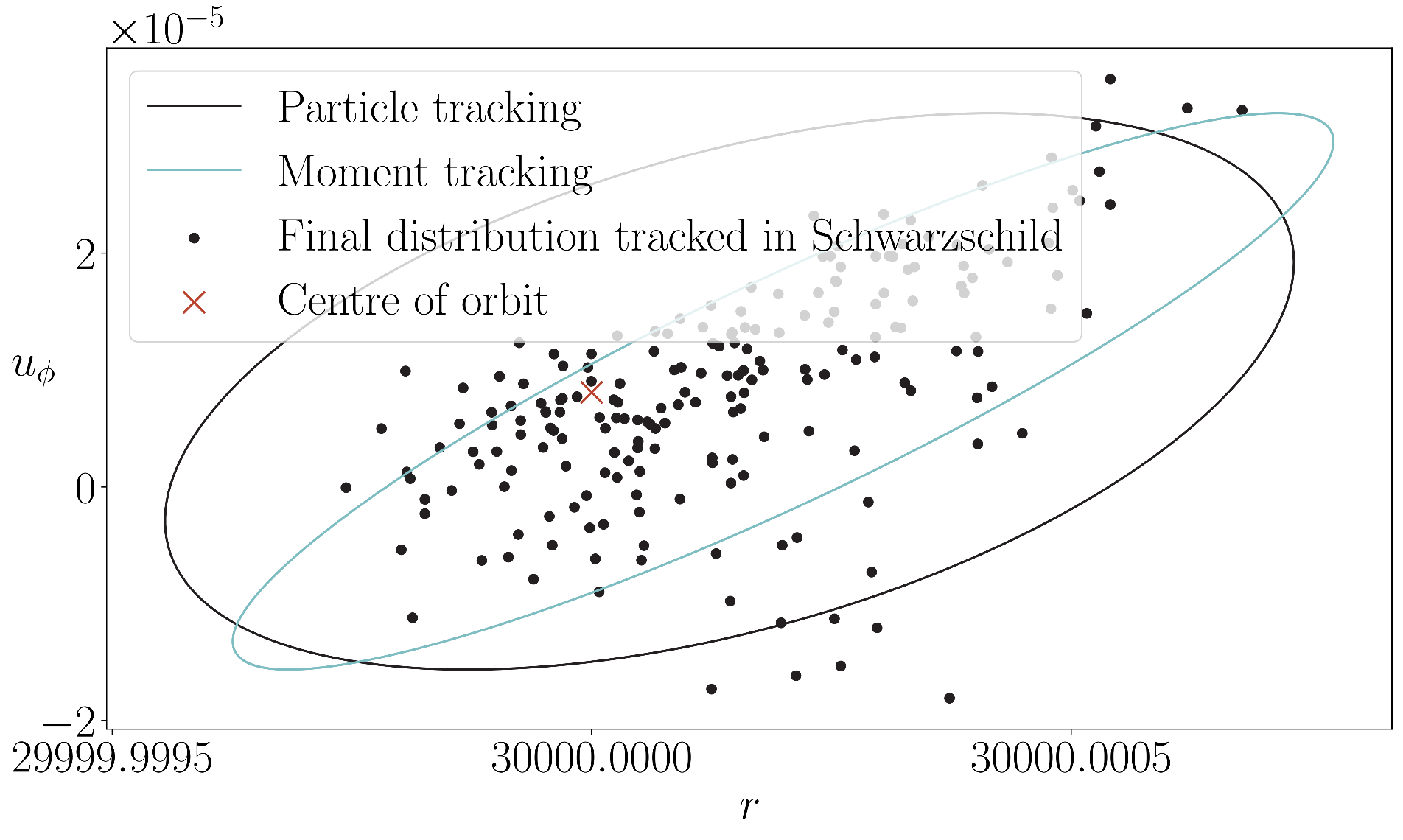}
			\caption{$t=10$ showing particles and moments tracked in Schwarzschild coordinates. }
			\label{subfig:PhaseSpacet2Schwarz}
		\end{subfigure}
    \begin{subfigure}[t]{.3\linewidth}
			\centering\includegraphics[width=\linewidth]{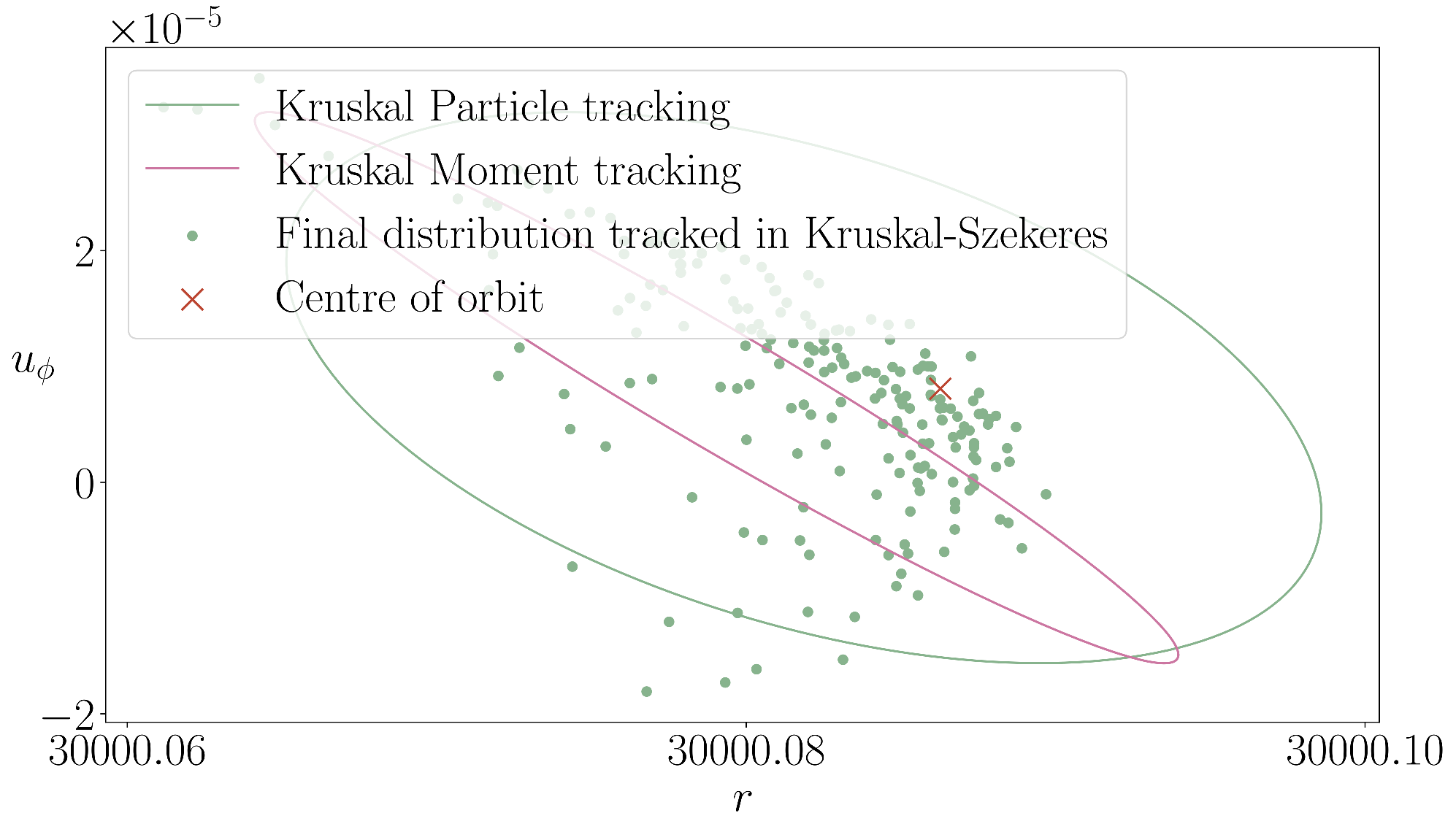}
			\caption{$t=10$ showing the same particles transformed to Kruskal-Szekeres coordinates, then the moments and particles tracked, then transformed back to Schwarzschild coordinates. }
			\label{subfig:PhaseSpacet2Kruskal}
		\end{subfigure}
		\caption{The $(r,u_\phi)$ phase space portraits in Schwarzschild coordinates for the individual particles, the centre of orbit with the path $\Map$, and the ellipses used to visualise the actual moments and tracked moments of these particles. Also shown are the ellipses generated from tracking the same particles and their moments in Kruskal-Szekeres coordinates, then the moments coordinate transformed back into Schwarzschild coordinates. These ellipses show the range that 95\% of particles will be within, if the particles are normally distributed. The reason that 95\% of particles are not within the ellipses in \ref{subfig:PhaseSpacet2Kruskal} and \ref{subfig:PhaseSpacet2Schwarz} is because the data is no longer normally distributed. Note that if just a standard macroparticle was tracked, only the centre of orbit would be known. [Associated dataset available at http://dx.doi.org/10.5281/zenodo.8082181] (Ref. \cite{Dataset}).}
		\label{fig:PhaseSpace}
	\end{center}
\end{figure*}		 

The results of the tracking are shown in figure \ref{fig:PhaseSpace}, with the moments used to calculate confidence ellipses in Schwarzschild spacetime. Figure \ref{fig:PhaseSpace} shows two things: firstly, whilst there is some deviation between the moment tracking and particle tracking ellipses, neither accurately reflect the underlying distribution of particles. This is because the data develops a large skew, as the faster particles orbits get thrown radially outwards. This means the particle distribution is no longer normally distributed, and as such, cannot be modelled accurately with just the first and second order moments. To improve this, higher order moments will also need to be tracked. All models correctly predict the spread in the radial coordinate, such that if the $r$ and $u_\phi$ axes were equally scaled, all models would correctly predict a long, horizontally thin ellipse. The second feature is that whilst the data is initially uncoupled, the velocity and position very quickly develop a covariance (a $V^{16}$ moment), coupling the $u_{\phi}$ and $r$ motion together. In this particular case, this is an expected result, as particles with a faster magnitude of $u_\phi$ than the one required for a circular orbit are spiralling outwards to a higher radius. \\

There is a noticeable displacement in the radial coordinate in the Kruskal-Szekeres coordinates tracking (figure \ref{subfig:PhaseSpacet2Schwarz}) from 30000 to 30000.08. This is because in Kruskal-Szekeres coordinates, small errors from the numerical integration will compound, and result in the ideal orbit drifting from its expected position. This compounding of errors is because Kruskal-Szekeres coordinates involves exponential functions, so small deviations can result in substantial offsets, which cannot be reduced by decreasing step size. This effect is small, and can be avoided by prescribing $\Map$ beforehand, if it is known. If $\Map$ is not known beforehand it is still not a major issue, as the small offset from these numerical factors is likely to be small compared to the other effects within a plasma. \\

\begin{table}[tb]
	\centering
	\begin{tabular}{|p{.9\linewidth}|}
\hline
\multicolumn{1}{|c|}{Error and description of the error}  \\ \hline 
 \multicolumn{1}{|c|}{$ \epsilon_{\text{\textbf{sp}}-\text{\textbf{sm}}} = \sqrt{\sum_{\OneToNa} \left|V^\OneToNa_{\text{\textbf{sp}}} - V^\OneToNa_{\text{\textbf{sm}}} \right|^2 + \sum_{\OneToNa \OneToNb} \left| V^{\OneToNa \OneToNb}_{\text{\textbf{sp}}} - V^{\OneToNa \OneToNb}_{\text{\textbf{sm}}} \right|}$} \\ The error between tracking particles then taking moments, compared to tracking moments, both in Schwarzschild coordinates. \\ \hline
\multicolumn{1}{|c|}{$\epsilon_{\text{\textbf{sp}}-\text{\textbf{kp}}} = \sqrt{\sum_{\OneToNa} \left|V^\OneToNa_{\text{\textbf{sp}}} - \hat{V}^\OneToNa_{\text{\textbf{kp}}} \right|^2 + \sum_{\OneToNa \OneToNb} \left| V^{\OneToNa \OneToNb}_{\text{\textbf{sp}}} - \hat{V}^{\OneToNa \OneToNb}_{\text{\textbf{kp}}} \right|}$} \\ The error between tracking particles and taking moments in Kruskal-Szekeres coordinates then transforming these into moments in Schwarzschild coordinates, compared to tracking particles and taking moments in Schwarzschild coordinates. \\ \hline
\multicolumn{1}{|c|}{$\epsilon_{\text{\textbf{sp}}-\text{\textbf{km}}} = \sqrt{\sum_{\OneToNa} \left|V^\OneToNa_{\text{\textbf{sp}}} - \hat{V}^\OneToNa_{\text{\textbf{km}}} \right|^2 + \sum_{\OneToNa \OneToNb} \left| V^{\OneToNa \OneToNb}_{\text{\textbf{sp}}} - \hat{V}^{\OneToNa \OneToNb}_{\text{\textbf{km}}} \right|}$} \\ The error between tracking moments in Kruskal-Szekeres coordinates then transforming these into moments in Schwarzschild coordinates, compared to tracking moments in Schwarzschild coordinates. \\ \hline
 \multicolumn{1}{|c|}{$\epsilon_{\text{\textbf{kp}}-\text{\textbf{km}}} = \sqrt{\sum_{\OneToNa} \left|V^\OneToNa_{\text{\textbf{kp}}} - V^\OneToNa_{\text{\textbf{km}}} \right|^2 + \sum_{\OneToNa \OneToNb} \left| V^{\OneToNa \OneToNb}_{\text{\textbf{kp}}} - V^{\OneToNa \OneToNb}_{\text{\textbf{km}}} \right|}$} \\ The error between tracking particles then taking moments, compared to tracking moments, both in Kruskal-Szekeres coordinates. \\ \hline
\multicolumn{1}{|c|}{$\epsilon_{\text{\textbf{kp}}-\text{\textbf{sp}}} = \sqrt{\sum_{\OneToNa} \left|V^\OneToNa_{\text{\textbf{kp}}} - \hat{V}^\OneToNa_{\text{\textbf{sp}}} \right|^2 + \sum_{\OneToNa \OneToNb} \left| V^{\OneToNa \OneToNb}_{\text{\textbf{kp}}} - \hat{V}^{\OneToNa \OneToNb}_{\text{\textbf{sp}}} \right|}$ } \\ The error between tracking particles and taking moments in Schwarzschild coordinates then transforming these into moments in Kruskal-Szekeres coordinates, compared to tracking particles and taking moments in Kruskal-Szekeres coordinates.\\ \hline
\multicolumn{1}{|c|}{$\epsilon_{\text{\textbf{kp}}-\text{\textbf{sm}}} = \sqrt{\sum_{\OneToNa} \left|V^\OneToNa_{\text{\textbf{kp}}} - \hat{V}^\OneToNa_{\text{\textbf{sm}}} \right|^2 + \sum_{\OneToNa \OneToNb} \left| V^{\OneToNa \OneToNb}_{\text{\textbf{kp}}} - \hat{V}^{\OneToNa \OneToNb}_{\text{\textbf{sm}}} \right|}$} \\ The error between tracking moments in Schwarzschild coordinates then transforming these into moments in Kruskal-Szekeres coordinates, compared to tracking moments in Kruskal-Szekeres coordinates. \\ \hline
\end{tabular}
	\caption{The types of error the numerical testing generates. These errors show the accuracy of both the moment tracking model and the coordinate transformations by comparing the results to a particle tracking model. Figure \ref{fig:TestingProcess} shows these errors diagrammatically.}
	\label{tab:ErrorAnalysis}
\end{table}

\begin{figure*}[tb]
	\centering
\begin{tikzcd}[row sep=4em, column sep=3em]
                                                                                 & {\substack{\text{Schwarzschild} \\ \text{spacetime}}} \arrow[rrrr, "\text{Transform coordinates}", leftrightarrow] \arrow[ld, "\substack{\text{Transport} \\ \text{particles}}"'] \arrow[rd, "\substack{\text{Take} \\ \text{moments}}"] &                                                                                                        & &                                        & {\substack{\text{Kruskal-} \\ \text{Szekeres} \\ \text{coordinates}}} \arrow[ld, "\substack{\text{Take} \\ \text{moments}}"'] \arrow[rd, "\substack{\text{Transport} \\ \text{particles}}"] &                              \\
	{} \arrow[dd, "\substack{\text{Take} \\ \text{moments}}"']                                                                    &                                                                                    & {} \arrow[d, "\substack{\text{Transport} \\ \text{ moments}}"]                                                                        & & {} \arrow[d, "\substack{\text{Transport} \\ \text{ moments}}"']        &                                                             & {} \arrow[dd, "\substack{\text{Take} \\ \text{moments}}"] \\
	&                                                                                    & V_{\text{\textbf{sm}}} \arrow[lld, "\epsilon_{\text{\textbf{sp}}-\text{\textbf{sm}}}"', rightarrow] \arrow[rrrrd, "\epsilon_{\text{\textbf{kp}}-\text{\textbf{sm}}}"', rightarrow] & & V_{\text{\textbf{km}}} \arrow[lllld, "\epsilon_{\text{\textbf{sp}}-\text{\textbf{km}}}",] \arrow[rrd, "\epsilon_{\text{\textbf{kp}}-\text{\textbf{km}}}", rightarrow] &                                                             &                              \\
	V_{\text{\textbf{sp}}} \arrow[rrrrrr, bend left = 5, "\epsilon_{\text{\textbf{kp}}-\text{\textbf{sp}}}"] \arrow[rrrrrr, leftarrow, bend right = 5, "\epsilon_{\text{\textbf{sp}}-\text{\textbf{kp}}}"'] &                                                                                    &                                                                                                        & &                                        &                                                             & V_{\text{\textbf{kp}}}                            
\end{tikzcd}
\caption{The model used to test the moment tracking and coordinate transformation theories. The error in the moment tracking model is the difference between transporting the moments and transporting the particles then taking moments. The error in coordinate transforming is the difference between transporting particles then taking moments in each frame. The combined error is the difference between transporting the moments in one frame, compared to tracking particle in the other. These errors are defined algebraically in table \ref{tab:ErrorAnalysis}.}
\label{fig:TestingProcess}
\end{figure*}
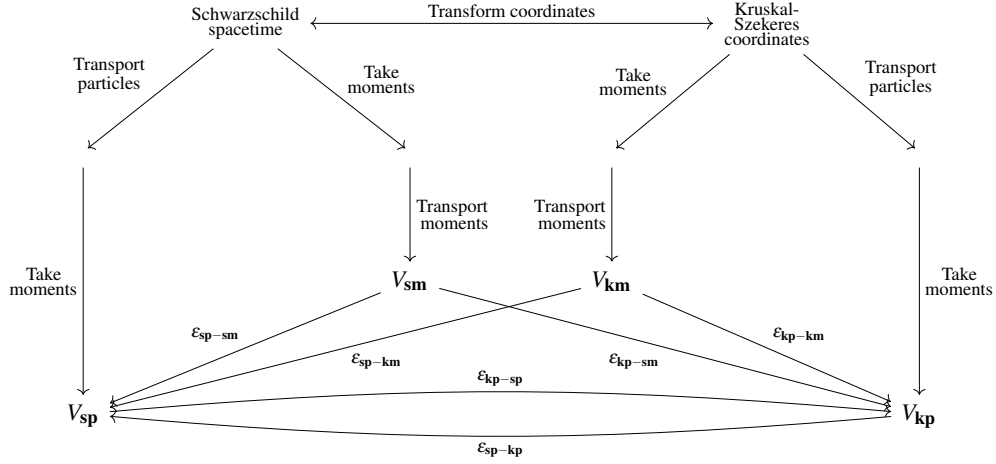

To quantify the error in the model, there are six different errors that are analyzed: $\epsilon_{\text{\textbf{sp}}-\text{\textbf{sm}}}, \epsilon_{\text{\textbf{sp}}-\text{\textbf{kp}}}, \epsilon_{\text{\textbf{sp}}-\text{\textbf{km}}}, \epsilon_{\text{\textbf{kp}}-\text{\textbf{km}}}, \epsilon_{\text{\textbf{kp}}-\text{\textbf{sp}}},$ and $\epsilon_{\text{\textbf{kp}}-\text{\textbf{sm}}}$. These errors are defined in table \ref{tab:ErrorAnalysis}. The subscript \textbf{sp} represents that particles were tracked, then the moments taken and the end of the simulation, all in Schwarzschild coordinates. Whilst a subscript \textbf{km} represents moments being tracked in Kruskal-Szekeres coordinates. These subscripts are defined pictorially in figure \ref{fig:TestingProcess}. A hatted $V$ in table \ref{tab:ErrorAnalysis} represents a coordinate transformation. As an example, $\epsilon_{\text{\textbf{sp}}-\text{\textbf{sm}}} $ represents the difference between tracking the group of particles and taking their moments at the end (the black ellipse in figure \ref{subfig:PhaseSpacet2Schwarz}), compared to tracking the moments using equation \eqref{eq:MomentTransportEquations} (the blue ellipse in figure \ref{subfig:PhaseSpacet2Schwarz}), all in Schwarzschild coordinates. These errors assess either the error in the moment tracking model, the error in the coordinate transformations, or the combined error of both. These six errors allow both the errors in moment tracking and coordinate transformations to be quantified and measured over time.\\

The error $\epsilon_{\textbf{km}-\textbf{sm}}$ could also be calculated, to show the error in the coordinate transformations between two different of moments that were tracked. The origin of this error would not be discernible, as it would be impossible to distinguish between errors caused by moment tracking in either coordinate system, or the error from the coordinate transformation. \\
	
\begin{figure}[tb]
	\begin{center}
		\begin{subfigure}[t]{.8\linewidth}
			\centering\includegraphics[width=\linewidth]{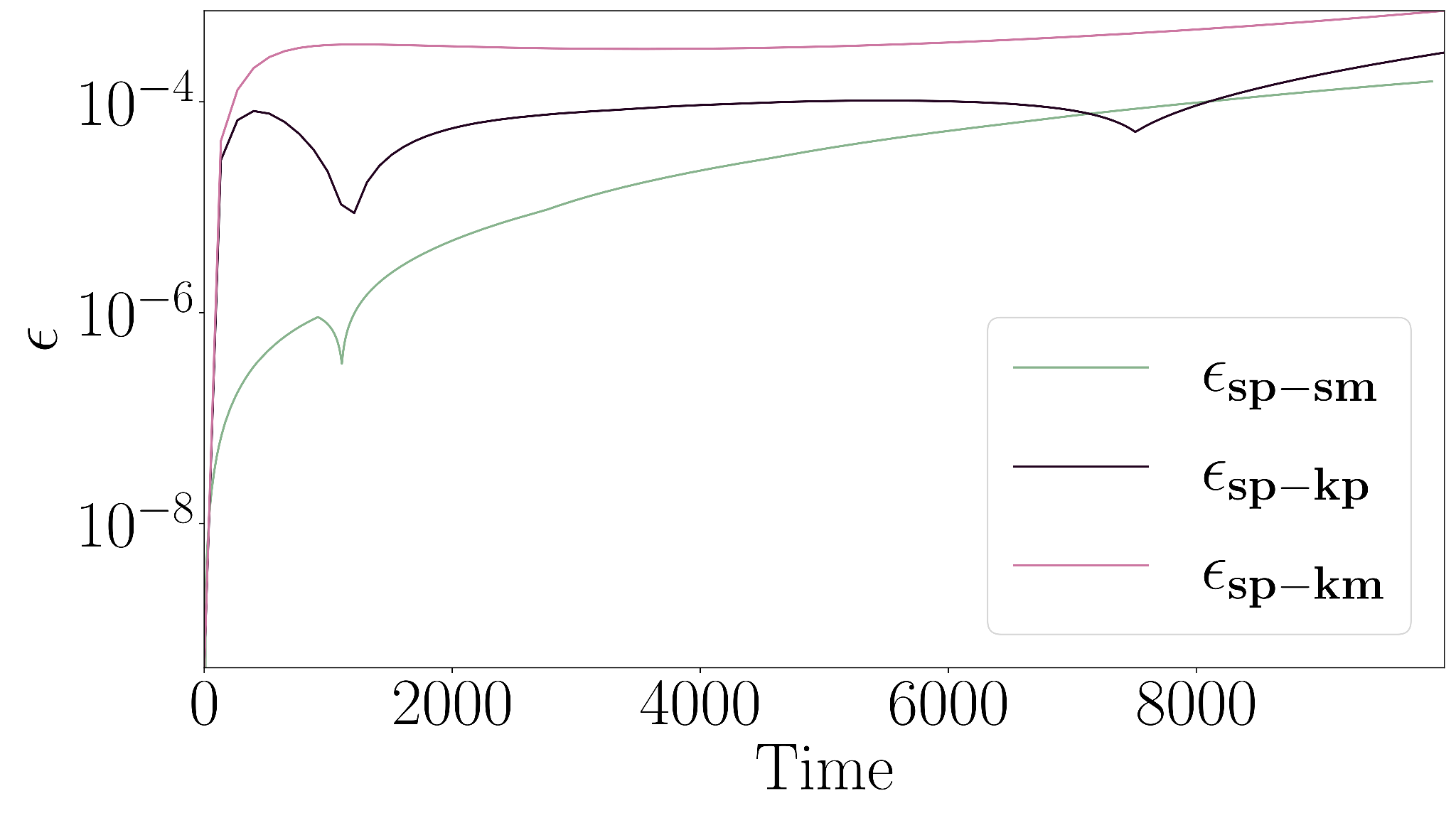}
			\caption{Schwarzschild coordinates.}
			\label{subfig:Schwarztimesteps}
		\end{subfigure}
		\qquad
		\begin{subfigure}[t]{.8\linewidth}
			\centering\includegraphics[width=\linewidth]{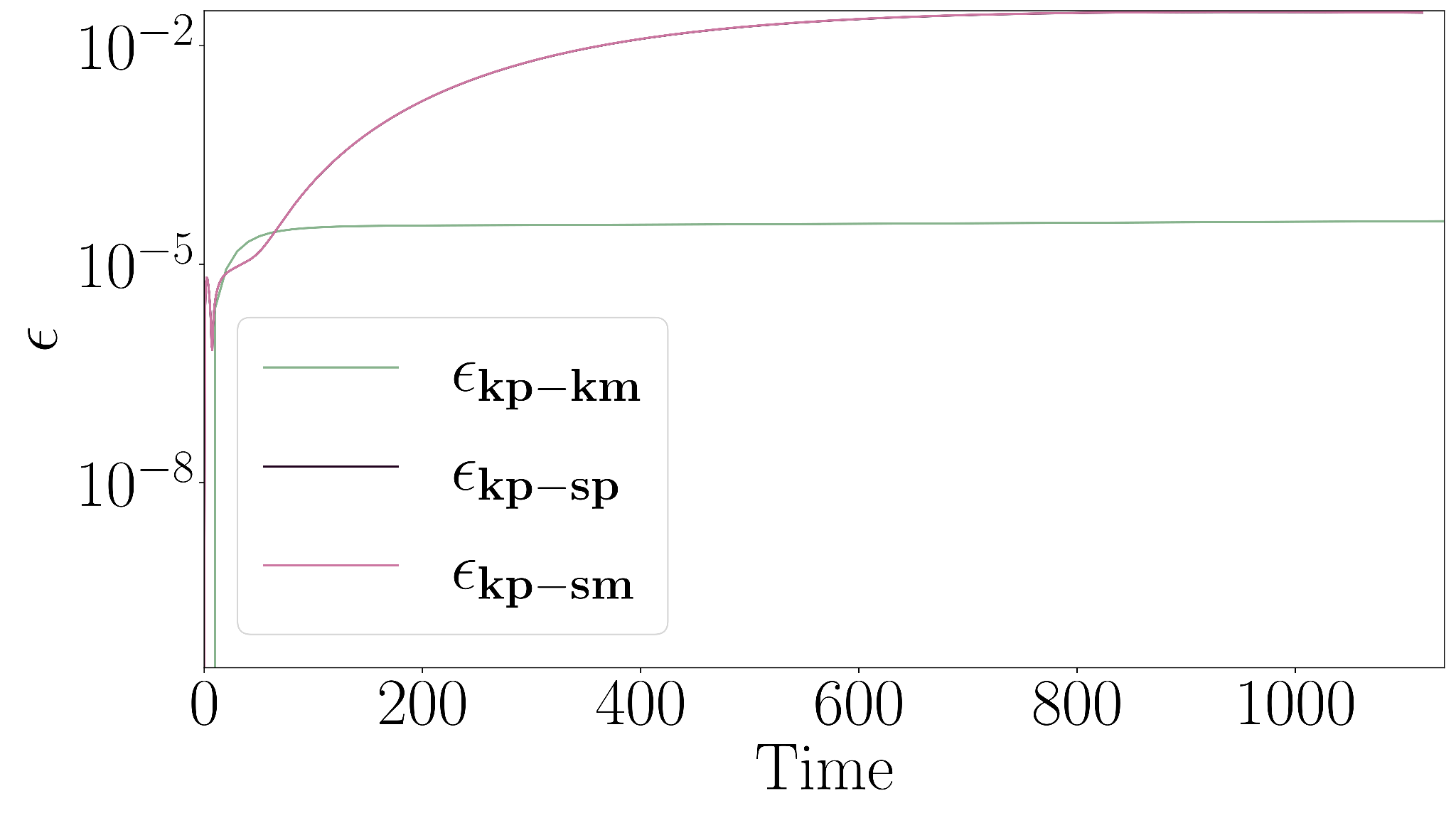}
			\caption{Kruskal-Szekeres coordinates. Note both blue and pink lines overlap.}
			\label{subfig:Kruskaltimesteps}
		\end{subfigure}
		\caption{The total error as a function of time for the different kinds of errors the theory can generate. [Associated dataset available at http://dx.doi.org/10.5281/zenodo.8082181] (Ref. \cite{Dataset}).}
		\label{fig:TotalErrorComparison}
	\end{center}
\end{figure}

To examine the error, another simulation was performed, again around a black hole with Schwarzschild radius $r_s = 3000$, and an ideal circular orbit at $r=30000$ in Schwarzschild coordinates. For these simulations the number of particles was decreased to $20$, and the time step used was increased to $\Delta t = 0.1$, with $10^6$ total iterations. This adjustment was made to allow the simulation to run for longer, to obtain better information about the long term behaviour of the model. Note the time step size is in the respective frame, so $t=10000$ in Schwarzschild coordinates corresponds to $T=1200$ in Kruskal-Szekeres coordinates. Figure \ref{fig:TotalErrorComparison} shows the three different total errors as functions of time, where the particles are normally distributed around the ideal orbit with variance $10^{-20}$ in all dimensions. \\

\begin{figure}[tb]
	\begin{center}
		\begin{subfigure}[t]{.8\linewidth}
			\centering\includegraphics[width=\linewidth]{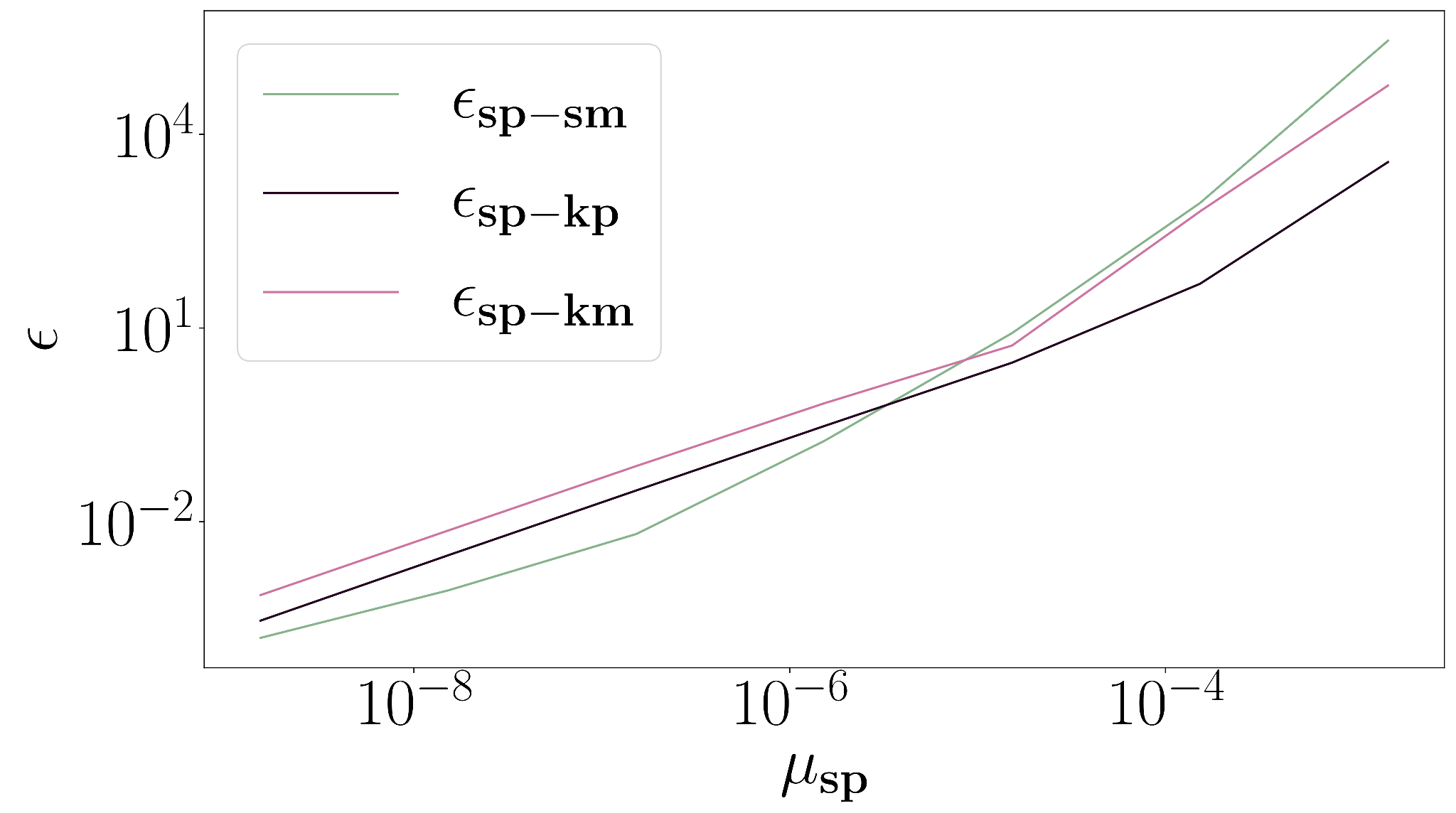}
			\caption{Schwarzschild coordinates.}
			\label{subfig:SchwarzErrorConvergence}
		\end{subfigure}
		\begin{subfigure}[t]{.8\linewidth}
			\centering\includegraphics[width=\linewidth]{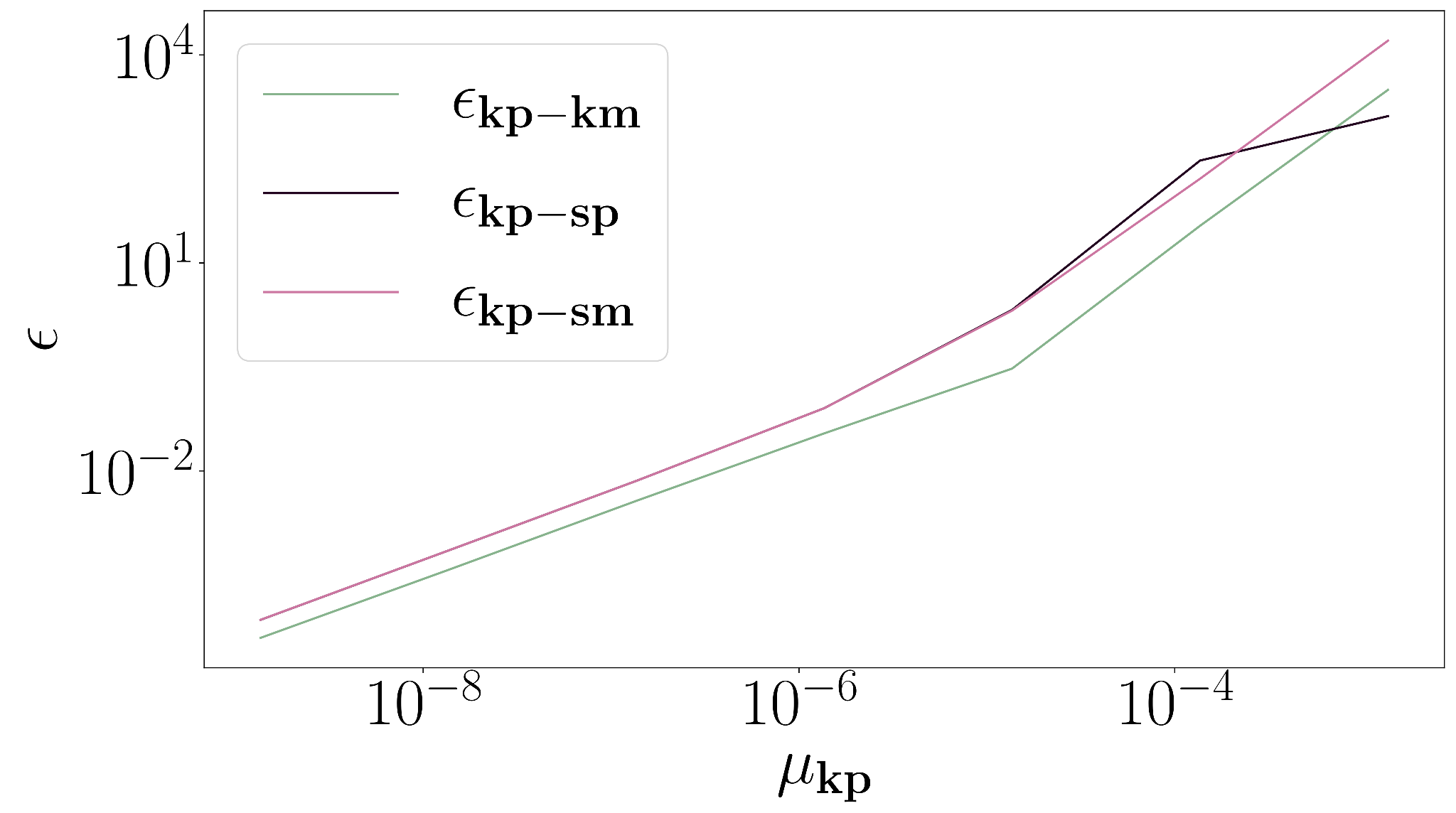}
			\caption{Kruskal-Szekeres coordinates. Note blue and pink lines overlap for $\mu_{\textbf{kp}} < 10^{-5}$.}
			\label{subfig:KruskalErrorConvergence}
		\end{subfigure}
		\caption{The total error as a function of $\mu$, the initial total moment, for the three different kinds of errors considered. The gradient of the lines is approximately linear in low $\mu$, and approximately 1.7 for larger $\mu$. [Associated dataset available at http://dx.doi.org/10.5281/zenodo.8082181] (Ref. \cite{Dataset}).}
		\label{fig:ErrorConvergence}
	\end{center}
\end{figure}

The error in all methods is small, although the error from the coordinate transformations is more substantial than the error from the moment tracking. It is postulated that this increased error is because the higher order moments affect the coordinate transformations twice, once during the coordinate transformation (during equation \eqref{eq:EllisCoordTransforms}), and once in the projection (during equation \eqref{eq:EllisProjection}). This suggests that for a moment tracking code that also incorporates a coordinate transformation, a higher order of moments will be needed. The bumps and discontinuities in the results are due to particles passing the macroparticle centre. If a particle is travelling faster than the macroparticle centre, the particle tracking moments (i.e. $V_{\text{\textbf{sp}}}$) will decrease, then increase once the particle passes the macroparticle centre. The moment tracking code will not see this behaviour, and will track the moments as always either decreasing or increasing, rather than the true mixture of both. These discontinuities are an intrinsic part of modelling moments. They can be avoided by sorting the particles before the modelling starts, so higher speed particles are ahead of the macroparticle centre, but this is no longer realistic. \\

To study the magnitude of the error, rather than just the shape, 3 main sources of error can be identified: floating point errors, numerical errors (the error arising from finite step size in numerical integration), and truncation errors (the error arising from truncating the multipole expansion at second order). These errors will dominate in different ways depending on the size of the total initial moments $\mu$, where
\begin{equation}
	\mu = \sum_{\OneToNa} \Big| V^{\OneToNa}|_{t=0} \Big|^2 + \sum_{\OneToNa \OneToNb} \left| V^{\OneToNa \OneToNb}|_{t=0} \right|.
\end{equation} 

Floating point errors can arise from a sufficiently small time step and small number of iterations in any numerical differential equation solver, but in the case of this work they also dominate if the moments are very small i.e. $\mu \approx 10^{-15}$. Numerical errors arise from the choice of integrator used, and in the case of forward Euler, are linear in $\Delta t$. The truncation errors arise from only running the moment tracking code up to quadrupole order. There is an infinite expansion of moments, which is truncated to quadrupole order in this paper. Including more moments will decrease the total error. At the quadrupole level, the truncation error is quadratic. This means total error $\epsilon(\mu) \approx \mu^2$. This is verified in figure \ref{fig:ErrorConvergence}. The error is linear in the low $\mu$ regime, where numerical errors dominate, and as $\mu$ increases, the total error increases at about $\epsilon(\mu) \approx \mu^{1.7}$. This is a combination of the predicted quadratic increase, and the numerical error. This quadratic behaviour suggests that if the moments are half the size, the total error will be quartered.

\section{Geometric interpretation of the multipole transport equations}
\label{sec:CoordinateFree}

\subsection{Introducing de Rham current distributions}

In this section we present the transport equations and coordinate transformations of multipoles in the language of differential geometry and de Rham currents \cite{gratusCorrectUnusualCoordinate2018, gratusDistributionalStressEnergy2020}. By using this method there is a more obvious split between the $V^\OneToNa$ and $X^{\OneToNa \OneToNb}$ components. This split means it is simpler to isolate each term when doing complicated calculations that mix the two terms, such as during coordinate transformations. It also means the evolution of the moments can be described in a coordinate free way. This is in contrast to working directly from equation \eqref{eq:GeneralMoment}, which is highly dependent on the coordinate system, in particular the choice of time slicing. The Ellis method also requires a coordinate system to define the action on a test form. \\

On a manifold $M$ with tangent bundle $TM$ the \FrameName frame hypersurface $E$ is defined such that the $Z^0$ component of any vector on $M$ is $1$,
\begin{equation}
	E = \left \{ Z \in TM \; | \; Z^0 = 1 \right\}. \label{eq:PhaseSpaceHypersurface}
\end{equation}

The bundle of $p$-forms is written $\Lambda^p E$ such that a specific $p$-form (field) is denoted as $\alpha \in \Gamma \Lambda^p E$. A vector field is denoted as $V \in \Gamma TE$.\\

A distributional $p$-form is defined by its action on a test $(7-p)$-form $\phi \in \Gamma \Lambda^{7-p} E$, this is a $(7-p)$-form with compact support. Given $\alpha \in \Gamma \Lambda^p E$ is a smooth $p$-form, a regular distribution $\alpha^D$ is constructed via
\begin{equation}
	\alpha^D [\phi] = \int_E \phi \wedge \alpha.
 \label{eq:DistributionOnTestForm}
\end{equation}

The definition of the wedge product, Lie derivatives, internal contractions and exterior derivatives for an arbitrary distribution $\Psi$ are defined as
\begin{equation}
	\begin{gathered}
		\left( \Psi_1 + \Psi_2 \right)[\phi] = \Psi_1[\phi] + \Psi_2[\phi], \\ \left(\beta \wedge \Psi\right)[\phi] = \Psi[\phi \wedge \beta], \\
		i_V \Psi [\phi] = -(-1)^{\text{deg} \phi} \Psi[i_V \phi], \quad d \Psi [\phi] = -(-1)^{\text{deg} \phi} \Psi[d \phi], \\ L_V \Psi [\phi] = -\Psi[L_V \phi] .
	\end{gathered}
 \label{eq:DistributionDefinitions}
\end{equation}
where $\beta \in \Gamma \Lambda^{q} E$ and $V \in \Gamma T E$. \\

The order of a $p$-form distribution is defined as follows. If
\begin{equation}
\begin{gathered}
	\Psi[\lambda^{k+1}\phi] = 0 \text{ for all } \phi \in \Gamma \Lambda^q E \text{ with compact support and } \\ \lambda \in \Gamma \Lambda^0 E \text{ such that } \Map^*(\lambda) = 0
\end{gathered}
\end{equation}
where $\Map^*$ is the pullback, then the order of $\Psi$ is at most $k$. Note that a quadrupole (a distribution of order 2) also includes the dipole and the monopole terms. Lie derivatives and exterior derivatives both increase order by one. Internal contractions do not affect the order of a distribution. \\

Given $\Map: \mathbb{R} \to E$ is a closed embedding parameterised by $t$, the de Rham pushforward with respect to $\Map$ of a $p$-form $\alpha \in \Gamma \Lambda^p \mathbb{R}$ is given by the distribution
\begin{equation}
	\Map_\varsigma(\alpha)[\phi] = \int_{\mathbb{R}} \Map^*(\phi) \wedge \alpha.
\end{equation}	
 The degree of the distribution $\Map_\varsigma(\alpha)$ is $6+ \text{deg}(\alpha)$. Since $\mathbb{R}$ is a curve, the degree of $\alpha$ is either $0$ or $1$, and the degree of $\Map_\varsigma(\alpha)$ is either $6$ or $7$. An internal contraction decreases the degree by one and an exterior derivative increases the degree by one. Lie derivatives do not affect the degree of a distribution. \\

A distribution $\Psi$ of degree $6$ is a semi-multipole of order at most $l$ if

\begin{equation}
\begin{gathered}
	\Psi [\lambda^l d\mu] = 0 \quad \text{for all } \lambda, \mu \in \Gamma \Lambda^0 E \\ \text{ such that } \Map^*(\lambda) = \Map^*(\mu) = 0.
\end{gathered}
\end{equation}

The integral curves of the Vlasov vector field are tangent to the de Rham pushforward, such that
\begin{equation}
	i_W \Map_\varsigma(\alpha) = \Map_\varsigma (i_{\frac{d}{dt}} \alpha).
\end{equation}

This work concerns the dynamics of a semi-quadrupole, which is a semi-multipole $\mathcal{J}$ of order 2 and degree 6. In coordinates, this is denoted as
\begin{multline}
	\mathcal{J} = \tfrac{1}{2} L_\OneToNa L_\OneToNb \Map_\varsigma (V^{\OneToNa \OneToNb}) - i_\OneToNa L_\OneToNb \Map_\varsigma (X^{\OneToNa \OneToNb} dt) \\ 
  - L_\OneToNa \Map_\varsigma (V^{\OneToNa} ) + i_\OneToNa \Map_\varsigma(X^\OneToNa dt) + \Map_\varsigma(q). 
\end{multline}
where $L_\OneToNa$ is the Lie derivative with respect to $\partial_\OneToNa$, and $i_\OneToNa$ is the internal contraction with respect to $\partial_\OneToNa$. Note in this representation of the semi-quadrupole, the $X^{\OneToNa \OneToNb}$ and $V^{\OneToNa}$ term are easier to distinguish by the additional internal contraction, as opposed to the Ellis representation (equation \eqref{eq:EllisSemiQuadrupole}), in which the separation between the terms was less clear. \\

The de Rham current representation of a multipole can be related to the Ellis representation through the relationship
\begin{equation}
  \mathcal{J} = \mathcal{J}^{\OToNa} i_\OToNa d^7 \xi,
  \label{eq:GeometricToEllis}
\end{equation}
and conversely
\begin{equation}
  \mathcal{J}^\OToNa = \mathcal{J} \wedge d\xi^\OToNa.
\end{equation}

In $\mathcal{J}$, there is no term of the form $i_\OneToNa L_\OneToNb L_\OneToNc \Map_{\varsigma} (X^{\OneToNa \OneToNb \OneToNc} dt)$ even though this contains two Lie derivatives. Similarly, there is no $X^{\OneToNa \OneToNb \OneToNc}$ term in the Ellis representation. This term is included if $\mathcal{J}$ is a full quadrupole, as opposed to a semi-quadrupole. The advantage of the coordinate free approach used in this section is that it can be shown that the $X^{\OneToNa \OneToNb \OneToNc}$ term vanishes in a coordinate system adapted to $\Map$, and hence $\mathcal{J}$ is a semi-quadrupole in this adapted coordinate system. Since the definition of a semi-multipole is coordinate free, $\mathcal{J}$ is a semi-quadrupole in every coordinate system. The coordinate free semi-multipoles written in this form correspond to the semi-multipoles of ref. \cite{gratusDistributionalStressEnergy2020}, and the electric multipoles of ref. \cite{gratusCorrectUnusualCoordinate2018}.

\subsection{The distributional transport equations}

By using the language of differential geometry, a more geometric approach to understanding the origin of the transport equations can be found. To find the dynamics of multipoles defined through distributions, the equivalent of equations \eqref{eq:ChargeConservation} and \eqref{eq:VlasovEquationCondition} for distributions are used,
\begin{equation}
	d \mathcal{J} = 0, \quad i_W \mathcal{J} = 0,
  \label{eq:deRhamTransportEquations}
\end{equation}
where $W = W^\OToNa \partial_\OToNa$. The $d\mathcal{J} = 0$ condition corresponds to the conservation of charge, and the $i_W \mathcal{J}$ condition says that the flow lines of $\mathcal{J}$ are the integral curves of the Vlasov field. These equations are the same as equations \eqref{eq:ChargeConservation} and \eqref{eq:VlasovEquationCondition}, i.e. it is equivalent to the Ellis representation.

\begin{proof}
  Using equation \eqref{eq:GeometricToEllis}
	\begin{align}
	  d\mathcal{J} &= d\big(\mathcal{J}^\OToNa i_\OToNa d^7 \xi \big) = d\big(\mathcal{J}^\OToNa \big) \wedge i_\OToNa d^7 \xi + \mathcal{J}^\OToNa L_\OToNa d^7 \xi \nonumber \\
    &= \partial_\OToNb \mathcal{J}^\OToNa d\xi^\OToNb i_\OToNa d^7 \xi + (0) = \partial_\OToNb \mathcal{J}^\OToNa \delta^\OToNb_{\OToNa} d^7 \xi \nonumber \\ &= \partial_\OToNa \mathcal{J}^\OToNa d^7 \xi. 
	\end{align}
This is zero if and only if $\partial_\OToNa \mathcal{J}^\OToNa = 0$, so $d\mathcal{J} = 0$ is equivalent to \eqref{eq:ChargeConservation}. For the $i_{W} \mathcal{J}$ term, 
\begin{equation}
  i_W \mathcal{J} = i_W \big(\mathcal{J}^\OToNa i_\OToNa d^7 \xi \big) = \mathcal{J}^\OToNa W^\OToNb i_\OToNb i_\OToNa d^7 \xi.
\end{equation}
Since $i_\OToNb i_\OToNa$ is antisymmetric, we can take the symmetric part of $\mathcal{J}^\OToNa W^\OToNb$,
\begin{equation}
  i_W \mathcal{J} = \left(\mathcal{J}^\OToNa W^\OToNb + \mathcal{J}^\OToNb W^\OToNa \right) i_\OToNb i_\OToNa d^7 \xi.
\end{equation}
This is zero if and only if $\mathcal{J}^\OToNa W^\OToNb + \mathcal{J}^\OToNb W^\OToNa = 0$, so $i_W \mathcal{J} = 0$ is equivalent to equation \eqref{eq:VlasovEquationCondition}.
\end{proof}

Since the conditions in the geometric approach are the same as the Ellis representation, both languages give the same differential equations for the moments.

\subsection{Coordinate transformations}

Coordinate transformations can also be found using the language of distributions. The coordinate transformations for internal contractions are given by
\begin{equation}
	i_{\OneToNa} \alpha = A_{\OneToNa}^{\hat{\OToNb}} i_{\hat{\OToNb}} \alpha = i_{\hat{\OToNb}} (A_{\OneToNa}^{\hat{\OToNb}} \alpha)
\end{equation} 
and for Lie derivatives the coordinate transformations are
\begin{equation}
	L_{\OneToNa} \alpha = A_{\OneToNa}^{\hat{\OToNb}} L_{\hat{\OToNb}} \alpha = L_{\hat{\OToNb}} (A_{\OneToNa}^{\hat{\OToNb}} \alpha) - \alpha \wedge L_{\hat{\OToNb}} A_{\OneToNa}^{\hat{\OToNb}}.
\end{equation}
Note that similarly to the transformation of $\partial_\OneToNa$, the indices in the transformed coordinate system run from $(0 \ldots 6)$, whilst the original indices only ran from $(1 \ldots 6)$. From these the coordinate transformations for the semi-quadrupole can be found. This is equivalent to the coordinate transformations found through the Ellis representation (equation \eqref{eq:EllisCoordTransforms}).

\begin{proof}
	The proof of this is in the appendix \ref{sec:DistributionalCoordTransform}.
\end{proof}

As before, the transformed quadrupole is still based on the original time slicing. In this case the projections to the new time slicing are based on the internal contraction and Lie derivatives along $\Map$. The projections are given by
\begin{equation}
	i_0 = i_{\dot{\Map}} - \dot{\Map}^\OneToNa i_\OneToNa, \quad L_0 = L_{\dot{\Map}} - \dot{\Map}^\OneToNa L_\OneToNa.
\end{equation}

Pushing these through the distributions, \eqref{eq:QuadrupoleFullTransform} and \eqref{eq:DipoleFullTransform} are found. These coordinate transformations are also the same as the ones found through the Ellis representation.

\begin{proof}
	The proof of this is in appendix \ref{sec:DistributionalProjections}.
\end{proof}

\section{Considerations for implementing macroparticles with moments in a full particle-in-cell code} 
\label{sec:FullPicCode}

With the work presented, it is possible to calculate all the dynamics for a single macroparticle in an arbitrary spacetime and external electromagnetic field. To convert this into a full PIC code, a process that uses the moments to reconstruct the distribution function and deposit the charge over multiple grid points must be developed. This will allow inter-macroparticle effects to be modelled. This is work currently being undertaken. Once this deposition process has been developed, a standard finite-difference time-domain (FDTD) algorithm can be used to update the fields, so a full PIC code can be developed.\\
 
One approach to use moments to deposit the charge and current onto the grid is to use a model function \cite{johnTechniquesReconstructionDistribution2007, hulburtProblemsParticleTechnology1964}. A model function is a function that is assumed to be of a similar shape to the actual distribution of particles, then shaped to have to have the required moments. This method works well in cases where the distribution of particles is similar to a top-hat function i.e. the distribution of particles in a plasma. 
In particle accelerators, the distribution of particles around the ideal orbit is not a top-hat, but rather a Gaussian. The model function depositing method does not work well for modelling these. An alternative approach in this case is to model the particle distribution function as a series of Hermite polynomials \cite{cramerMathematicalMethodsStatistics1946}. This can be related to the moments, to approximate the distribution function using the moments and Hermite polynomials. It is possible to approximate the conventional plasma moments (the pressure tensor, the energy flux density etc.) used in magnetohydrodynamics by using the higher order moments to approximate the underlying distribution. This is through the same process as depositing charge and current onto the grid, and requires tracking moments of a higher order than the plasma moment e.g. finding the pressure tensor would need the hexadecapole moments to be tracked. The reconstructed model function is normalised to have the same monopole moment as the macroparticle, this ensures no extra charge is added into the system through deposition. As this method does not use shape functions that conserve charge exactly \cite{villasenorRigorousChargeConservation1992, esirkepovExactChargeConservation2001} it may be necessary to perform divergence cleaning to stop accumulation of charge through the FDTD Maxwell solver (See ref. \cite{birdsallPlasmaPhysicsComputer1985} for more detail on this procedure). It should be noted that the exactly charge conserving shape functions can only be used in Cartesian coordinates, and so cannot be applied to the curved spacetimes presented in this work. \\

As mentioned in the introduction, this work only considered particles interacting with external fields, and as such, it was not meaningful to talk about cell sizes for the system. In a full PIC code, further work will need to be done to find the appropriate cell size for a macroparticle with moments. As shown in this work, the dominating source of error in the moment tracking is when the field cannot be accurately modelled by a small number of derivatives at the macroparticle centre. This means that one limit for the cell size will be based on the density of grid points required to accurately find the derivatives of the fields. Because the macroparticle has an internal structure, it needs to span over several cells to allow the extra structure from tracking the moments to be reflected in the deposited charge and current. This is in contrast to existing methods, which deposit over several cells to reduce numerical instability \cite{birdsallPlasmaPhysicsComputer1985, hockneyMeasurementsCollisionHeating1971}. It may be that in cases where the fields can be approximated using a low resolution grid, that it is possible for a single macroparticle to be larger than the Debye length of the plasma. By tracking the moments, information about the structure within the macroparticle is known, and this affects how the charge and current will be deposited onto the grid points. Because of this, it may be that the effects of instabilities generated from using a finite grid (aliasing instabilities) \cite{brackbillEnergyMomentumConservation2016, barnesFiniteSpatialgridEffects2021, birdsallPlasmaPhysicsComputer1985, langdonEffectsSpatialGrid1970} are not as significant. \\

\section{Conclusion and discussion}
\label{sec:Conclusion}
 
This paper found the dynamics of the moments of a macroparticle obeying the Vlasov equation (equation \eqref{eq:MomentTransportEquations}), and the coordinate transformations of the quadrupole moments (equations \eqref{eq:QuadrupoleFullTransform} and \eqref{eq:DipoleFullTransform}). By using the Ellis representation or the de Rham current representation of the moments, coordinate transformations can be found between frames that mix the space and time coordinates. By representing a group of particles as a macroparticle and its moments, this can be used in PIC codes. These results were validated numerically for the case of particles orbiting a black hole. This was done by transporting particles and moments in both Schwarzschild and Kruskal-Szekeres coordinates and comparing results. The numerical results show that a large number of particles can be successfully modelled by a single macroparticle with moments, although for a full PIC code it is likely moments will need to be calculated to an order larger than the quadrupole. In addition, more macroparticles will need to be modelled, and a method to use the structure of the moments to deposit the charge and current onto the grid will need to be developed. By doing these, a full PIC code can be developed, after which it is possible to assess how the full process of moment tracking benefits PIC codes.\\
	
The dynamics of the moments depend on the Vlasov field and its derivatives. The focus of this paper was on how the number of moments taken affects the accuracy of the moment tracking. For a full PIC code, the derivatives of the Vlasov field will also be important. If the electromagnetic or gravitational fields quickly vary in space (i.e. the higher order derivatives of the fields are large), then this will mean more moments need to be tracked. The moment tracking model is likely to work well in situations where both the distribution of particles represented by a macroparticle can be described by only a small number of moments, and the variation in electromagnetic and gravitational fields across the volume the macroparticle represents is small, such that the Vlasov field across the extent of the macroparticle can be modelled by just the Vlasov field and a small number of its derivatives at the macroparticle centre. Since numerically calculating the derivatives of the electromagnetic field requires a high density grid, the moment tracking method may also work in cases where a high resolution grid is already needed, such as laser-solid interactions \cite{ridgersDenseElectronpositronPlasmas2013, wuParticleincellSimulationsLaser2018}. In such cases, the moment tracking method will be able to model much larger macroparticles, with less particles per cell, even if more cells are needed to compensate.\\

The modelling in this work was done at 10 Schwarzschild radii away from the black hole. This was to model a stable accretion disc. In the development of the theory, there was no assumption that the macroparticle is far away from the black hole, so there is nothing in the theory that stops a macroparticle being close to the singularity. In practice, close to the Schwarzschild radius, the error from only using a finite number of moments will become significant. There are two reasons for this, firstly, the particles being represented by the macroparticle and its moments will begin to undergo spaghettification, and this will cause the higher order moments to become large. Secondly, close to the singularity, the Schwarzschild Christoffel symbols become very large, and their derivatives will become even larger. This means the rate of change of the moments will be dominated by the higher order moments which are being neglected in the code. Because of this, the accuracy of the model is yet to be determined in the extreme environments close to the Schwarzschild radius. \\

Another potential application of the moment tracking method is to model inter-bunch forces within particle accelerators. It is possible to calculate the Li\'{e}nard-Wiechert fields directly from the moments of a moving quadrupole \cite{ellisElectromagneticFieldsMoving1966}. By using this method the electromagnetic field, and its derivatives, can be calculated without the need to deposit the charge onto a grid. This is particularly useful for modelling coherent synchrotron radiation in particle accelerators, where macroparticles are close together compared to the radius of the beam pipe, such that the effects of boundary conditions on the electromagnetic fields can be ignored. \\

It may be possible to add internal dynamics into the moment equations. Whilst this has been previously done in refs. \cite{shadwickmodelingMuonCooling1999} and \cite{channellMomentApproachCharged1983} for specific applications of muon cooling and space-charge respectively, it may be possible to come up with general internal dynamics equations by modifying the transport equations (equation \eqref{eq:MomentTransportEquations}). This would allow the moment tracking method to model intra-bunch effects within particle accelerators. \\

The applications of the coordinate transformations in astrophysical scenarios are wide. A particularly useful case is the transformation from the global time to the backwards light cone frame. This is particularly useful in cases where black holes are modelled in the fiducial observer (FIDO) frame, where the global time coordinate is significantly different to the backwards light cone frame global time. By doing this the difficulty of calculating the backwards light cone through ray tracing only needs to be done once, rather than each time step. This transformation would allow the moments observed by the observer a finite distance away from the black hole to be calculated. \\

There are also applications of the coordinate transformation formulae in circular particle accelerators. Results from accelerators are often presented in Frenet-Serret coordinates, where the parameter is the position along the beamline, rather than time. By finding either the Vlasov equation in Frenet-Serret coordinates, or finding the coordinate transformation between Cartesian and Frenet-Serret coordinate systems for a given beamline, the moment tracking can be applied to circular accelerators. Additionally by using the spacetime coordinate transformations presented in this article, the coordinate transformation into the frame of an accelerating bunch can be found (this is a similar transformation to the one presented in figure \ref{fig:differentfoliations2}). \\

Lastly the use of the Vlasov equation to model dynamics may be extended to modelling stress-energy-momentum quadrupoles as a source for linearised gravity. In both refs. \cite{gratusDistributionalStressEnergy2020} and \cite{gratusTensorialRepresentationDistributional2023} it was shown that the dynamics of stress-energy-momentum quadrupoles contain a number of free components, known as \emph{constitutive relations}. In the case of a plasma these constitutive relations may be determined by the Vlasov equation. In this case the dynamics will be governed by the divergenceless of the stress-energy-momentum tensor combined with the Vlasov equation. \\

\begin{acknowledgments}
AW would like to thank the Faculty of Science and Technology, Lancaster University for their support. JG is grateful for the support provided by STFC (the Cockcroft Institute ST/V001612/1).
\end{acknowledgments}

\section*{Author declarations}
\subsection{Conflict of interest}

The authors have no conflicts to disclose. \\

\section*{Data Availability Statement}

The data that support the findings of this study are openly available in Zenodo at http://dx.doi.org/10.5281/zenodo.8082181.

\bibliography{references}

\begin{widetext}
\appendix

\section{Proofs}

\subsection{Proof of unique components}
\label{sec:UniqueComponentsProof}

This section shows that by acting an Ellis distribution $\mathcal{J}^\OToNa$ on specific test forms, the components of a multipole can be extracted. If it is possible to do this, the components of a multipole are unique. The components are isolated using specific tests forms:
\begin{equation}
	\begin{gathered}
		q(t_0) = \lim_{\epsilon \to 0} \; \frac{1}{\epsilon} \; \int_{E} \mathcal{J}^\OToNa \; \delta^0_{\OToNa} \; \psi\left(\frac{t-t_0}{\epsilon} \right) \prod_{i=1}^6 \psi \left( \xi^i - \Map^i \right) d^7 \xi\\
		V^{\OneToNb}(t_0) = \lim_{\epsilon \to 0} \; \frac{1}{\epsilon} \; \int_{E} \mathcal{J}^\OToNa \; \delta^0_{\OToNa} \; \psi \left(\frac{t-t_0}{\epsilon} \right) \Big( \xi^\OneToNb - \Map^\OneToNb \Big) \prod_{i=1}^6 \psi \left( \xi^i - \Map^i \right) d^7 \xi\\
		V^{\OneToNb \OneToNc}(t_0) = \lim_{\epsilon \to 0} \; \frac{2}{\epsilon} \; \int_{E} \mathcal{J}^\OToNa \; \delta^0_{\OToNa} \; \psi\left(\frac{t-t_0}{\epsilon} \right) \Big( \xi^\OneToNb - \Map^\OneToNb \Big) \Big( \xi^\OneToNc - \Map^\OneToNc \Big) \prod_{i=1}^6 \psi \left( \xi^i - \Map^i \right) d^7 \xi\\
		X^\OneToNb(t_0) + q(t_0) \dot{\Map}^\OneToNb(t_0) = \lim_{\epsilon \to 0} \; \frac{1}{\epsilon} \; \int_{E} \mathcal{J}^\OToNa \; \delta^\OneToNb_{\OToNa} \; \psi\left(\frac{t-t_0}{\epsilon} \right) \prod_{i=1}^6 \psi \left( \xi^i - \Map^i \right)d^7 \xi\\ 
		X^{\OneToNb \OneToNc} (t_0) + V^{\OneToNb} (t_0) \dot{\Map}^\OneToNc(t_0) = \lim_{\epsilon \to 0} \; \frac{1}{\epsilon} \; \int_{E} \mathcal{J}^\OToNa \; \delta^\OneToNc_{\OToNa} \; \psi\left(\frac{t-t_0}{\epsilon} \right) \left( \xi^\OneToNb - \Map^\OneToNb \right) \; \prod_{i=1}^6 \psi \left( \xi^i - \Map^i \right) d^7 \xi
	\end{gathered}
\label{eq:SpecificTestForms}
\end{equation}	
where $\psi: \mathbb{R} \to \mathbb{R}$ is a test function such that $\psi_1 (0) = 1$, it is flat about zero and $\int \psi_1 (t) dt = 1$, $t_0$ is the point at which the moments are evaluated, and $E$, the \FrameName frame hypersurface, is defined by equation \eqref{eq:PhaseSpaceHypersurface} in section \ref{sec:CoordinateFree}.

\begin{proof}
	Only the $V^{\OneToNb \OneToNc}$ term and $X^{\OneToNb \OneToNc} + V^{\OneToNb} \dot{\Map}^{\OneToNc}$ term will be shown as the other terms follow trivially. \\
	
	Consider $\mathcal{J}^\OToNa$ acting on the $V^{\OneToNb \OneToNc}$ equation of \eqref{eq:SpecificTestForms}. The only non-zero derivatives are the $\xi^{\OneToNb}$ and $\xi^{\OneToNc}$ terms. There are three possibilities, each $\xi^{\OneToNb}$ or $\xi^{\OneToNc}$ can either not be differentiated, differentiated once, or differentiated twice. If it is not differentiated, then the evaluation at $\Map$ gives $ (\xi^\OneToNb - \Map^{\OneToNb})|_{\Map} = \Map^\OneToNb - \Map^\OneToNb = 0$. If it is differentiated exactly once, then $\partial_\OneToNa (\xi^{\OneToNb} - \Map^{\OneToNb}) = \delta^\OneToNb_{\OneToNa}$, a Kronecker delta. If this is differentiated twice, then the derivative of a Kronecker delta will vanish. Thus the only non zero term when acting $\mathcal{J}^\OToNa$ on the $V^{\OneToNb \OneToNc}$ equation of \eqref{eq:SpecificTestForms} is the term where the number of derivatives matches the number of $\xi^{\OneToNb}$ terms. In this case this happens when there are exactly two partial derivatives. This gives
	\begin{equation}
		\lim_{\epsilon \to 0} \; \frac{2}{\epsilon} \; \int_{E} \mathcal{J}^\OToNa \; \delta^0_{\OToNa} \; \psi\left(\frac{t-t_0}{\epsilon} \right) \Big( \xi^\OneToNb - \Map^\OneToNb \Big) \Big( \xi^\OneToNc - \Map^\OneToNc \Big) \prod_{i=1}^6 \psi \left( \xi^i - \Map^i \right) d^7 \xi = 
		 \lim_{\epsilon \to 0} \; \frac{1}{\epsilon} \; \int_{\mathbb{R}} \dot{\Map}^0 V^{\OneToNb \OneToNc} \psi\left(\frac{t-t_0}{\epsilon} \right) \prod_{i=1}^6 \psi(\xi^i|_{\Map} - \Map^i) dt.
	\end{equation}
	Noting $\xi^{\OToNa}|_{\Map} = \Map^\OToNa$, $\dot{\Map}^0 = 1$, $V^{\OneToNb \OneToNc} = V^{\OneToNb \OneToNc} (t)$ and introducing the substitution $t = t_0 + \epsilon t'$ gives
	\begin{equation}
		\lim_{\epsilon \to 0} \; \frac{2}{\epsilon} \; \int_{E} \mathcal{J}^\OToNa \; \delta^0_{\OToNa} \; \psi\left(\frac{t-t_0}{\epsilon} \right) \Big( \xi^\OneToNb - \Map^\OneToNb \Big) \Big( \xi^\OneToNc - \Map^\OneToNc \Big) \prod_{i=1}^6 \psi \left( \xi^i - \Map^i \right) d^7 \xi = \lim_{\epsilon \to 0} \int_{\mathbb{R}} V^{\OneToNb \OneToNc}(t_0 + \epsilon t') \psi\left(t' \right) (\psi(0))^6 dt'.
	\end{equation}
Integrating and taking the limit, noting $\psi(0) = 1$ gives
	\begin{equation}
		\lim_{\epsilon \to 0} \; \frac{2}{\epsilon} \; \int_{E} \mathcal{J}^\OToNa \; \delta^0_{\OToNa} \; \psi\left(\frac{t-t_0}{\epsilon} \right) \Big( \xi^\OneToNb - \Map^\OneToNb \Big) \Big( \xi^\OneToNc - \Map^\OneToNc \Big) \prod_{i=1}^6 \psi \left( \xi^i - \Map^i \right) d^7 \xi = V^{\OneToNb \OneToNc}(t_0)
	\end{equation}
as required. In the extension to a higher order multipole, this still works as the $\delta^0_\OToNa$ term isolates only the $V^{\OneToNa \OneToNb}$ term.\\
	
To isolate the $X^{\OneToNb \OneToNc} + V^{\OneToNb} \dot{\Map}^{\OneToNc}$ term of $\mathcal{J}^\OToNa$, consider $\mathcal{J}^\OToNa$ acting on the $X^{\OneToNb \OneToNc} + V^{\OneToNb} \dot{\Map}^{\OneToNc}$ equation of \eqref{eq:SpecificTestForms},	
	\begin{multline}
	\lim_{\epsilon \to 0} \; \frac{1}{\epsilon} \; \int_{E} \mathcal{J}^\OToNa \; \delta^\OneToNc_{\OToNa} \; \psi\left(\frac{t-t_0}{\epsilon} \right) \left( \xi^\OneToNb - \Map^\OneToNb \right) \; \prod_{i=1}^6 \psi \left( \xi^i - \Map^i \right) d^7 \xi = \lim_{\epsilon \to 0} \; \frac{1}{\epsilon} \; \int_{\mathbb{R}} \dot{\Map}^\OneToNa V^{\OneToNb} \psi\left(\frac{t-t_0}{\epsilon} \right) \prod_{i=1}^6 \psi(\xi^i|_{\Map} - \Map^i) dt
	\\ + \lim_{\epsilon \to 0} \; \frac{1}{\epsilon} \int_{\mathbb{R}} X^{\OneToNa \OneToNb} \psi\left(\frac{t-t_0}{\epsilon} \right) \prod_{i=1}^6 \psi(\xi^i|_{\Map} - \Map^i) dt.
	\end{multline}	
Repeating the previous process gives
\begin{equation}
		\lim_{\epsilon \to 0} \; \frac{1}{\epsilon} \; \int_{E} \mathcal{J}^\OToNa \; \delta^\OneToNc_{\OToNa} \; \psi\left(\frac{t-t_0}{\epsilon} \right) \left( \xi^\OneToNb - \Map^\OneToNb \right) \; \prod_{i=1}^6 \psi \left( \xi^i - \Map^i \right) d^7 \xi = X^{\OneToNb \OneToNc} (t_0) + V^{\OneToNb} (t_0) \dot{\Map}^\OneToNc(t_0).
\end{equation}	
$X^{\OneToNa \OneToNb}$ can be isolated by finding $V^{\OneToNa}$ using the appropriate test form. 	
\end{proof}

Since the components $V^{\OneToNa}, V^{\OneToNa \OneToNb}, X^{\OneToNa}$ and $X^{\OneToNa \OneToNb}$ can all be extracted using test forms, the components of $\mathcal{J}^\OToNa$ are unique.

\subsection{Proof of squeezed forms}
\label{sec:SqueezedFormProof}

This section shows that the components of the distribution $\mathcal{J}^\OToNa$ (equation \eqref{eq:EllisSemiQuadrupole}) are closely related to the moments of $f$, defined by equation \eqref{eq:VandX}. Consider a smooth 6-form that describes the flow of particles in a collisionless plasma,
\begin{equation}
	\theta = f \, i_W \, d^7 \xi = f \, W^\OToNa i_\OToNa \, d^7 \xi = f d \xi^{1 \ldots 6} - f \, W^{\OneToNa} dt \wedge i_\OneToNa d \xi^{1 \ldots 6}
\end{equation}
where $\wedge$ is the wedge product, $i_W$ is an internal contraction with respect to $W$, $i_\OneToNa$ is an internal contraction with respect to $\partial_\OneToNa$,
\begin{equation}
	d \xi^{1 \ldots 6} = d\xi^1 \wedge d\xi^2 \wedge d\xi^3 \wedge d\xi^4 \wedge d\xi^5 \wedge d\xi^6
\end{equation}
and $f$ is a solution to the Vlasov equation. This 6-form obeys the transport equations $d\theta=0$, $i_W \theta = 0$.\\

The one parameter family of smooth 6-forms $\theta_\varepsilon$ is defined as 
\begin{equation} 
	\theta_\varepsilon|_{t,\bm{\xi}} = \left. \frac{1}{\varepsilon^6} \, f \left(t, \frac{\bm{\xi} - \Map}{\varepsilon} \right) d \xi^{1 \ldots 6} \right|_{(t,\bm{\xi})} \\ - \left. \frac{1}{\varepsilon^6} \, \Big( f \left(t, {\bm{\xi} - \Map}{\varepsilon} \right) \, W^\OneToNa \Big) \, dt \wedge i_\OneToNa d \xi^{1 \ldots 6} \right|_{(t,\bm{\xi})}
	\label{eq:SquuezedTheta}
\end{equation}
where $\bm{\xi}$ refers to the combination of all spatial coordinates. \\

By expanding $\theta_\varepsilon$ about $\epsilon=0$, 
\begin{equation}
 \phi \wedge \theta_\varepsilon = \mathcal{\bar{J}}^\OToNa \phi_\OToNa + \mathcal{O}(\varepsilon^3)
\end{equation}
where $\phi$ is a test form (a form with compact support) $\phi_\OToNa d\xi^\OToNa$, and 
\begin{multline}
	\mathcal{\bar{J}}^\OToNa = \frac{1}{2} \int_{\mathbb{R}} \dot{\Map}^\OToNa \, \bar{V}^{\OneToNb \OneToNc} \left( \partial_\OneToNb \partial_\OneToNc \delta^{(6)} (\xi - \Map) \right) dt - \int_{\mathbb{R}} \delta^\OToNa_{\OneToNb} \, \bar{X}^{\OneToNb \OneToNc} \left( \partial_\OneToNc \delta^{(6)} (\xi-\Map) \right) dt 
	- \int_{\mathbb{R}} \dot{\Map}^\OToNa \, \bar{V}^{\OneToNb} \left( \partial_\OneToNb \delta^{(6)}(\xi-\Map) \right) dt \\ + \int_{\mathbb{R}} \delta^\OToNa_{\OneToNb} \, \bar{X}^{\OneToNb} \, \delta^{(6)}(\xi-\Map) dt + \int_{\mathbb{R}} \dot{\Map}^\OToNa \, q \, \delta^{(6)} (\xi-\Map) dt
	\label{eq:EllisbarredSemiQuadrupole}
\end{multline}
where 
\begin{equation}
	\begin{gathered}
	q = \int_{\Sigma} f(t, \bm{\bar{\xi}}) d\bar{\xi}^{1 \ldots 6}, \quad \bar{V}^{\OneToNa} = \varepsilon \int_{\Sigma} f(t, \bm{\bar{\xi}}) \Big( \bar{\xi}^\OneToNa - \Map^\OneToNa \Big) d\bar{\xi}^{1 \ldots 6}, \quad \bar{V}^{\OneToNa \OneToNb} = \varepsilon^2 \int_{\Sigma} f(t, \bm{\bar{\xi}}) \Big( \bar{\xi}^\OneToNa - \Map^\OneToNa \Big) \left( \bar{\xi}^\OneToNb - \Map^\OneToNb \right) d\bar{\xi}^{1 \ldots 6}, \\ \bar{X}^{\OneToNa} = \bar{V}^{\OneToNb} (\partial_{{\OneToNb}} W^{\OneToNa})|_{\Map} + \frac{1}{2} \bar{V}^{\OneToNb \OneToNc} (\partial_{{\OneToNb}} \partial_{{\OneToNc}} W^{\OneToNa})|_{\Map}, \quad \bar{X}^{\OneToNa \OneToNb} = \bar{V}^{\OneToNb \OneToNc} (\partial_{{\OneToNc}} W^\OneToNa)|_{\Map}.
	\end{gathered}
 \label{eq:VbarAndXbar}
\end{equation}
where $\Sigma$ is the spatial part of $E$. Note $\bar{X}^\OneToNa$ is inhomogeneous in $\varepsilon$, since $V^{\OneToNa}$ contains an $\varepsilon$ term whilst $V^{\OneToNa \OneToNb}$ contains an $\varepsilon^2$ term.

\begin{proof}
	
Begin by wedging $\theta_\varepsilon$ against a test form $\phi = \phi_\OToNa d\xi^\OToNa$,
\begin{align}
	\phi \wedge \theta_\varepsilon|_{(t, \bm{\xi})} &= \begin{multlined}[t] \left. \frac{1}{\varepsilon^6} f \left(t, \frac{\bm{\xi} - \Map}{\varepsilon} \right) \phi_0 dt \wedge d \xi^{1 \ldots 6} \right|_{(t,\bm{\xi})} - \left. \frac{1}{\varepsilon^6} f \left(t, \frac{\bm{\xi} - \Map}{\varepsilon} \right) W^\OneToNa \phi_\OneToNb d\xi^\OneToNb \wedge dt \wedge i_\OneToNa d \xi^{1 \ldots 6} \right|_{(t,\bm{\xi})} \end{multlined} \\
	&= \begin{multlined}[t] \left. \frac{1}{\varepsilon^6} f \left(t, \frac{\bm{\xi} - \Map}{\varepsilon} \right) \phi_0 dt \wedge d \xi^{1 \ldots 6} \right|_{(t,\bm{\xi})} + \left. \frac{1}{\varepsilon^6} f \left(t, \frac{\bm{\xi} - \Map}{\varepsilon} \right) W^\OneToNa \phi_\OneToNb dt \wedge d\xi^\OneToNb \wedge i_\OneToNa d \xi^{1 \ldots 6} \right|_{(t,\bm{\xi})} \end{multlined} \\
	&= \begin{multlined}[t] \left. \frac{1}{\varepsilon^6} f \left(t, \frac{\bm{\xi} - \Map}{\varepsilon} \right) \phi_0( \bm{\xi}) dt \wedge d \xi^{1 \ldots 6} \right|_{(t,\bm{\xi})} + \left. \frac{1}{\varepsilon^6} f \left(t, \frac{\bm{\xi} - \Map}{\varepsilon} \right) W^\OneToNa \phi_\OneToNa dt \wedge d \xi^{1 \ldots 6} \right|_{(t,\bm{\xi})}. \end{multlined}
\end{align}	
Making the substitution $\bm{\bar{\xi}} = (\bm{\xi}-\Map)/\varepsilon$, and making the evaluation at $(t, \bm{\xi})$ implicitly,
\begin{align}
	\phi \wedge \theta_\varepsilon|_{t, \bm{\xi}} &= \begin{multlined}[t] \frac{1}{\varepsilon^6} f(t, \bm{\bar{\xi}}) \phi_0 \left(t, \Map + \varepsilon \bm{\bar{\xi}} \right) \varepsilon^6 dt \wedge d \bar{\xi}^{1 \ldots 6}
	\\	+ \frac{1}{\varepsilon^6} f(t, \bm{\bar{\xi}}) W^\OneToNa \left( \Map + \varepsilon \bm{\bar{\xi}} \right) \phi_\OneToNa \left(t, \Map + \varepsilon \bm{\bar{\xi}} \right) \varepsilon^6 dt \wedge d \bar{\xi}^{1 \ldots 6} \end{multlined} \\
	&= \begin{multlined}[t] f(t, \bm{\bar{\xi}}) \phi_0 \left(t, \Map + \varepsilon \bm{\bar{\xi}} \right) dt \wedge d \bar{\xi}^{1 \ldots 6}	\\ + f(t, \bm{\bar{\xi}}) W^\OneToNa \left(t, \Map + \varepsilon \bm{\bar{\xi}} \right) \phi_\OneToNa \left(t, \Map + \varepsilon \bm{\bar{\xi}} \right) dt \wedge d \bar{\xi}^{1 \ldots 6}.
	\end{multlined}
\end{align}
Taylor expanding $\phi_0, \phi_\OneToNa$ and $W^{\OneToNa}$ around $\Map$, noting there are only spatial derivatives as $\xi^0|_{\Map} - \Map^0 = t-t = 0$,
\begin{multline}
		\phi \wedge \theta_\varepsilon|_{t, \bm{\xi}} = f(t, \bm{\bar{\xi}}) \phi_0 |_{\Map} dt \wedge d \bar{\xi}^{1 \ldots 6} + f(t, \bm{\bar{\xi}}) W^\OneToNa|_{\Map} \phi_\OneToNa|_{\Map} dt \wedge d \bar{\xi}^{1 \ldots 6} \\
		+ f(t, \bm{\bar{\xi}}) \varepsilon \Big( \bar{\xi}^\OneToNb - \Map^\OneToNb \Big) \partial_{{\OneToNb}} \phi_0|_{\Map}dt \wedge d \bar{\xi}^{1 \ldots 6}
		+ f(t, \bm{\bar{\xi}}) \varepsilon \left( \bar{\xi}^\OneToNb - \Map^\OneToNb \right) \partial_{{\OneToNb}} (\phi_\OneToNa W^{\OneToNa}) |_{\Map} dt \wedge d \bar{\xi}^{1 \ldots 6} \\
		+ \frac{1}{2} f(t, \bm{\bar{\xi}}) \varepsilon^2 \left( \bar{\xi}^\OneToNb - \Map^\OneToNb \right) \Big( \bar{\xi}^\OneToNc - \Map^\OneToNc \Big) \partial_{{\OneToNb}} \partial_{{\OneToNc}} \phi_0|_{\Map}dt \wedge d \bar{\xi}^{1 \ldots 6} \\
		+ \frac{1}{2} f(t, \bm{\bar{\xi}}) \varepsilon^2 \left( \bar{\xi}^\OneToNb - \Map^\OneToNb \right) \Big( \bar{\xi}^\OneToNc - \Map^\OneToNc \Big) \partial_{{\OneToNb}} \partial_{{\OneToNc}} (\phi_\OneToNa W^\OneToNa)|_{\Map}dt \wedge d \bar{\xi}^{1 \ldots 6} + \mathcal{O}(\varepsilon^3).
\end{multline}	
Next, integrate this over $E$, splitting $E$ into $\mathbb{R} \times \Sigma$. The terms only depending on $\bm{\bar{\xi}}$ can be integrated out,
\begin{multline}
	\int_{E} \phi \wedge \theta_\varepsilon|_{t, \bm{\xi}} = \int_{\mathbb{R}} \left( \int_{\Sigma} f(t, \bm{\bar{\xi}}) d \bar{\xi}^{1 \ldots 6} \right) \phi_0 |_{\Map} dt + \int_{\mathbb{R}} \left( \int_{\Sigma} f(t, \bm{\bar{\xi}}) d \bar{\xi}^{1 \ldots 6} \right) W^\OneToNa|_{\Map} \phi_\OneToNa|_{\Map} dt \\
	+ \int_{\mathbb{R}} \left( \varepsilon \int_{\Sigma} f(t, \bm{\bar{\xi}}) \left( \bar{\xi}^\OneToNb - \Map^\OneToNb \right) d \bar{\xi}^{1 \ldots 6} \right) \partial_{{\OneToNb}} \phi_0|_{\Map}dt
	+ \int_{\mathbb{R}} \left( \varepsilon \int_{\Sigma} f(t, \bm{\bar{\xi}}) \left( \bar{\xi}^\OneToNb - \Map^\OneToNb \right) d \bar{\xi}^{1 \ldots 6} \right) \partial_{{\OneToNb}} (\phi_\OneToNa W^{\OneToNa}) |_{\Map} dt \\
	+ \frac{1}{2} \int_{\mathbb{R}} \left( \varepsilon^2 \int_{\Sigma} f(t, \bm{\bar{\xi}}) \left( \bar{\xi}^\OneToNb - \Map^\OneToNb \right) \Big( \bar{\xi}^\OneToNc - \Map^\OneToNc \Big) d \bar{\xi}^{1 \ldots 6} \right) \partial_{{\OneToNb}} \partial_{{\OneToNc}} \phi_0|_{\Map}dt \\
	+ \frac{1}{2} \int_{\mathbb{R}} \left( \varepsilon^2 \int_{\Sigma} f(t, \bm{\bar{\xi}}) \left( \bar{\xi}^\OneToNb - \Map^\OneToNb \right) \Big( \bar{\xi}^\OneToNc - \Map^\OneToNc \Big) d \bar{\xi}^{1 \ldots 6} \right) \partial_{{\OneToNb}} \partial_{{\OneToNc}} (\phi_\OneToNa W^\OneToNa)|_{\Map}dt + \mathcal{O}(\varepsilon^3).
  \label{eq:IntegratingOutMoments}
\end{multline}
Using the definitions of $q, \bar{V}^\OneToNa$, and $\bar{V}^{\OneToNa \OneToNb}$ from equation \eqref{eq:VbarAndXbar},
\begin{multline}
	\int_{E} \phi \wedge \theta_\varepsilon|_{t, \bm{\xi}} = \int_{\mathbb{R}} q \phi_0 |_{\Map} dt + \int_{\mathbb{R}} q W^\OneToNa|_{\Map} \phi_\OneToNa|_{\Map} dt 
	+ \int_{\mathbb{R}} \bar{V}^{\OneToNb} \partial_{{\OneToNb}} \phi_0|_{\Map}dt
	+ \int_{\mathbb{R}} \bar{V}^{\OneToNb} \partial_{{\OneToNb}} (\phi_\OneToNa W^{\OneToNa}) |_{\Map} dt \\
	+ \frac{1}{2} \int_{\mathbb{R}} \bar{V}^{\OneToNb \OneToNc} \partial_{{\OneToNb}} \partial_{{\OneToNc}} \phi_0|_{\Map}dt 
	+ \frac{1}{2} \int_{\mathbb{R}} \bar{V}^{\OneToNb \OneToNc} \partial_{{\OneToNb}} \partial_{{\OneToNc}} (\phi_\OneToNa W^\OneToNa)|_{\Map}dt + \mathcal{O}(\varepsilon^3).
\end{multline}	
Next, expand the partial derivatives, and note $W^{\OneToNa}|_{\Map} = \dot{\Map}^{\OneToNa}$, 
\begin{multline}
	\int_{E} \phi \wedge \theta_\varepsilon|_{t, \bm{\xi}} = \int_{\mathbb{R}} q \phi_0 |_{\Map} dt + \int_{\mathbb{R}} q \dot{\Map}^{\OneToNa} \phi_\OneToNa|_{\Map} dt 
	+ \int_{\mathbb{R}} \bar{V}^{\OneToNb} \partial_{{\OneToNb}} \phi_0|_{\Map} dt
	\\ + \int_{\mathbb{R}} \bar{V}^{\OneToNb} \dot{\Map}^{\OneToNa} \partial_{{\OneToNb}} \phi_\OneToNa|_{\Map} dt + \int_{\mathbb{R}} \bar{V}^{\OneToNb} \phi_\OneToNa|_{\Map} (\partial_{{\OneToNb}} W^{\OneToNa})|_{\Map} dt 
	+ \frac{1}{2} \int_{\mathbb{R}} \bar{V}^{\OneToNb \OneToNc} \partial_{{\OneToNb}} \partial_{{\OneToNc}} \phi_0|_{\Map}dt 
	\\ + \frac{1}{2} \int_{\mathbb{R}} \bar{V}^{\OneToNb \OneToNc} \dot{\Map}^{\OneToNa} \partial_{{\OneToNb}} \partial_{{\OneToNc}} \phi_\OneToNa|_{\Map} dt + \int_{\mathbb{R}} \bar{V}^{\OneToNb \OneToNc} (\partial_{{\OneToNc}} W^{\OneToNa})|_{\Map}  (\partial_{{\OneToNb}} \phi_\OneToNa)|_{\Map} dt + \frac{1}{2} \int_{\mathbb{R}} \bar{V}^{\OneToNb \OneToNc} \phi_\OneToNa|_{\Map} (\partial_{{\OneToNb}} \partial_{{\OneToNc}} W^{\OneToNa})|_{\Map} dt + \mathcal{O}(\varepsilon^3).
\end{multline}	
Inserting $\bar{X}^\OneToNa$ and $\bar{X}^{\OneToNa \OneToNb}$ from equation \eqref{eq:VbarAndXbar} into this, 
\begin{multline}
	\int_{E} \phi \wedge \theta_\varepsilon|_{t, \bm{\xi}} = \int_{\mathbb{R}} q \phi_0 |_{\Map} dt + \int_{\mathbb{R}} q \dot{\Map}^{\OneToNa} \phi_\OneToNa|_{\Map} dt 
	+ \int_{\mathbb{R}} \bar{V}^{\OneToNb} \partial_{{\OneToNb}} \phi_0|_{\Map} dt
	\\ + \int_{\mathbb{R}} \bar{V}^{\OneToNb} \dot{\Map}^{\OneToNa} \partial_{{\OneToNb}} \phi_\OneToNa|_{\Map} dt + \int_{\mathbb{R}} \bar{X}^{\OneToNa} \phi_\OneToNa|_{\Map} dt 
	+ \frac{1}{2} \int_{\mathbb{R}} \bar{V}^{\OneToNb \OneToNc} \partial_{{\OneToNb}} \partial_{{\OneToNc}} \phi_0|_{\Map}dt 
	\\ + \frac{1}{2} \int_{\mathbb{R}} \bar{V}^{\OneToNb \OneToNc} \dot{\Map}^{\OneToNa} \partial_{{\OneToNb}} \partial_{{\OneToNc}} \phi_\OneToNa|_{\Map} dt + \frac{1}{2} \int_{\mathbb{R}} \bar{X}^{\OneToNa \OneToNc} (\partial_{{\OneToNc}} \phi_\OneToNa)|_{\Map} dt + \mathcal{O}(\varepsilon^3).
\end{multline}	
Recalling $\dot{\Map}^0 = 1$, this can be further simplified,
\begin{multline}
	\int_{E} \phi \wedge \theta_\varepsilon|_{t, \bm{\xi}} = \int_{\mathbb{R}} q \phi_0 |_{\Map} dt + \int_{\mathbb{R}} q \dot{\Map}^{\OneToNa} \phi_\OneToNa|_{\Map} dt 
	+ \int_{\mathbb{R}} \bar{V}^{\OneToNb} \dot{\Map}^{\OToNa} \partial_{{\OneToNb}} \phi_\OToNa|_{\Map} dt + \int_{\mathbb{R}} \bar{X}^{\OneToNa} \phi_\OneToNa|_{\Map} dt 
	\\ + \frac{1}{2} \int_{\mathbb{R}} \bar{V}^{\OneToNb \OneToNc} \dot{\Map}^{\OToNa} \partial_{{\OneToNb}} \partial_{{\OneToNc}} \phi_\OToNa|_{\Map} dt + \int_{\mathbb{R}} \bar{X}^{\OneToNa \OneToNc} (\partial_{{\OneToNc}} \phi_\OneToNa)|_{\Map} dt + \mathcal{O}(\varepsilon^3).
\end{multline}	
This is the same as $\mathcal{\bar{J}}^\OToNa \phi_\OToNa$, as required.
\end{proof}

Thus there is a close relationship between the components of a multipole and the moments of $f$.
 
\subsection{Proof of Ellis multipole transport equations solutions}	
\label{sec:EllisTransportEquationSolution}	
	
In this section, the solutions to the transport equations \eqref{eq:ChargeConservation} and \eqref{eq:VlasovEquationCondition} will be found. Begin by considering the conservation of charge (equation \eqref{eq:ChargeConservation}), and act on a test form $\lambda$,
\begin{equation}
	\int_{E} \mathcal{J}^\OToNa \partial_\OToNa \lambda d^7 \xi = \frac{1}{2} \int_{\mathbb{R}} V^{\OneToNa \OneToNb} \dot{\Map}^\OToNc \partial_\OneToNa \partial_\OneToNb \partial_\OToNc \lambda dt
	+ \int_\mathbb{R} X^{\OneToNa \OneToNb} \partial_\OneToNb \partial_\OneToNa \lambda dt 
	+ \int_{\mathbb{R}} V^{\OneToNa} \dot{\Map}^\OToNb \partial_\OneToNa \partial_\OToNb \lambda dt 
	+ \int_\mathbb{R} X^\OneToNa \partial_\OneToNa \lambda dt
	+ \int_{\mathbb{R}} q \dot{\Map}^{\OToNa} \partial_\OToNa \lambda dt = 0.
\end{equation}		
From here, note
\begin{equation}
\dot{\Map}^\OToNa \frac{\partial \lambda(t)}{\partial \xi^\OToNa} = \left. \frac{d \lambda}{dt} \right|_{\Map}
\end{equation}
so integration by parts can be used to pass a derivative onto the $V^{\OneToNa}$ and $V^{\OneToNa \OneToNb}$ terms. This gives
\begin{equation}
	\int_{E} \mathcal{J}^\OToNa \partial_\OToNa \lambda d^7 \xi = - \frac{1}{2} \int_{\mathbb{R}} \frac{dV^{\OneToNa \OneToNb}}{dt} \partial_\OneToNa \partial_\OneToNb \lambda dt
	+ \int_\mathbb{R} X^{\OneToNa \OneToNb} \partial_\OneToNb \partial_\OneToNa \lambda dt 
	- \int_{\mathbb{R}} \frac{dV^{\OneToNa}}{dt} \partial_\OneToNa \lambda dt 
	+ \int_\mathbb{R} X^\OneToNa \partial_\OneToNa \lambda dt
	+ \int_{\mathbb{R}} \frac{dq}{dt} \lambda dt = 0.
\end{equation}		
Collecting terms based on derivatives of $\lambda$ gives the first line of \eqref{eq:EllisTransportSolutions}. \\
	
To consider the effects of the Vlasov equation (equation \eqref{eq:VlasovEquationCondition}), act on $W^\OToNb \alpha_{\OToNa \OToNb}$, and expand the partial derivatives,
\begin{multline}
	\int_E \mathcal{J}^\OToNa W^\OToNb \alpha_{\OToNa \OToNb} d^7 \xi = \frac{1}{2} \int_\mathbb{R} V^{\OneToNc \OneToNb} \dot{\Map}^\OToNa \left(\alpha_{\OToNa \OToNd} \partial_\OneToNb \partial_\OneToNc W^\OToNd + \partial_\OneToNb \alpha_{\OToNa \OToNd} \partial_\OneToNc W^\OToNd + \partial_\OneToNc \alpha_{\OToNa \OToNd} \partial_\OneToNb W^\OToNd + W^\OToNd \partial_\OneToNb \partial_\OneToNc \alpha_{\OToNa \OToNd} \right) dt \\
	+ \int_{\mathbb{R}} X^{\OneToNa \OneToNb} \left( W^\OToNc \partial_\OneToNb \alpha_{\OneToNa \OToNc} + \alpha_{\OneToNa \OToNc} \partial_\OneToNb W^\OToNc \right) dt 
	+ \int_{\mathbb{R}} V^{\OneToNb} \dot{\Map}^{\OToNa} \left( W^\OToNc \partial_\OneToNb \alpha_{\OToNa \OToNc} + \alpha_{\OToNa \OToNc} \partial_\OneToNb W^\OToNc \right) dt \\
	+ \int_\mathbb{R} X^{\OneToNa} W^\OToNb \alpha_{\OneToNa \OToNb} dt + \int_{\mathbb{R}} q W^\OToNb \dot{\Map}^{\OToNc} \alpha_{\OToNb \OToNc} dt.
\end{multline}
Next, look at terms of the form
\begin{equation}
	W^\OToNa \dot{\Map}^{\OToNb} \alpha_{\OToNa \OToNb},
\end{equation}
recalling the implicit evaluation at $\Map$, $W|_{\Map} = \dot{\Map}^{\OToNa}$. Since $\alpha_{\OToNa \OToNb}$ is antisymmetric,
\begin{equation}
	W^\OToNa \dot{\Map}^{\OToNb} \alpha_{\OToNa \OToNb} = 0.
\end{equation}
This is true for the derivatives of $\alpha_{\OToNa \OToNb}$ as well, since the derivatives are also antisymmetric. Rearranging the remaining terms in derivatives of $\alpha_{\OToNa \OToNb}$ gives 
\begin{multline}
	\int_E \mathcal{J}^\OToNa W^\OToNb \alpha_{\OToNa \OToNb} d^7 \xi 
	= \int_{\mathbb{R}} \left(V^{\OneToNc \OneToNb} \partial_\OneToNc W^\OneToNa - X^{\OneToNa \OneToNb} \right) \dot{\Map}^{\OToNd} \partial_\OneToNb \alpha_{\OToNd \OneToNa} dt 
	+ \int_{\mathbb{R}} \left( V^{\OneToNc \OneToNb} \partial_{\OneToNc} W^0 \right) \dot{\Map}^{\OToNd} \partial_\OneToNb \alpha_{\OToNd 0} dt 
	+ \int_{\mathbb{R}} X^{\OneToNa \OneToNb} \partial_\OneToNb W^{\OToNd} \alpha_{\OneToNa \OToNd} dt\\
	+ \int_{\mathbb{R}} \left( V^{\OneToNb} \partial_\OneToNb W^\OneToNa + \frac{1}{2} V^{\OneToNb \OneToNc} \partial_{\OneToNb} \partial_{\OneToNc} W^{\OneToNa} - X^\OneToNa \right) \dot{\Map}^\OToNd \alpha_{\OToNd \OneToNa} dt 
	+\int_{\mathbb{R}} \left( \frac{1}{2} V^{\OneToNb \OneToNc} \partial_\OneToNb \partial_\OneToNc W^0 + V^{\OneToNb} \partial_{\OneToNb} W^0 \right) \dot{\Map}^{\OToNd} \alpha_{\OToNd 0} dt
	\label{eq:NearlyDifferentialEquations}
\end{multline}
where the minus sign in the $X^\OneToNa$ and $X^{\OneToNa \OneToNb}$ terms comes from flipping the $\alpha_{\OToNa \OneToNb}$ indices. Next, note that calculations are in a frame where $W^0 = 1$, so derivatives of $W^0$ vanish. Two of the remaining terms give the required differential equations,
\begin{align}
	X^{\OneToNa \OneToNb} &= V^{\OneToNb \OneToNc} \partial_\OneToNc W^\OneToNa, \\
	X^{\OneToNa} &= V^{\OneToNb} \partial_\OneToNb W^\OneToNa + \frac{1}{2} V^{\OneToNb \OneToNc} \partial_\OneToNb \partial_\OneToNc W^\OneToNa.	
\end{align}	
Lastly, for the remaining integral, the third integral of equation \eqref{eq:NearlyDifferentialEquations}, taking the antisymmetric part of $X^{\OneToNa \OneToNb} \partial_\OneToNb W^\OneToNd$ gives
\begin{align}
	X^{\OneToNa \OneToNb} \partial_{\OneToNb} W^\OneToNd \alpha_{\OneToNa \OneToNd} &= \left(X^{\OneToNa \OneToNb} \partial_{\OneToNb} W^\OneToNd - X^{\OneToNd \OneToNb} \partial_{\OneToNb} W^\OneToNa \right) \alpha_{\OneToNa \OneToNd} \\
	&= \left(V^{\OneToNb \OneToNc} \partial_\OneToNb W^\OneToNa \partial_\OneToNc W^\OneToNd - V^{\OneToNc \OneToNb} \partial_\OneToNc W^{\OneToNa} \partial_\OneToNb W^{\OneToNd} \right) \alpha_{\OneToNa \OneToNd} \\
	&= 0. 		
\end{align}	
So the differential equations in the bottom line of \eqref{eq:EllisTransportSolutions} uniquely solve the system, giving the equations of motion for the moments through the Ellis representation of multipoles.

\subsection{Proof of the Ellis multipole coordinate transformation}
\label{sec:EllisCoordTransform}

In this section the first step of the coordinate transformation for the moments is performed, to obtain non-unique components in terms of $U^\OToNa, U^{\OToNa \OToNb}, Y^{\OToNa}$ and $Y^{\OToNa \OToNb}$. Proceeding term by term using the known transformation rules:
For the $V^{\OToNa \OToNb}$ term, 
\begin{align}
	\int_{{{\mathbb{R}}}} V^{\OneToNb \OneToNc} \dot{\Map}^\OToNa \partial_\OneToNb \partial_\OneToNc \phi_\OToNa dt &= \int_{{{\mathbb{R}}}} \hat{V}^{\OneToNb \OneToNc} \dot{\hat{\Map}}^\OToNa \partial_{\hat{\OneToNb}} \partial_{\hat{\OneToNc}} \hat{\phi}_\OToNa d \hat{t} \\
	&= \int_{{{\mathbb{R}}}} V^{\OneToNb \OneToNc} \dot{\hat{\Map}}^{\hat{\OToNd}} A^{\OToNa}_{\hat{\OToNd}} \frac{dt}{d\hat{t}} A^{\hat{\OToNe}}_{\OneToNb} \partial_{\hat{\OToNe}} \left( A^{\hat{\OToNf}}_{\OneToNc} \partial_{\hat{\OToNf}} \left(A^{\hat{\OToNg}}_{\OToNa} \hat{\phi}_{\hat{\OToNg}} \right) \right) \frac{d t}{d \hat{t}} d \hat{t}.
\end{align}
Performing these derivatives gives
\begin{multline}
	\int_{{{\mathbb{R}}}} V^{\OneToNb \OneToNc} \dot{\Map}^\OToNa \partial_\OneToNb \partial_\OneToNc \phi_\OToNa dt = \int_{{{\mathbb{R}}}} V^{\OneToNb \OneToNc} \dot{\hat{\Map}}^{\hat{\OToNd}} A^{\OToNa}_{\hat{\OToNd}} \frac{dt}{d\hat{t}} A^{\hat{\OToNg}}_{\OToNa \OneToNb \OneToNc} \hat{\phi}_{\hat{\OToNg}} \frac{d t}{d \hat{t}} d \hat{t} + \int_{{{\mathbb{R}}}} V^{\OneToNb \OneToNc} \dot{\hat{\Map}}^{\hat{\OToNd}} A^{\OToNa}_{\hat{\OToNd}} \frac{dt}{d\hat{t}} A^{\hat{\OToNe}}_{\OneToNb} A^{\hat{\OToNg}}_{\OToNa \OneToNc} \partial_{\hat{\OToNe}} (\hat{\phi}_{\hat{\OToNg}}) \frac{d t}{d \hat{t}} d \hat{t} \\
	+ \int_{{{\mathbb{R}}}} V^{\OneToNb \OneToNc} \dot{\hat{\Map}}^{\hat{\OToNd}} A^{\OToNa}_{\hat{\OToNd}} \frac{dt}{d\hat{t}} A^{\hat{\OToNe}}_{\OneToNc} A^{\hat{\OToNf}}_{\OToNa \OneToNb} \partial_{\hat{\OToNe}} (\hat{\phi}_{\hat{\OToNf}}) \frac{d t}{d \hat{t}} d \hat{t} + \int_{{{\mathbb{R}}}} V^{\OneToNb \OneToNc} \dot{\hat{\Map}}^{\hat{\OToNd}} A^{\OToNa}_{\hat{\OToNd}} \frac{dt}{d\hat{t}} A^{\hat{\OToNe}}_{\OneToNb} A^{\hat{\OToNf}}_{\OneToNc} A^{\hat{\OToNg}}_{\OToNa} \partial_{\hat{\OToNe}} \partial_{\hat{\OToNf}} ( \hat{\phi}_{\hat{\OToNg}}) \frac{d t}{d \hat{t}} d \hat{t} \\
	+ \int_{{{\mathbb{R}}}} V^{\OneToNb \OneToNc} \dot{\hat{\Map}}^{\hat{\OToNd}} A^{\OToNa}_{\hat{\OToNd}} \frac{dt}{d\hat{t}} A^{\hat{\OToNf}}_{\OneToNc \OneToNb} A^{\OToNh}_{\hat{\OToNf}} A^{\hat{\OToNg}}_{\OToNa \OToNh} \hat{\phi}_{\hat{\OToNg}} \frac{d t}{d \hat{t}} d \hat{t} + \int_{{{\mathbb{R}}}} V^{\OneToNb \OneToNc} \dot{\hat{\Map}}^{\hat{\OToNd}} A^{\OToNa}_{\hat{\OToNd}} \frac{dt}{d\hat{t}} A^{\hat{\OToNf}}_{\OneToNc \OneToNb} A^{\hat{\OToNg}}_{\OToNa} \partial_{\hat{\OToNf}} (\hat{\phi}_{\hat{\OToNg}}) \frac{d t}{d \hat{t}} d \hat{t}.
\end{multline}
Simplifying these terms down
\begin{multline}
	\int_{{{\mathbb{R}}}} V^{\OneToNb \OneToNc} \dot{\Map}^\OToNa \partial_\OneToNb \partial_\OneToNc \phi_\OToNa dt = \int_{{{\mathbb{R}}}} V^{\OneToNb \OneToNc} \dot{{\Map}}^{\OToNa} \partial_\OToNa \left( A^{\hat{\OToNg}}_{\OneToNb \OneToNc} \right) \hat{\phi}_{\hat{\OToNg}} \frac{d t}{d \hat{t}} d \hat{t} + \int_{{{\mathbb{R}}}} V^{\OneToNb \OneToNc} A^{\hat{\OToNe}}_{\OneToNb} \dot{{\Map}}^{\OToNa} \partial_\OToNa \left(A^{\hat{\OToNg}}_{\OneToNc} \right) \partial_{\hat{\OToNe}} (\hat{\phi}_{\hat{\OToNg}}) \frac{d t}{d \hat{t}} d \hat{t} \\
	+ \int_{{{\mathbb{R}}}} V^{\OneToNb \OneToNc}  A^{\hat{\OToNe}}_{\OneToNc} \dot{{\Map}}^{\hat{\OToNa}} \partial_\OToNa \left(A^{\hat{\OToNf}}_{\OneToNb}\right) \partial_{\hat{\OToNe}} (\hat{\phi}_{\hat{\OToNf}}) \frac{d t}{d \hat{t}} d \hat{t} + \int_{{{\mathbb{R}}}} V^{\OneToNb \OneToNc} \dot{\hat{\Map}}^{\hat{\OToNd}} A^{\hat{\OToNe}}_{\OneToNb} A^{\hat{\OToNf}}_{\OneToNc} \partial_{\hat{\OToNe}} \partial_{\hat{\OToNf}} ( \hat{\phi}_{\hat{\OToNd}}) d \hat{t}
	+ \int_{{{\mathbb{R}}}} V^{\OneToNb \OneToNc} \dot{\hat{\Map}}^{\hat{\OToNd}} A^{\hat{\OToNf}}_{\OneToNc \OneToNb} \partial_{\hat{\OToNf}} (\hat{\phi}_{\hat{\OToNd}}) d \hat{t}.
\end{multline}
For the $X^{\OneToNa \OneToNb}$ term,
\begin{align}
	\int_{{{\mathbb{R}}}} X^{\OneToNa \OneToNb} \partial_\OneToNb \phi_\OneToNa dt &= \int_{{{\mathbb{R}}}} \hat{X}^{\OneToNa \OneToNb} \partial_{\hat{\OneToNb}} \hat{\phi}_\OneToNa d\hat{t} \\
	&= \int_{{{\mathbb{R}}}} X^{\OneToNa \OneToNb} A^{\hat{\OToNc}}_{\OneToNb} \partial_{\hat{\OToNc}} \left( A^{\hat{\OToNd}}_{\OneToNa} \hat{\phi}_{\hat{\OToNd}} \right) \frac{dt}{d \hat{t}} d \hat{t} \\
	&= \int_{{{\mathbb{R}}}} X^{\OneToNa \OneToNb} A^{\hat{\OToNd}}_{\OneToNa \OneToNb} \hat{\phi}_{\hat{\OToNd}} \frac{dt}{d \hat{t}} d \hat{t} + \int_{{{\mathbb{R}}}} X^{\OneToNa \OneToNb} A^{\hat{\OToNc}}_{\OneToNb} A^{\hat{\OToNd}}_{\OneToNa} \partial_{\hat{\OToNc}} (\hat{\phi}_{\hat{\OToNd}}) \frac{dt}{d \hat{t}} d \hat{t}.
\end{align}
For the $V^{\OToNa}$ term,
\begin{align}
	\int_{{\mathbb{R}}} {\dot{\Map}}^\OToNa V^{\OneToNb} \frac{\partial \phi_{\OToNa}}{\partial \xi^\OneToNb} d t &= \int_{{\mathbb{R}}} \hat{\dot{\Map}}^\OToNa \hat{V}^{\OneToNb} \frac{\partial \hat{\phi}_{\OToNa}}{\partial \hat{\xi}^\OneToNb} d \hat{t} \\
	&= \int_{{\mathbb{R}}} \dot{\hat{\Map}}^{\hat{\OToNc}} A^{\OToNa}_{\hat{\OToNc}} \frac{d \hat{t}}{dt} {V}^{\OneToNb} A^{\hat{\OToNd}}_{\OneToNb} \frac{\partial}{\partial \hat{\xi}^\OToNd} \left( A^{\hat{\OToNe}}_{\OToNa} {\hat{\phi}}_{\hat{\OToNe}} \right) \frac{d \hat{t}}{dt} dt \\
	&= \int_{{\mathbb{R}}} \dot{\Map}^{\OToNa} {V}^{\OneToNb} A^{\hat{\OToNe}}_{\OneToNb \OToNa} {\hat{\phi}}_{\hat{\OToNe}} \frac{d \hat{t}}{dt} dt + \int_{{\mathbb{R}}} \dot{\hat{\Map}}^{\hat{\OToNc}} A^{\OToNa}_{\hat{\OToNc}} {V}^{\OneToNb} A^{\hat{\OToNd}}_{\OneToNb} A^{\hat{\OToNe}}_{\OToNa} \partial_{\hat{\OToNd}} \hat{\phi}_{\hat{\OToNe}} dt \\
	&= \int_{{\mathbb{R}}} {V}^{\OneToNb} \dot{\Map}^{\OToNa} \partial_\OToNa (A^{\hat{\OToNe}}_{\OneToNc}) {\hat{\phi}}_{\hat{\OToNe}} \frac{d \hat{t}}{dt} dt + \int_{{\mathbb{R}}} \dot{\hat{\Map}}^{\hat{\OToNc}} {V}^{\OneToNb} A^{\hat{\OToNd}}_{\OneToNb} \partial_{\hat{\OToNd}} \hat{\phi}_{\hat{\OToNc}} dt.
\end{align}
For the $X^{\OneToNa}$ term,
\begin{equation}
	\int_{{\mathbb{R}}} {X}^\OneToNd {\phi}_{\OneToNd} dt = \int_{\hat{\mathbb{R}}} \hat{X}^\OneToNd \hat{\phi}_{\OneToNd} d \hat{t}
	= \int_{\hat{\mathbb{R}}} X^\OneToNd A^{\hat{\OToNc}}_{\OneToNd} {\hat{\phi}}_{\hat{\OToNc}} \frac{d \hat{t}}{dt} d t.
\end{equation}
Summing all these terms together gives the transformed quadrupole,
\begin{multline}
	\int \mathcal{J}^\OToNa \phi_\OToNa d^7 \xi = \int \hat{\mathcal{J}}^\OToNa \hat{\phi}_\OToNa d^7 \hat{\xi} = \frac{1}{2} \int_{{{\mathbb{R}}}} V^{\OneToNb \OneToNc} \dot{\hat{\Map}}^{\hat{\OToNd}} A^{\hat{\OToNe}}_{\OneToNb} A^{\hat{\OToNf}}_{\OneToNc} \partial_{\hat{\OToNe}} \partial_{\hat{\OToNf}} ( \hat{\phi}_{\hat{\OToNd}}) d \hat{t} \\
	+ \int_{{{\mathbb{R}}}} \left(X^{\OneToNa \OneToNb} A^{\hat{\OToNc}}_{\OneToNb} A^{\hat{\OToNd}}_{\OneToNa} + \frac{1}{2} V^{\OneToNe \OneToNf} A^{\hat{\OToNc}}_{\OneToNe} \dot{{\Map}}^{\OToNa} \partial_\OToNa \left(A^{\hat{\OToNd}}_{\OneToNf} \right) + \frac{1}{2} V^{\OneToNe \OneToNf} A^{\hat{\OToNd}}_{\OneToNf} \dot{{\Map}}^{\OToNa} \partial_\OToNa \left(A^{\hat{\OToNc}}_{\OneToNe} \right) \right) \partial_{\hat{\OToNc}} (\hat{\phi}_{\hat{\OToNd}}) \frac{dt}{d \hat{t}} d \hat{t}\\
	+ \int_{{\mathbb{R}}} \dot{\hat{\Map}}^{\hat{\OToNa}} \left({V}^{\OneToNb} A^{\hat{\OToNd}}_{\OneToNb} + \frac{1}{2} V^{\OneToNb \OneToNc} A^{\hat{\OToNa}}_{\OneToNc \OneToNb} \right) \partial_{\hat{\OToNd}} \hat{\phi}_{\hat{\OToNa}} dt \\
	+ \int_{\hat{\mathbb{R}}} \left(X^\OneToNd A^{\hat{\OToNc}}_{\OneToNd} + {V}^{\OneToNb} \dot{\Map}^{\OToNa} \partial_\OToNa (A^{\hat{\OToNc}}_{\OneToNb}) + X^{\OneToNa \OneToNb} A^{\hat{\OToNc}}_{\OneToNa \OneToNb} + \frac{1}{2} V^{\OneToNd \OneToNe} \dot{{\Map}}^{\OToNa} \partial_\OToNa \left( A^{\hat{\OToNc}}_{\OneToNd \OneToNe} \right) \right) {\hat{\phi}}_{\hat{\OToNc}} \frac{d t}{d\hat{t}} d \hat{t}.
\end{multline}
This gives the coordinate transformations for the quadrupole components,
\begin{equation}
	\begin{gathered}
		U^{\OToNb \OToNc} = V^{\OneToNd \OneToNe} A^{\hat{\OToNb}}_{\OneToNd} A^{\hat{\OToNc}}_{\OneToNe}, \\ \begin{multlined}[t] Y^{\OToNc \OToNd} = \left(X^{\OneToNa \OneToNb} A^{\hat{\OToNc}}_{\OneToNb} A^{\hat{\OToNd}}_{\OneToNa} + \frac{1}{2} V^{\OneToNe \OneToNf} A^{\hat{\OToNc}}_{\OneToNe} \dot{{\Map}}^{\OToNa} \partial_\OToNa \left(A^{\hat{\OToNd}}_{\OneToNf} \right) \right. \left. + \frac{1}{2} V^{\OneToNe \OneToNf} A^{\hat{\OToNd}}_{\OneToNf} \dot{{\Map}}^{\OToNa} \partial_\OToNa \left(A^{\hat{\OToNc}}_{\OneToNe} \right) \right) \frac{dt}{d \hat{t}}, \end{multlined}\\
		U^{\OToNa} = V^{\OneToNb} A^{\hat{\OToNa}}_{\OneToNb} + \frac{1}{2} V^{\OneToNb \OneToNc} A^{\hat{\OToNa}}_{\OneToNb \OneToNc}, \\ Y^{\OToNc} = \left(X^\OneToNd A^{\hat{\OToNc}}_{\OneToNd} + {V}^{\OneToNb} \dot{\Map}^{\OToNa} \partial_\OToNa (A^{\hat{\OToNc}}_{\OneToNb}) + X^{\OneToNa \OneToNb} A^{\hat{\OToNc}}_{\OneToNa \OneToNb} \right. \left. + \frac{1}{2} V^{\OneToNd \OneToNe} \dot{{\Map}}^{\OToNa} \partial_\OToNa \left( A^{\hat{\OToNc}}_{\OneToNd \OneToNe} \right) \right) \frac{d t}{d\hat{t}}.
	\end{gathered}
\end{equation}
where the components of the quadrupole in the new coordinate system are no longer unique.
\subsection{Proof of the Ellis multipole projection}
\label{sec:EllisCoordProjection}

This section performs the projection $\partial_0 = \partial_{\dot{\Map}} - \dot{\Map}^\OneToNa \partial_\OneToNa$ into the quadrupole with non unique components defined by equation \eqref{eq:EllisCoordTransforms}. By performing this projection $U^\OToNa, U^{\OToNa \OToNb}, Y^{\OToNa}$ and $Y^{\OToNa \OToNb}$ can be written in a form where the components are unique, giving the full coordinate transformation for the quadrupole. \\

Inserting \eqref{eq:EllisProjection} term by term into the non-unique semi-quadrupole
\begin{equation} 
	\int \mathcal{J}^\OToNa \phi_\OToNa d^6 \xi = \frac{1}{2} \int_{\mathbb{R}} \dot{\Map}^\OToNa U^{\OToNb \OToNc} \partial_\OToNb \partial_\OToNc \phi_\OToNa dt + \int_{\mathbb{R}} Y^{\OToNa \OToNb} \partial_\OneToNb \phi_\OneToNa dt \\+ \int_{\mathbb{R}} \dot{\Map}^\OToNa U^{\OneToNb} \partial_\OneToNb \phi_\OToNa dt + \int_{\mathbb{R}} Y^{\OneToNa} \phi_\OneToNa dt + \int_{\mathbb{R}} \dot{\Map}^\OToNa q \phi_\OToNa dt.
\end{equation}
For the $U^{\OToNa \OToNb}$ term, noting the symmetry of $U^{\OToNa \OToNb}$,
  \begin{align}
    \int_{{\mathbb{R}}} {\dot{\Map}}^\OToNa {U}^{\OToNb \OToNc} \partial_\OToNb \partial_\OToNc \phi_{\OToNa} dt &= \int_{{\mathbb{R}}} {\dot{\Map}}^\OToNa {U}^{\OneToNb \OneToNc} \partial_\OneToNb \partial_\OneToNc \phi_{\OToNa} dt + 2 \int_{{\mathbb{R}}} {\dot{\Map}}^\OToNa {U}^{\OneToNb 0} \partial_\OneToNb \partial_0 \phi_{\OToNa} dt + \int_{{\mathbb{R}}} {\dot{\Map}}^\OToNa {U}^{0 0} \partial_0 \partial_0 \phi_{\OToNa} dt \\
    &= \begin{multlined}[t] \int_{{\mathbb{R}}} {\dot{\Map}}^\OToNa {U}^{\OneToNb \OneToNc} \partial_\OneToNb \partial_\OneToNc \phi_{\OToNa} dt + 2 \int_{{\mathbb{R}}} {\dot{\Map}}^\OToNa {U}^{\OneToNb 0} \partial_{\dot{\Map}} \partial_\OneToNb \phi_{\OToNa} dt - 2 \int_{{\mathbb{R}}} {\dot{\Map}}^\OToNa \dot{\Map}^\OneToNd {U}^{\OneToNb 0} \partial_\OneToNb \partial_\OneToNd \phi_{\OToNa} dt \\ + \int_{{\mathbb{R}}} {\dot{\Map}}^\OToNa {U}^{0 0} \partial_{\dot{\Map}} \partial_0 \phi_{\OToNa} dt - \int_{{\mathbb{R}}} {\dot{\Map}}^\OToNa \dot{\Map}^\OneToNb {U}^{0 0} \partial_0 \partial_\OneToNb \phi_{\OToNa} dt \end{multlined} \\
    &= \begin{multlined}[t] \int_{{\mathbb{R}}} {\dot{\Map}}^\OToNa \left( {U}^{\OneToNb \OneToNc} - 2 \dot{\Map}^\OneToNc U^{\OneToNc 0} \right) \partial_\OneToNb \partial_\OneToNc \phi_{\OToNa} dt - 2 \int_{{\mathbb{R}}} \frac{d}{dt} \left( {\dot{\Map}}^\OToNa {U}^{\OneToNb 0} \right) \partial_\OneToNb \phi_{\OToNa} dt \\ + \int_{{\mathbb{R}}} \frac{d}{dt} \left( {\dot{\Map}}^\OToNa {U}^{0 0} \right) \partial_0 \phi_{\OToNa} dt - \int_{{\mathbb{R}}} {\dot{\Map}}^\OToNa \dot{\Map}^\OneToNb {U}^{0 0} \partial_\OneToNb \partial_0 \phi_{\OToNa} dt \end{multlined} \\
    &= \begin{multlined}[t] \int_{{\mathbb{R}}} {\dot{\Map}}^\OToNa \left( {U}^{\OneToNb \OneToNc} - 2 \dot{\Map}^\OneToNc U^{\OneToNc 0} \right) \partial_\OneToNb \partial_\OneToNc \phi_{\OToNa} dt - 2 \int_{{\mathbb{R}}} \frac{d}{dt} \left( {\dot{\Map}}^\OToNa {U}^{\OneToNb 0} \right) \partial_\OneToNb \phi_{\OToNa} dt + \int_{{\mathbb{R}}} \frac{d}{dt} \left( {\dot{\Map}}^\OToNa {U}^{0 0} \right) \partial_{\dot{\Map}} \phi_{\OToNa} dt \\ + \int_{{\mathbb{R}}} \frac{d}{dt} \left( {\dot{\Map}}^\OToNa {U}^{0 0} \right) \dot{\Map}^{\OneToNb} \partial_\OneToNb \phi_{\OToNa} dt 
	  - \int_{{\mathbb{R}}} {\dot{\Map}}^\OToNa \dot{\Map}^\OneToNb {U}^{0 0} \partial_{\dot{\Map}} \partial_\OneToNb \phi_{\OToNa} dt + \int_{{\mathbb{R}}} {\dot{\Map}}^\OToNa \dot{\Map}^\OneToNb \dot{\Map}^\OneToNc {U}^{0 0} \partial_\OneToNb \partial_\OneToNc \phi_{\OToNa} dt \end{multlined} \\
	  &= \begin{multlined}[t] \int_{{\mathbb{R}}} {\dot{\Map}}^\OToNa \left( {U}^{\OneToNb \OneToNc} - 2 \dot{\Map}^\OneToNc U^{\OneToNc 0} + \dot{\Map}^\OneToNb \dot{\Map}^\OneToNc U^{00} \right) \partial_\OneToNb \partial_\OneToNc \phi_{\OToNa} dt 
	  - 2 \int_{{\mathbb{R}}} \frac{d}{dt} \left( {\dot{\Map}}^\OToNa {U}^{\OneToNb 0} \right) \partial_\OneToNb \phi_{\OToNa} dt \\ + \int_{{\mathbb{R}}} \frac{d^2}{dt^2} \left( {\dot{\Map}}^\OToNa {U}^{0 0} \right) \phi_{\OToNa} dt + \int_{{\mathbb{R}}} \frac{d}{dt} \left( {\dot{\Map}}^\OToNa {U}^{0 0} \right) \dot{\Map}^{\OneToNb} \partial_\OneToNb \phi_{\OToNa} dt - \int_{{\mathbb{R}}} \frac{d}{dt} \left( {\dot{\Map}}^\OToNa \dot{\Map}^\OneToNb {U}^{0 0} \right) \partial_\OneToNb \phi_{\OToNa} dt. \end{multlined}
\end{align} 	
For the $Y^{\OToNa \OToNb}$ term,
\begin{align}
	\int_{{{\mathbb{R}}}} {Y}^{\OToNa \OToNb} \partial_\OToNb \phi_{\OToNa} dt &= \int_{{{\mathbb{R}}}} {Y}^{\OToNa \OneToNb} \partial_\OneToNb \phi_{\OToNa} dt + \int_{{{\mathbb{R}}}} {Y}^{\OToNa 0} \partial_0 \phi_{\OToNa} dt \\
	&= \begin{multlined}[t] \int_{{{\mathbb{R}}}} {Y}^{\OToNa \OneToNb} \partial_\OneToNb \phi_{\OToNa} dt + \int_{{{\mathbb{R}}}} {Y}^{\OToNa 0} \partial_{\dot{\Map}} \phi_{\OToNa} dt - \int_{{{\mathbb{R}}}} {Y}^{\OToNa 0} \dot{\Map}^\OneToNb \partial_\OneToNb \phi_{\OToNa} dt \end{multlined} \\
	&= \int_{{{\mathbb{R}}}} \left( {Y}^{\OToNa \OneToNb} - {Y}^{\OToNa 0} \dot{\Map}^\OneToNb \right) \partial_\OneToNb \phi_{\OToNa} dt - \int_{{{\mathbb{R}}}} \frac{d}{dt} {Y}^{\OToNa 0} \phi_{\OToNa} dt.
\end{align}
For the $U^{\OToNa}$ term,
\begin{align} 
	\int_{{{\mathbb{R}}}} \dot{\Map}^\OToNa {U}^{\OToNb} \partial_\OToNb \phi_{\OToNa} dt &= \int_{\mathbb{R}} \dot{\Map}^\OToNa U^{0} \partial_{0} \phi_{\OToNa} dt
	+ \int_{\mathbb{R}} \dot{\Map}^\OToNa U^{\OneToNb} \partial_{\OneToNb} \phi_{\OToNa} dt \\
	&= \begin{multlined}[t] \int_{\mathbb{R}} \dot{\Map}^\OToNa U^{0} \partial_{\dot{\Map}} \phi_{\OToNa} dt - \int_{\mathbb{R}} \dot{\Map}^\OToNa U^{0} \dot{\Map}^\OneToNa \partial_\OneToNb \phi_{\OToNa} dt + \int_{\mathbb{R}} \dot{\Map}^\OToNa U^{\OneToNb} \partial_{\OneToNb} \phi_{\OToNa} dt \end{multlined} \\
	&= \begin{multlined}[t] - \int_{\mathbb{R}} \frac{d}{dt} \left(\dot{\Map}^\OToNa U^{0} \right) \phi_{\OToNa} dt
	+ \int_{\mathbb{R}} \dot{\Map}^\OToNa \left(U^{\OneToNb} - U^0 \dot{\Map}^{\OneToNb} \right) \partial_{\OneToNb} \phi_{\OToNa} dt . \end{multlined}
\end{align}	
Summing all these terms together,
\begin{multline}
	\int \mathcal{J}^\OToNa \phi_\OToNa d^7 \xi = \frac{1}{2} \int_{{\mathbb{R}}} {\dot{\Map}}^\OToNa \left( {U}^{\OneToNb \OneToNc} - 2 \dot{\Map}^\OneToNc U^{\OneToNc 0} + \dot{\Map}^\OneToNb \dot{\Map}^\OneToNc U^{00} \right) \partial_\OneToNb \partial_\OneToNc \phi_{\OToNa} dt 
	- \int_{{\mathbb{R}}} \frac{d}{dt} \left( {\dot{\Map}}^\OToNa {U}^{\OneToNb 0} \right) \partial_\OneToNb \phi_{\OToNa} dt + \frac{1}{2} \int_{{\mathbb{R}}} \frac{d^2}{dt^2} \left( {\dot{\Map}}^\OToNa {U}^{0 0} \right) \phi_{\OToNa} dt \\ + \frac{1}{2} \int_{{\mathbb{R}}} \frac{d}{dt} \left( {\dot{\Map}}^\OToNa {U}^{0 0} \right) \dot{\Map}^{\OneToNb} \partial_\OneToNb \phi_{\OToNa} dt -\frac{1}{2} \int_{{\mathbb{R}}} \frac{d}{dt} \left( {\dot{\Map}}^\OToNa \dot{\Map}^\OneToNb {U}^{0 0} \right) \partial_\OneToNb \phi_{\OToNa} dt 
	+ \int_{{{\mathbb{R}}}} \left( {Y}^{\OToNa \OneToNb} - {Y}^{\OToNa 0} \dot{\Map}^\OneToNb \right) \partial_\OneToNb \phi_{\OToNa} dt - \int_{{{\mathbb{R}}}} \frac{d}{dt} {Y}^{\OToNa 0} \phi_{\OToNa} dt \\
	- \int_{\mathbb{R}} \frac{d}{dt} \left(\dot{\Map}^\OToNa U^{0} \right) \phi_{\OToNa} dt
	+ \int_{\mathbb{R}} \dot{\Map}^\OToNa \left(U^{\OneToNb} - U^0 \dot{\Map}^{\OneToNb} \right) \partial_{\OneToNb} \phi_{\OToNa} dt
	+ \int_{\mathbb{R}} Y^{\OneToNa} \phi_\OneToNa dt + \int_{\mathbb{R}} \dot{\Map}^\OToNa q \phi_\OToNa dt.
\end{multline}
Grouping terms together,
\begin{multline}
	\int \mathcal{J}^\OToNa \phi_\OToNa d^7 \xi = \frac{1}{2} \int_{{\mathbb{R}}} {\dot{\Map}}^\OToNa \left( {U}^{\OneToNb \OneToNc} - 2 \dot{\Map}^\OneToNc U^{\OneToNc 0} + \dot{\Map}^\OneToNb \dot{\Map}^\OneToNc U^{00} \right) \partial_\OneToNb \partial_\OneToNc \phi_{\OToNa} dt \\
	+ \int_{{{\mathbb{R}}}} \left( Y^{\OToNa \OneToNb} - {Y}^{\OToNa 0} \dot{\Map}^\OneToNb - \frac{d}{dt} \left(\dot{\Map}^\OToNa U^{\OneToNb 0} \right) + \frac{1}{2} \frac{d}{dt} \left( {\dot{\Map}}^\OToNa \dot{\Map}^\OneToNb {U}^{0 0} \right) \right) \partial_\OneToNb \phi_{\OToNa} dt
	\\
	+ \frac{1}{2} \int_{{\mathbb{R}}} \frac{d^2}{dt^2} \left( {\dot{\Map}}^\OToNa {U}^{0 0} \right) \phi_{\OToNa} dt + \frac{1}{2} \int_{{\mathbb{R}}} \frac{d}{dt} \left( {\dot{\Map}}^\OToNa {U}^{0 0} \right) \dot{\Map}^{\OneToNb} \partial_\OneToNb \phi_{\OToNa} dt - \int_{{{\mathbb{R}}}} \frac{d}{dt} {Y}^{\OToNa 0} \phi_{\OToNa} dt 
	+ \int_{\mathbb{R}} \dot{\Map}^\OToNa \left(U^{\OneToNb} - U^0 \dot{\Map}^{\OneToNb} \right) \partial_{\OneToNb} \phi_{\OToNa} dt \\
	+ \int_{\mathbb{R}} \left( Y^{\OneToNa} - \frac{d}{dt} \left(\dot{\Map}^\OToNa U^{0} \right) \right) \phi_\OneToNa dt + \int_{\mathbb{R}} \dot{\Map}^\OToNa q \phi_\OToNa dt.
\end{multline}
Calculating some derivatives and symmetrising $U^{0 \OneToNb}$ gives 
\begin{multline}
	\int {\mathcal{J}}^\OToNa_{Q} {\phi}_{\OToNa} d^7 {\xi} = \frac{1}{2} \int_{{\mathbb{R}}} {\dot{\Map}}^\OToNa \left( {U}^{\OneToNb \OneToNc} - \dot{\Map}^\OneToNb U^{\OneToNc 0} - \dot{\Map}^\OneToNc U^{\OneToNb 0} + \dot{\Map}^\OneToNb \dot{\Map}^\OneToNc U^{00} \right) \partial_\OneToNb \partial_\OneToNc \phi_{\OToNa} dt \\ 
	+ \int_{{{\mathbb{R}}}} \left( Y^{\OToNa \OneToNb} - {Y}^{\OToNa 0} \dot{\Map}^\OneToNb - \dot{\Map}^\OToNa \frac{d}{dt} \left(U^{\OneToNb 0} \right) + \frac{1}{2} {\dot{\Map}}^\OToNa \dot{\Map}^\OneToNb \frac{d}{dt} \left( {U}^{0 0} \right) \right) \partial_\OneToNb \phi_{\OToNa} dt 
	+ \int_{{{\mathbb{R}}}} \left(-U^{\OneToNb 0} \frac{d}{dt} \left( \dot{\Map}^\OToNa \right) + \frac{1}{2} {U}^{0 0} \frac{d}{dt} \left( {\dot{\Map}}^\OToNa \dot{\Map}^\OneToNb \right) \right) \partial_\OneToNb \phi_{\OToNa} dt \\
	+ \frac{1}{2} \int_{{\mathbb{R}}} \frac{d^2}{dt^2} \left( {\dot{\Map}}^\OToNa {U}^{0 0} \right) \phi_{\OToNa} dt + \frac{1}{2} \int_{{\mathbb{R}}} \frac{d}{dt} \left( {\dot{\Map}}^\OToNa {U}^{0 0} \right) \dot{\Map}^{\OneToNb} \partial_\OneToNb \phi_{\OToNa} dt - \int_{{{\mathbb{R}}}} \frac{d}{dt} {Y}^{\OToNa 0} \phi_{\OToNa} dt \\
	+ \int_{\mathbb{R}} \dot{\Map}^\OToNa \left(U^{\OneToNb} - U^0 \dot{\Map}^{\OneToNb} \right) \partial_{\OneToNb} \phi_{\OToNa} dt
	+ \int_{\mathbb{R}} \left( Y^{\OneToNa} - \frac{d}{dt} \left(\dot{\Map}^\OToNa U^{0} \right) \right) \phi_\OneToNa dt + \int_{\mathbb{R}} \dot{\Map}^\OToNa q \phi_\OToNa dt.
\end{multline}
Noting $dU^{\OToNa \OToNb}/dt = Y^{\OToNa \OToNb} + Y^{\OToNb \OToNa}$ gives	
\begin{multline}
	\int {\mathcal{J}}^\OToNa_{Q} {\phi}_{\OToNa} d^7 {\xi} = \frac{1}{2} \int_{{\mathbb{R}}} {\dot{\Map}}^\OToNa \left( {U}^{\OneToNb \OneToNc} - \dot{\Map}^\OneToNb U^{\OneToNc 0} - \dot{\Map}^\OneToNc U^{\OneToNb 0} + \dot{\Map}^\OneToNb \dot{\Map}^\OneToNc U^{00} \right) \partial_\OneToNb \partial_\OneToNc \phi_{\OToNa} dt \\
	+ \int_{{{\mathbb{R}}}} \left( Y^{\OToNa \OneToNb} - {Y}^{\OToNa 0} \dot{\Map}^\OneToNb - \dot{\Map}^\OToNa Y^{\OneToNb 0} - \dot{\Map}^\OToNa Y^{0 \OneToNb} + {\dot{\Map}}^\OToNa \dot{\Map}^\OneToNb Y^{00} \right) \partial_\OneToNb \phi_{\OToNa} dt 
	+ \int_{{{\mathbb{R}}}} \left(\frac{1}{2} {U}^{0 0} \frac{d}{dt} \left( {\dot{\Map}}^\OToNa \dot{\Map}^\OneToNb \right) -U^{\OneToNb 0} \frac{d}{dt} \left( \dot{\Map}^\OToNa \right) \right) \partial_\OneToNb \phi_{\OToNa} dt \\
	+ \frac{1}{2} \int_{{\mathbb{R}}} \frac{d^2}{dt^2} \left( {\dot{\Map}}^\OToNa {U}^{0 0} \right) \phi_{\OToNa} dt + \frac{1}{2} \int_{{\mathbb{R}}} \frac{d}{dt} \left( {\dot{\Map}}^\OToNa {U}^{0 0} \right) \dot{\Map}^{\OneToNb} \partial_\OneToNb \phi_{\OToNa} dt 
	- \int_{{{\mathbb{R}}}} \frac{d}{dt} {Y}^{\OToNa 0} \phi_{\OToNa} dt \\
	+ \int_{\mathbb{R}} \dot{\Map}^\OToNa \left(U^{\OneToNb} - U^0 \dot{\Map}^{\OneToNb} \right) \partial_{\OneToNb} \phi_{\OToNa} dt
	+ \int_{\mathbb{R}} \left( Y^{\OneToNa} - \frac{d}{dt} \left(\dot{\Map}^\OToNa U^{0} \right) \right) \phi_\OneToNa dt + \int_{\mathbb{R}} \dot{\Map}^\OToNa q \phi_\OToNa dt.
\end{multline}
Rearranging for clarity,
\begin{multline}
	\int {\mathcal{J}}^\OToNa_{Q} {\phi}_{\OToNa} d^7 {\xi} = \frac{1}{2} \int_{{\mathbb{R}}} {\dot{\Map}}^\OToNa \left( {U}^{\OneToNb \OneToNc} - \dot{\Map}^\OneToNb U^{\OneToNc 0} - \dot{\Map}^\OneToNc U^{\OneToNb 0} + \dot{\Map}^\OneToNb \dot{\Map}^\OneToNc U^{00} \right) \partial_\OneToNb \partial_\OneToNc \phi_{\OToNa} dt 
	+ \int_{{{\mathbb{R}}}} \left( Y^{\OToNa \OneToNb} - {Y}^{\OToNa 0} \dot{\Map}^\OneToNb - \dot{\Map}^\OToNa Y^{0 \OneToNb} + {\dot{\Map}}^\OToNa \dot{\Map}^\OneToNb Y^{00} \right) \partial_\OneToNb \phi_{\OToNa} dt \\
	+ \int_{{{\mathbb{R}}}} \left({U}^{0 0} \dot{\Map}^\OneToNb \frac{d}{dt} \left( {\dot{\Map}}^\OToNa \right) - U^{\OneToNb 0} \frac{d}{dt} \left( \dot{\Map}^\OToNa \right) \right) \partial_\OneToNb \phi_{\OToNa} dt 
	+\int_{{\mathbb{R}}} \dot{\Map}^\OToNa \left( U^{\OneToNb} - U^0 \dot{\Map}^{\OneToNb} + \frac{1}{2} \frac{d}{dt} \left(\dot{\Map}^{\OneToNb} {U}^{0 0} \right) - Y^{\OneToNb 0} \right) \partial_\OneToNb \phi_{\OToNa} dt \\ 
	+ \int_{{\mathbb{R}}} \left( Y^{\OneToNa} - \frac{d}{dt} \left(\dot{\Map}^\OToNa U^{0} \right) + \frac{1}{2} \frac{d^2}{dt^2} \left( {\dot{\Map}}^\OToNa {U}^{0 0} \right) - \frac{d}{dt} {Y}^{\OToNa 0} \right) \phi_{\OToNa} dt.
\end{multline}			
By noting $\dot{\Map}^0 = 1$ the $Y^{0\OneToNa}, Y^{\OneToNa 0}, Y^{00}$ and $Y^{0}$ components vanish. Thus the projected moments are given by
\begin{equation}
	\begin{gathered}
		\hat{V}^{\OneToNb \OneToNc} = {U}^{\OneToNb \OneToNc} - \dot{\Map}^\OneToNb U^{\OneToNc 0} - \dot{\Map}^\OneToNc U^{\OneToNb 0} + \dot{\Map}^\OneToNb \dot{\Map}^\OneToNc U^{00}, \\
		\hat{X}^{\OneToNa \OneToNb} = \begin{multlined}[t] Y^{\OneToNa \OneToNb} - {Y}^{\OneToNa 0} \dot{\Map}^\OneToNb - \dot{\Map}^\OneToNa Y^{0 \OneToNb} + {\dot{\Map}}^\OneToNa \dot{\Map}^\OneToNb Y^{00} + {U}^{0 0} \dot{\Map}^\OneToNb \frac{d}{dt} \left( {\dot{\Map}}^\OneToNa \right) - U^{\OneToNb 0} \frac{d}{dt} \left( \dot{\Map}^\OneToNa \right), \end{multlined}\\
		\hat{V}^\OneToNb = U^{\OneToNb} - U^0 \dot{\Map}^{\OneToNb} + \frac{1}{2} \frac{d}{dt} \left(\dot{\Map}^{\OneToNb} {U}^{0 0} \right) - Y^{\OneToNb 0}, \\
		\hat{X}^\OneToNa = Y^{\OneToNa} - \frac{d}{dt} \left(\dot{\Map}^\OneToNa U^{0} \right) + \frac{1}{2} \frac{d^2}{dt^2} \left( {\dot{\Map}}^\OneToNa {U}^{0 0} \right) - \frac{d}{dt} {Y}^{\OneToNa 0}.
 \label{eq:FullEllisProjection}
	\end{gathered}
\end{equation}
Combining this with \eqref{eq:EllisCoordTransforms} give the full coordinate transformations for the quadrupole.	

\subsection{De Rham current representation of the coordinate transformation}
\label{sec:DistributionalCoordTransform}

This section finds the first step of the coordinate transformation for the de Rham current representation of a quadrupole, obtaining non-unique components in terms of $U^\OToNa, U^{\OToNa \OToNb}, Y^\OToNa$ and $Y^{\OToNa\OToNb}$. \\ 

Consider a unique distributional quadrupole,
\begin{equation}
	\Psi_Q = \tfrac{1}{2} L^{(\xi)}_\OneToNa L^{(\xi)}_\OneToNb \Map_\varsigma(V^{\OneToNa \OneToNb}) - L^{(\xi)}_\OneToNa i^{(\xi)}_\OneToNb \Map_\varsigma(X^{\OneToNa \OneToNb} dt) \\ - L^{(\xi)}_\OneToNa \Map_\varsigma(V^\OneToNa) + i^{(\xi)}_\OneToNa \Map_\varsigma(X^\OneToNa dt) + \Map_\varsigma(q).
\end{equation}
This is transformed by noting
\begin{equation}
	\frac{\partial}{\partial \xi^\OneToNa} = \frac{\partial \hat{\xi}^\OToNb}{\partial \xi^{\OneToNa}} \frac{\partial}{\partial \hat{\xi}^\OToNb}
\end{equation}
so
\begin{equation}
	i^{(\xi)}_{\OneToNa} = i^{(\hat{\xi})}_\OToNb \frac{\partial \hat{\xi}^\OToNb}{\partial \xi^\OneToNa}.
\end{equation}
Using Cartan's identity the Lie derivative transformation can be found,
\begin{align}
	L_\OneToNa(\alpha) &= d i_\OneToNa \alpha + i_\OneToNa d \alpha \\
	&= d i_{\hat{\OToNb}} (A^{\hat{\OToNb}}_{\OneToNa} \wedge \alpha) + i_{\hat{\OToNb}} (A^{\hat{\OToNb}}_{\OneToNa} \wedge d \alpha) \\
	&= d i_{\hat{\OToNb}} (A^{\hat{\OToNb}}_{\OneToNa} \wedge \alpha) + i_{\hat{\OToNb}} d \left( A^{\hat{\OToNb}}_{\OneToNa} \wedge \alpha \right) - i_{\hat{\OToNb}} \left( dA^{\hat{\OToNb}}_{\OneToNa} \wedge \alpha \right) \\
	&= L_{\hat{\OToNb}} \left( A^{\hat{\OToNb}}_{\OneToNa} \wedge \alpha \right) - i_{\hat{\OToNb}} \left(dA^{\hat{\OToNb}}_{\OneToNa} \wedge \alpha \right)
\end{align}
For the double Lie derivative term, use the standard differential geometry result
\begin{equation}
	L_U (fg) = f L_U g + g L_U f \quad \forall f,g \in \Gamma \Lambda^0 E, U \in \Gamma T E
\end{equation}
to find
\begin{align} 
	L_\OneToNa L_\OneToNb \alpha &= L_\OneToNa \left( L_{\hat{\OToNd}} (A^{\hat{\OToNd}}_\OneToNb \wedge \alpha) - i_{\hat{\OToNd}} (d A^{\hat{\OToNd}}_\OneToNb \wedge \alpha)\right) \\
	&= \begin{multlined}[t] L_{\hat{\OToNc}} \left( A^{\hat{\OToNc}}_\OneToNa \wedge \left( L_{\hat{\OToNd}} (A^{\hat{\OToNd}}_\OneToNb \wedge \alpha) - i_{\hat{\OToNd}} (d A^{\hat{\OToNd}}_\OneToNb \wedge \alpha)\right) \right) - i_{\hat{\OToNc}} \left(d A^{\hat{\OToNc}}_\OneToNa \wedge \left( L_{\hat{\OToNd}} (A^{\hat{\OToNd}}_\OneToNb \wedge \alpha) - i_{\hat{\OToNd}} (d A^{\hat{\OToNd}}_\OneToNb \wedge \alpha)\right) \right) \end{multlined} \\
	&= \begin{multlined}[t] L_{\hat{\OToNc}} L_{\hat{\OToNd}} \left( A^{\hat{\OToNc}}_\OneToNa A^{\hat{\OToNd}}_\OneToNb \wedge \alpha \right) - L_{\hat{\OToNc}} \left( i_{\hat{\OToNd}} (d A^{\hat{\OToNc}}_\OneToNa) \wedge A^{\hat{\OToNd}}_\OneToNb \wedge \alpha \right) - L_{\hat{\OToNc}} i_{\hat{\OToNd}} (A^{\hat{\OToNc}}_\OneToNa \wedge d A^{\hat{\OToNd}}_\OneToNb \wedge \alpha)
	- i_{\hat{\OToNc}} L_{\hat{\OToNd}} \left(d A^{\hat{\OToNc}}_\OneToNa \wedge A^{\hat{\OToNd}}_\OneToNb \wedge \alpha \right) \\ + i_{\hat{\OToNc}} \left( (d i_{\hat{\OToNd}} d A^{\hat{\OToNc}}_\OneToNa) \wedge A^{\hat{\OToNd}}_\OneToNb \wedge \alpha \right) - i_{\hat{\OToNc}} i_{\hat{\OToNd}} \left( d A^{\hat{\OToNc}}_\OneToNa \wedge dA^{\hat{\OToNd}}_\OneToNb \wedge \alpha \right) + i_{\hat{\OToNc}} \left( i_{\hat{\OToNd}} dA^{\hat{\OToNc}}_\OneToNa \wedge dA^{\hat{\OToNd}}_\OneToNb \wedge \alpha \right) \end{multlined} \\
	&= \begin{multlined}[t] L_{\hat{\OToNc}} L_{\hat{\OToNd}} \left( A^{\hat{\OToNc}}_\OneToNa A^{\hat{\OToNd}}_\OneToNb \wedge \alpha \right) - L_{\hat{\OToNc}} i_{\hat{\OToNd}} \left( d\left(A^{\hat{\OToNc}}_\OneToNa A^{\hat{\OToNd}}_\OneToNb \right) \wedge \alpha \right) 
	- L_{\hat{\OToNc}} \left( \partial_{\hat{\OToNd}} A^{\hat{\OToNc}}_\OneToNa \wedge A^{\hat{\OToNd}}_\OneToNb \wedge \alpha \right)
	 \\+ i_{\hat{\OToNc}} \left( d \left(\partial_{\hat{\OToNd}} A^{\hat{\OToNc}}_\OneToNa \wedge A^{\hat{\OToNd}}_\OneToNb \right) \wedge \alpha \right) - i_{\hat{\OToNc}} i_{\hat{\OToNd}} \left( d A^{\hat{\OToNc}}_\OneToNa \wedge dA^{\hat{\OToNd}}_\OneToNb \wedge \alpha \right). \end{multlined}
\end{align}
Using $\partial_{\hat{\OToNd}} A^{\hat{\OToNc}}_{\OneToNa} = A^{\OToNe}_{\hat{\OToNd}} A^{\hat{\OToNc}}_{\OneToNa \OneToNb}$, and noting the $i_{\hat{\OToNc}} i_{\hat{\OToNd}}$ term vanishes as the term inside the brackets is symmetric, gives
\begin{equation}
	L_\OneToNa L_\OneToNb \alpha = L_{\hat{\OToNc}} L_{\hat{\OToNd}} \left( A^{\hat{\OToNc}}_\OneToNa A^{\hat{\OToNd}}_\OneToNb \wedge \alpha \right) - L_{\hat{\OToNc}} i_{\hat{\OToNd}} \left( d\left(A^{\hat{\OToNc}}_\OneToNa A^{\hat{\OToNd}}_\OneToNb \right) \wedge \alpha \right) \\
	- L_{\hat{\OToNc}} \left( A^{\hat{\OToNc}}_{\OneToNa \OneToNb} \wedge \alpha \right) + i_{\hat{\OToNc}} \left( dA^{\hat{\OToNc}}_{\OneToNa \OneToNb} \wedge \alpha \right).
\end{equation}
This gives the coordinate transformation for the $V^{\OneToNa \OneToNb}$ term,
\begin{equation}
		L_\OneToNa L_\OneToNb \Map_{\varsigma}(V^{\OneToNa \OneToNb}) = L_{\hat{\OToNc}} L_{\hat{\OToNd}} \Map_{\varsigma} \left( A^{\hat{\OToNc}}_\OneToNa A^{\hat{\OToNd}}_\OneToNb V^{\OneToNa \OneToNb} \right) \\ - L_{\hat{\OToNc}} i_{\hat{\OToNd}} \Map_{\varsigma} \left( d\left(A^{\hat{\OToNc}}_\OneToNa A^{\hat{\OToNd}}_\OneToNb \right) \wedge V^{\OneToNa \OneToNb} \right) 
		- L_{\hat{\OToNc}} \Map_{\varsigma} \left( A^{\hat{\OToNc}}_{\OneToNa \OneToNb} V^{\OneToNa \OneToNb} \right) \\ + i_{\hat{\OToNc}} \Map_{\varsigma} \left( dA^{\hat{\OToNc}}_{\OneToNa \OneToNb} \wedge V^{\OneToNa \OneToNb} \right).
\end{equation}
For the $X^{\OneToNa \OneToNb}$ term,
\begin{align}
	 i_\OneToNa L_\OneToNb \Map_\varsigma(X^{\OneToNa \OneToNb} dt) &= \begin{multlined}[t] i_\OneToNa L_{\hat{\OToNc}}\Map_\varsigma(A^{\hat{\OToNc}}_{\OneToNb} \wedge X^{\OneToNa \OneToNb} dt) - i_{\OneToNa} i_{\hat{\OToNc}} \left( \Map_{\varsigma}(dA^{\hat{\OToNc}}_{\OneToNb} \wedge X^{\OneToNa \OneToNb} dt) \right) \end{multlined} \\
	 &= i_{\hat{\OToNd}} A^{\hat{\OToNd}}_{\OneToNa} L_{\hat{\OToNc}} \Map_\varsigma(A^{\hat{\OToNc}}_{\OneToNb} \wedge X^{\OneToNa \OneToNb} dt) \\
	 &= \begin{multlined}[t]
	i_{\hat{\OToNd}} L_{\hat{\OToNc}} \Map_\varsigma( A^{\hat{\OToNd}}_{\OneToNa} A^{\hat{\OToNc}}_{\OneToNb} \wedge X^{\OneToNa \OneToNb} dt) - i_{\hat{\OToNd}} \Map_\varsigma \Big( L_{\hat{\OToNc}}(A^{\hat{\OToNd}}_{\OneToNa}) \wedge A^{\hat{\OToNc}}_{\OneToNb} \wedge X^{\OneToNa \OneToNb} dt \Big)   \end{multlined} \\
	 &= \begin{multlined}[t] i_{\hat{\OToNd}} L_{\hat{\OToNc}} \Map_\varsigma( A^{\hat{\OToNd}}_{\OneToNa} A^{\hat{\OToNc}}_{\OneToNb} \wedge X^{\OneToNa \OneToNb} dt) - i_{\hat{\OToNd}} \Map_\varsigma( A^{\hat{\OToNd}}_{\OneToNa \OneToNb} \wedge X^{\OneToNa \OneToNb} dt) \end{multlined} \\
	 &= \begin{multlined}[t]
	i_{\hat{\OToNd}} L_{\hat{\OToNc}} \Map_\varsigma \left( A^{\hat{\OToNd}}_{\OneToNa} A^{\hat{\OToNc}}_{\OneToNb} \wedge X^{\OneToNa \OneToNb} \frac{dt}{d\hat{t}} d\hat{t} \right) - i_{\hat{\OToNd}} \Map_\varsigma \left( A^{\hat{\OToNd}}_{\OneToNa \OneToNb} \wedge X^{\OneToNa \OneToNb} \frac{dt}{d\hat{t}} d\hat{t} \right)   
	 \end{multlined} 	 
\end{align}
where $dA^{\hat{\OToNc}}_{\OneToNb} \wedge X^{\OneToNa \OneToNb}dt = 0$ is used. The $V^{\OneToNa}$ and $X^{\OneToNa}$ terms are more straightforward,
\begin{gather}
	L_{\OneToNa} \Map_{\varsigma} (V^{\OneToNa}) = L_{\hat{\OToNb}} \Map_{\varsigma} \left( A^{\hat{\OToNb}}_{\OneToNa} \wedge V^{\OneToNa} \right) - i_{\hat{\OToNb}} \Map_{\varsigma} \left(dA^{\hat{\OToNb}}_{\OneToNa} \wedge V^{\OneToNa} \right) \\
	i_{\OneToNa} \Map_{\varsigma} (X^{\OneToNa} dt) = i_{\hat{\OToNb}} \Map_{\varsigma} \left( A^{\hat{\OToNb}}_{\OneToNa} X^{\OneToNa} \frac{dt}{d\hat{t}} d\hat{t} \right).
\end{gather}
Summing these together (noting that the monopole term is invariant under transformation), gives
\begin{multline} \mathcal{J} = \hat{\mathcal{J}} = \frac{1}{2} L_\OneToNa L_\OneToNb \Map_\varsigma (V^{\OneToNa \OneToNb}) - i_\OneToNa L_\OneToNb \Map_\varsigma (X^{\OneToNa \OneToNb} dt) - L_\OneToNa \Map_\varsigma (V^{\OneToNa} ) + i_\OneToNa \Map_\varsigma(X^\OneToNa dt) + \Map_\varsigma(q) \\
= \frac{1}{2} L_{\hat{\OToNc}} L_{\hat{\OToNd}} \Map_{\varsigma} \left( A^{\hat{\OToNc}}_\OneToNa A^{\hat{\OToNd}}_\OneToNb V^{\OneToNa \OneToNb} \right) - i_{\hat{\OToNd}} L_{\hat{\OToNc}} \Map_{\varsigma} \left( A^{\hat{\OToNd}}_{\OneToNa} A^{\hat{\OToNc}}_{\OneToNb} X^{\OneToNa \OneToNb} \frac{dt}{d\hat{t}} d\hat{t} + \frac{1}{2} d \left(A^{\hat{\OToNc}}_\OneToNa A^{\hat{\OToNd}}_\OneToNb \right) V^{\OneToNa \OneToNb} \right) 
- L_{\hat{\OToNc}} \Map_{\varsigma} \left(A^{\hat{\OToNc}}_{\OneToNa} V^{\OneToNa} + \frac{1}{2} A^{\hat{\OToNc}}_{\OneToNa \OneToNb} V^{\OneToNa \OneToNb} \right) \\
+ i_{\hat{\OToNc}} \Map_{\varsigma} \left( A^{\hat{\OToNc}}_{\OneToNa} X^{\OneToNa} \frac{dt}{d\hat{t}} d\hat{t} + (d(A^{\hat{\OToNc}}_{\OneToNa}) V^{\OneToNa} + d(A^{\hat{\OToNc}}_{\OneToNa \OneToNb} V^{\OneToNa \OneToNb}) + A^{\hat{\OToNc}}_{\OneToNa \OneToNb} X^{\OneToNa \OneToNb}) \frac{dt}{d\hat{t}} d\hat{t} \right) + \Map_\varsigma(q).
\end{multline}
By taking the external derivatives, this is equivalent to \eqref{eq:EllisCoordTransforms}.

\subsection{De Rham current representation of the projections}
\label{sec:DistributionalProjections}

This section performs the projections $L_0 = L_{\dot{\Map}} - \dot{\Map}^\OneToNa L_\OneToNa$ and $i_0 = i_{\dot{\Map}} - \dot{\Map}^\OneToNa i_\OneToNa$ into the de Rham current representation of the quadrupole with non-unqiue components. By performing this projection $U^\OToNa, U^{\OToNa \OToNb}, Y^{\OToNa}$ and $Y^{\OToNa \OToNb}$ can be written in a form where the components are unique, finding the full coordinate transformation for the quadrupole. \\

The non-unique quadrupole is given by
\begin{equation}
  \mathcal{J} = \tfrac{1}{2} L_\OToNa L_\OToNb \Map_\varsigma (U^{\OToNa \OToNb}) - i_{\OToNa} L_{\OToNb} \Map_\varsigma (Y^{\OToNa \OToNb} d t) - L_\OToNa \Map_\varsigma (U^\OToNa) + i_\OToNa \Map_\varsigma (Y^\OToNa dt) + \Map_\varsigma (q).
\end{equation}
To find the projections, proceed term by term, beginning with the $L_\OToNa L_\OToNb$ term,
\begin{align}
	L_\OToNa L_\OToNb \Map_\varsigma (U^{\OToNa \OToNb}) &= L_\OToNa L_\OneToNb \Map_\varsigma (U^{\OToNa \OneToNb}) + L_\OToNa L_0 (U^{\OToNa 0}) \\
	&= L_\OToNa L_\OneToNb \Map_\varsigma (U^{\OToNa \OneToNb}) + L_\OToNa L_{\dot{\Map}} (U^{\OToNa 0}) - L_\OToNa \dot{\Map}^\OneToNb L_\OneToNb \Map_\varsigma(U^{\OToNa 0}) \\
	&= L_\OToNa L_\OneToNb \Map_\varsigma (U^{\OToNa \OneToNb} - \dot{\Map}^\OneToNb U^{\OToNa 0}) + L_\OToNa \Map_\varsigma \left(\frac{d{U}^{\OToNa 0}}{dt} \right) + L_\OToNa i_\OneToNb \Map_{\varsigma} \left(U^{\OToNa 0} d\dot{\Map}^\OneToNb \right) \\
	&= \begin{multlined}[t] 
	L_{\OneToNa} L_\OneToNb \Map_\varsigma (U^{\OneToNa \OneToNb} - \dot{\Map}^\OneToNb U^{\OneToNa 0})
		+ L_\OneToNb L_{\dot{\Map}} \Map_\varsigma (U^{0 \OneToNb} - \dot{\Map}^\OneToNb U^{0 0}) 
		- L_\OneToNb \dot{\Map}^\OneToNa L_\OneToNa \Map_\varsigma (U^{0 \OneToNb} - \dot{\Map}^\OneToNb U^{0 0})
		+ L_\OneToNa \Map_\varsigma \left(\frac{d{U}^{\OneToNa 0}}{dt} \right) \\
		+ L_{\dot{\Map}} \Map_\varsigma \left(\frac{d{U}^{0 0}}{dt} \right) - \dot{\Map}^\OneToNa L_\OneToNa \Map_\varsigma \left(\frac{d{U}^{0 0}}{dt} \right)
		+ L_\OneToNa i_\OneToNb \Map_\varsigma \left(U^{\OneToNa 0} d\dot{\Map}^\OneToNb \right) \\
  + i_\OneToNb L_{\dot{\Map}} \Map_\varsigma \left(U^{0 0} d \dot{\Map}^\OneToNb \right) - i_\OneToNb \dot{\Map}^\OneToNa L_\OneToNa \Map_\varsigma \left(U^{0 0} d \dot{\Map}^\OneToNb \right) \end{multlined} \\
	&= \begin{multlined}[t] 
		L_{\OneToNa} L_\OneToNb \Map_\varsigma \left(U^{\OneToNa \OneToNb} - \dot{\Map}^\OneToNb U^{\OneToNa 0} - \dot{\Map}^\OneToNa U^{\OneToNb 0} + \dot{\Map}^\OneToNa \dot{\Map}^\OneToNb U^{00} \right)
		+ L_\OneToNb i_\OneToNa \Map_\varsigma \left((2U^{0 \OneToNb} - 2\dot{\Map}^\OneToNb U^{0 0}) d\dot{\Map}^\OneToNa \right) \\
		+ L_\OneToNa \Map_\varsigma \left(2 \frac{d{U}^{\OneToNa 0}}{dt} - \dot{\Map}^\OneToNa \frac{d U^{00}}{dt} - \frac{d}{dt} \left( \dot{\Map}^\OneToNa U^{0 0} \right) \right) 
		+ i_\OneToNa \Map_\varsigma \left(\frac{d{U}^{0 0}}{dt} d\dot{\Map}^\OneToNa + \frac{d}{dt} \left(U^{0 0} d \dot{\Map}^\OneToNa \right) \right) +
		\Map_\varsigma \left(\frac{d^2{U}^{0 0}}{dt^2} \right) 
		\\ + i_\OneToNb i_\OneToNa \Map_\varsigma \left(U^{0 0} d \dot{\Map}^\OneToNa \wedge d \dot{\Map}^\OneToNb \right) \end{multlined}
\end{align}
and note the last term vanishes as $\mathbb{R}$ is only 1-dimensional. For the $L_\OToNa i_\OToNb$ term,
\begin{align}
	L_\OToNb i_\OToNa \Map_\varsigma(Y^{\OToNa \OToNb} dt) &= L_{\OToNb} i_\OneToNa \Map_{\varsigma}(Y^{\OneToNa \OToNb} dt) + L_\OToNa i_{\dot{\Map}} \Map_\varsigma(Y^{0 \OToNb} dt) - L_{\OToNa} \dot{\Map}^\OneToNb i_{\OneToNb} \Map_{\varsigma} (Y^{0 \OToNb} dt) \\
	&= L_{\OToNb} i_\OneToNa \Map_{\varsigma}(Y^{\OneToNa \OToNb} dt - \dot{\Map}^\OneToNa Y^{0 \OToNb} dt) + L_\OToNb \Map_\varsigma(Y^{0 \OToNb}) \\
	&= L_{\OneToNb} i_\OneToNa \Map_{\varsigma}(Y^{\OneToNa \OneToNb} dt - \dot{\Map}^\OneToNa Y^{0 \OneToNb} dt) + L_{0} i_\OneToNa \Map_{\varsigma}(Y^{\OneToNa 0} dt - \dot{\Map}^\OneToNa Y^{0 0} dt) + L_\OneToNb \Map_\varsigma(Y^{0 \OneToNb}) + L_0 \Map_\varsigma(Y^{0 0}) \\
	&= 	\begin{multlined}[t] L_{\OneToNb} i_\OneToNa \Map_{\varsigma}(Y^{\OneToNa \OneToNb} dt - \dot{\Map}^\OneToNa Y^{0 \OneToNb} dt) + L_{\dot{\Map}} i_\OneToNa \Map_{\varsigma}(Y^{\OneToNa 0} dt - \dot{\Map}^\OneToNa Y^{0 0} dt) -\dot{\Map}^{\OneToNb} L_{\OneToNb} i_\OneToNa \Map_{\varsigma}(Y^{\OneToNa 0} dt - \dot{\Map}^\OneToNa Y^{0 0} dt) \\ + L_\OneToNb \Map_\varsigma(Y^{0 \OneToNb}) + L_{\dot{\Map}} \Map_\varsigma(Y^{0 0}) - \dot{\Map}^\OneToNa L_{\OneToNa} \Map_\varsigma(Y^{0 0}) \end{multlined} \\
	&= \begin{multlined}[t] L_{\OneToNb} i_\OneToNa \Map_{\varsigma}(Y^{\OneToNa \OneToNb} dt - \dot{\Map}^\OneToNa Y^{0 \OneToNb} dt) + L_{\dot{\Map}} i_\OneToNa \Map_{\varsigma}(Y^{\OneToNa 0} dt - \dot{\Map}^\OneToNa Y^{0 0} dt) \\ - L_{\OneToNb} \dot{\Map}^{\OneToNb} i_\OneToNa \Map_{\varsigma}(Y^{\OneToNa 0} dt - \dot{\Map}^\OneToNa Y^{0 0} dt) + L_\OneToNb \Map_\varsigma(Y^{0 \OneToNb}) 
	+ L_{\dot{\Map}} \Map_\varsigma(Y^{0 0}) - L_{\OneToNa} \dot{\Map}^\OneToNa \Map_\varsigma(Y^{0 0})
	\\ + i_\OneToNa \Map_{\varsigma}(Y^{00} d\dot{\Map}^\OneToNa) + i_\OneToNb i_\OneToNa \Map_{\varsigma} \left( d\dot{\Map}^\OneToNb \wedge( Y^{\OneToNa 0} - \dot{\Map} Y^{00}) dt \right) \end{multlined} \\
	&= \begin{multlined}[t] L_{\OneToNb} i_\OneToNa \Map_{\varsigma}(Y^{\OneToNa \OneToNb} dt - \dot{\Map}^\OneToNa Y^{0 \OneToNb} dt - \dot{\Map}^{\OneToNb} Y^{\OneToNa 0} dt + \dot{\Map}^\OneToNa \dot{\Map}^{\OneToNb} Y^{0 0} dt) + L_\OneToNb \Map_\varsigma(Y^{0 \OneToNb} - \dot{\Map}^{\OneToNb} Y^{00}) \\ + i_\OneToNa \Map_{\varsigma} \left(\frac{dY^{\OneToNa 0}}{dt} dt - \frac{d}{dt} \left( \dot{\Map}^\OneToNa Y^{0 0} \right) dt + Y^{00} d\dot{\Map}^\OneToNa \right) + \Map_\varsigma \left(\frac{dY^{0 0}}{dt} \right). \end{multlined}
\end{align}
The $U^{\OToNa}$ term,
\begin{align}
	L_\OToNa \Map_{\varsigma} (U^{\OToNa}) &= L_\OneToNa \Map_{\varsigma} (U^{\OneToNa}) + L_{\dot{\Map}} \Map_\varsigma (U^{0}) - \dot{\Map}^\OneToNa L_{\OneToNa} \Map_{\varsigma} (U^0) \\
	&= \begin{multlined}[t] L_\OneToNa \Map_{\varsigma} (U^{\OneToNa} - \dot{\Map}^\OneToNa U^0) + i_\OneToNa \Map_{\varsigma} (U^0 d \dot{\Map}^\OneToNa) + \Map_\varsigma \left(\frac{dU^{0}}{dt} \right). \end{multlined}
\end{align}
Lastly the $Y^{\OToNa}$ term,
\begin{align}
	i_\OToNa \Map_{\varsigma} (Y^{\OToNa} dt) &= i_\OneToNa \Map_{\varsigma} (Y^{\OneToNa} dt) + i_{\dot{\Map}} \Map_\varsigma (Y^{0} dt) - i_{\OneToNa} \dot{\Map}^\OneToNa \Map_{\varsigma} (Y^0 dt) \\
	&= i_\OneToNa \Map_{\varsigma} (Y^{\OneToNa} dt - \dot{\Map}^\OneToNa Y^0 dt) + \Map_\varsigma (Y^0).
\end{align}
Summing these together,
\begin{multline}
	\Psi_Q = \frac{1}{2} L_{\OneToNa} L_\OneToNb \Map_\varsigma \left(U^{\OneToNa \OneToNb} - \dot{\Map}^\OneToNb U^{\OneToNa 0} - \dot{\Map}^\OneToNa U^{\OneToNb 0} + \dot{\Map}^\OneToNa \dot{\Map}^\OneToNb U^{00} \right)
	\\ - L_{\OneToNb} i_\OneToNa \Map_{\varsigma}\Big(Y^{\OneToNa \OneToNb} dt - \dot{\Map}^\OneToNa Y^{0 \OneToNb} dt - \dot{\Map}^{\OneToNb} Y^{\OneToNa 0} dt + \dot{\Map}^\OneToNa \dot{\Map}^{\OneToNb} Y^{0 0} dt - (U^{0 \OneToNb} - \dot{\Map}^\OneToNb U^{0 0}) d\dot{\Map}^\OneToNa \Big) \\
	- L_\OneToNa \Map_{\varsigma} (U^{\OneToNa} - \dot{\Map}^\OneToNa U^0 + Y^{0 \OneToNa} - \dot{\Map}^\OneToNa Y^{00}) 	
	- \frac{1}{2} L_\OneToNa \Map_\varsigma \left( \dot{\Map}^\OneToNa \frac{d U^{00}}{dt} + \frac{d}{dt} \left( \dot{\Map}^\OneToNa U^{0 0} \right) - 2 \frac{d{U}^{\OneToNa 0}}{dt} \right) \\
	+ i_\OneToNa \Map_{\varsigma} \left(Y^{\OneToNa} dt - \dot{\Map}^\OneToNa Y^0 dt - \frac{dY^{\OneToNa 0}}{dt} dt - U^{0} d\dot{\Map}^{\OneToNa} + \frac{d}{dt} \left( \dot{\Map}^\OneToNa Y^{0 0} \right) dt - Y^{00} d\dot{\Map}^\OneToNa \right) \\
	+ \frac{1}{2} i_\OneToNa \Map_\varsigma \left(\frac{d{U}^{0 0}}{dt} d\dot{\Map}^\OneToNa + \frac{d}{dt} \left(U^{0 0} d \dot{\Map}^\OneToNb \right) \right) +
	\frac{1}{2} \Map_\varsigma \left(\frac{d^2{U}^{0 0}}{dt^2} \right)
 	 - \Map_\varsigma \left(\frac{dY^{0 0}}{dt} \right)
 	- \Map_\varsigma \left(\frac{dU^{0}}{dt} \right) + \Map_\varsigma (Y^0) + \Map_\varsigma(q).
\end{multline}
To simplify this, recall \eqref{eq:EllisTransportSolutions}, giving
\begin{multline}
	\Psi_Q = \frac{1}{2} L_{\OneToNa} L_\OneToNb \Map_\varsigma \left(U^{\OneToNa \OneToNb} - \dot{\Map}^\OneToNb U^{\OneToNa 0} - \dot{\Map}^\OneToNa U^{\OneToNb 0} + \dot{\Map}^\OneToNa \dot{\Map}^\OneToNb U^{00} \right)
	\\ - L_{\OneToNb} i_\OneToNa \Map_{\varsigma}\Big(Y^{\OneToNa \OneToNb} dt - \dot{\Map}^\OneToNa Y^{0 \OneToNb} dt - \dot{\Map}^{\OneToNb} Y^{\OneToNa 0} dt + \dot{\Map}^\OneToNa \dot{\Map}^{\OneToNb} Y^{0 0} dt + (U^{0 \OneToNb} - \dot{\Map}^\OneToNb U^{0 0}) d\dot{\Map}^\OneToNa \Big) \\
	- L_\OneToNa \Map_{\varsigma} \left(U^{\OneToNa} - \dot{\Map}^\OneToNa U^0 + \frac{1}{2} \frac{d}{dt} \left( \dot{\Map}^\OneToNa U^{0 0} \right) - Y^{\OneToNa 0} \right) 
	+ i_\OneToNa \Map_{\varsigma} \left(Y^{\OneToNa} dt - \dot{\Map}^\OneToNa Y^0 dt - U^{0} d\dot{\Map}^{\OneToNa} - \frac{dY^{\OneToNa 0}}{dt} dt + \frac{1}{2} \frac{d^2}{dt^2} \left( \dot{\Map}^\OneToNa U^{0 0} \right) dt \right) \\ + \Map_\varsigma(q).
 \label{eq:GeometricFullProjection}
\end{multline}
This is the full coordinate transformation for the quadrupole in the de Rham current representation.
\end{widetext}

\end{document}